\documentclass[apj]{emulateapj}

\usepackage[usenames,dvipsnames]{color}
\usepackage[normalem]{ulem}
\usepackage{subfigure,amsmath}
\usepackage{xspace}
\usepackage{footnote}
\usepackage{flexisym}
\usepackage[colorlinks,urlcolor=blue,citecolor=black,linkcolor=blue]{hyperref}

\newcommand\msun{M\ensuremath{_{\odot}}\xspace}                   
\newcommand{\Msun}{\msun}                                         
\newcommand\semi{\ensuremath{\alpha_{\rm{sc}}}\xspace}                 
\newcommand\thermo{\ensuremath{\alpha_{\rm{th}}}\xspace}               
\newcommand\overshoot{\ensuremath{f_{\rm{ov}}}\xspace}            
\newcommand\mesh{\ensuremath{\delta_{\rm{mesh}}}\xspace}          
\newcommand\var{\ensuremath{w_{\rm{t}}}\xspace}                          
\newcommand{\rhos}{\ensuremath{r_{\mathrm{s}}}}                     
\newcommand{\Tc}{\ensuremath{T_{\mathrm{\!c}}}}                   
\newcommand{\rhoc}{\ensuremath{\rho_{\mathrm{c}}}}                
\newcommand{\net}[1]{{\tt mesa\_{#1}.net}}                        

\newcommand{\code}[1]{\texttt{#1}}
\newcommand{\mesa}{\code{MESA}\xspace}
\newcommand{\MESA}{\mesa}

\begin{document}

\title{Properties of Carbon-Oxygen White Dwarfs From Monte Carlo Stellar Models}

\author{
C.~E.~Fields\altaffilmark{1,2},
R.~Farmer\altaffilmark{1},
I.~Petermann\altaffilmark{1,2}, 
C.~Iliadis\altaffilmark{3,4}, and
F.~X.~Timmes\altaffilmark{1,2}
}

\altaffiltext{1}{School of Earth and Space Exploration, Arizona State University, Tempe, AZ}
\altaffiltext{2}{Joint Institute for Nuclear Astrophysics}
\altaffiltext{3}{University of North Carolina at Chapel Hill, Chapel Hill, NC 27599-3255, USA}
\altaffiltext{4}{Triangle Universities Nuclear Laboratory, Durham, North Carolina 27708-0308, USA}

\email{cef@asu.edu}
\slugcomment{Accepted for publication in The Astrophysical Journal, March 21, 2016}

\begin{abstract}
We investigate properties of carbon-oxygen white dwarfs with respect to 
the composite uncertainties in the reaction rates using the 
stellar evolution toolkit, Modules for Experiments in Stellar Astrophysics 
(\texttt{MESA}) and the probability density functions in the reaction
rate library STARLIB.
These are the first Monte Carlo stellar evolution studies that use
complete stellar models. Focusing on \mbox{3 M$_{\odot}$} models evolved 
from the pre main-sequence to the first thermal pulse, 
we survey the remnant core mass, composition, and structure properties as a 
function of 26 STARLIB reaction rates covering hydrogen and helium 
burning using a Principal Component Analysis and Spearman Rank-Order Correlation.
Relative to the arithmetic mean value, we find the width of the
95\% confidence interval to be
\mbox{$\Delta M_{{\rm 1TP}}$ $\approx$ 0.019 M$_{\odot}$} for the core mass
at the first thermal pulse, 
\mbox{$\Delta$$t_{\rm{1TP}}$ $\approx$ 12.50} Myr for the age,
$\Delta \log(T_{{\rm c}}/{\rm K}) \approx$ 0.013 for the central temperature,
$\Delta \log(\rho_{{\rm c}}/{\rm g \ cm}^{-3}) \approx$ 0.060 for the central density,
$\Delta Y_{\rm{e,c}} \approx$ 2.6$\times$10$^{-5}$ for the central electron fraction,
\mbox{$\Delta X_{\rm c}(^{22}\rm{Ne}) \approx$ 5.8$\times$10$^{-4}$}, 
\mbox{$\Delta X_{\rm c}(^{12}\rm{C}) \approx$ 0.392}, and
\mbox{$\Delta X_{\rm c}(^{16}\rm{O}) \approx$ 0.392}.
Uncertainties in the experimental $^{12}$C($\alpha,\gamma)^{16}\rm{O}$, triple-$\alpha$, 
and $^{14}$N($p,\gamma)^{15}\rm{O}$ 
reaction rates dominate these variations.
We also consider a grid of 1 to 6 M$_{\odot}$ 
models evolved from the pre main-sequence to the final white dwarf
to probe the sensitivity of  the initial-final mass relation to experimental uncertainties 
in the hydrogen and helium reaction rates. 

\end{abstract}

\keywords{stars: evolution --- stars: interiors --- stars: abundances --- supernovae: general}

\section{Introduction}
\label{sec:introduction}
Thermonuclear reaction rates are at the core of every stellar model.
Experimental rates are complex quantities derived from several nuclear
physics properties meticulously extracted from laboratory measurements
of resonance energies and strengths, non-resonant cross sections,
spectroscopic factors, and others
\citep{caughlan_1988_aa,angulo_1999_aa,iliadis_2001_aa,cyburt_2010_ab}.  Although
stellar reaction rate libraries champion experimental rates over
theoretical rates and cull their experimental data from the above
evaluations, all of the common libraries currently exclude an estimate
of the probability density function for each reaction rate.

However, the probability density functions are essential for defining
the ``theoretical error bars'' on the nuclear energy generation,
nucleosynthesis, and stellar structure profiles in models of stellar
phenomena.  In the past, compiling reaction rates without
uncertainties was standard practice in part because the additional
resources needed to explore the impacts of the reaction rate
uncertainties in stellar models was prohibitive and in part because it
was unclear how to proceed in evaluating a limited number of stellar
models in a statistically rigorous manner.  To produce realistic 
nucleosynthesis and stellar structure it is
necessary for stellar models to access a reaction rate library that
incorporates probability density functions for each reaction rate at a
given temperature \citep[e.g.,][]{iliadis_2015_aa}.

STARLIB is a tabular reaction rate library that provides the necessary
probability density functions \citep{sallaska_2013_aa}.  
All of the measured nuclear physics properties entering into a
reaction rate calculation are randomly sampled according to their
individual probability density functions. The sampling is repeated
many times and thus provides a Monte Carlo (MC) reaction rate probability
density. The associated cumulative distribution is used to derive
reaction rates and their uncertainties with a quantifiable coverage
probability \citep{longland_2010_aa,iliadis_2010_aa,iliadis_2010_ab}.
For example, for a coverage probability of 68\%, the low, recommended,
and high Monte Carlo based reaction rates can be defined as the 16th,
50th, and 84th percentiles, respectively, of the cumulative reaction
rate distribution.

Monte Carlo nucleosynthesis studies have been performed previously
\citep{iliadis_2002_aa,stoesz_2003_aa,roberts_2006_aa,parikh_2008_aa,parikh_2013_aa},
but did not use statistically meaningful rate probability density
functions derived from experimental nuclear physics input. Instead,
arbitrary and temperature independent multiplicative factors are assigned
to a reaction rate or a group of reaction rates
\citep[e.g.,][]{magkotsios_2010_aa}.  However, reaction
rate uncertainties display a strong temperature-dependence arising
from different resonance contributions and hence the temperature
dependence must be considered carefully in the reaction rate sampling
procedure \citep{longland_2012_aa}.

This paper is novel in three ways. One, we construct the first Monte Carlo
stellar evolution studies of a 3 \msun model evolved from the pre
main-sequence (pre-MS) to a carbon-oxygen white dwarf (CO WD).  
We focus on a solar metallicity 3 \msun model because they produce CO
WD masses near the \mbox{M $\approx$ 0.6 \msun} peak of the observed
DA and DB WD mass distributions
\citep{eisenstein_2006_aa,kepler_2007_aa,kepler_2015_aa,kepler_2016_aa}.
Two, we
quantify the variations in the WD mass, composition, and structure properties as a
function of 26 STARLIB reaction rates covering hydrogen (H) and helium (He)
burning using a Principal Component Analysis (PCA)
and Spearman Rank-Order Correlation (SRC).
Three, 
we investigate the evolution of 1 to 6 \msun 
zero-age main-sequence (ZAMS) 
models to
probe the sensitivity of the initial-final mass relation (IFMR)
to experimental uncertainties in the H and He nuclear reaction rates.
The IFMR maps the ZAMS mass 
to the WD remnant mass
\citep[e.g,][]{weidemann_2000_aa,dobbie_2006_aa,catalan_2008_aa,williams_2009_aa}.
The IFMR encapsulates the amount of material recycled back to the
interstellar medium
\citep[e.g.,][]{zhao_2012_aa,cummings_2016_aa,raddi_2016_aa} and provides 
constraints 
on star formation scenarios in galaxies
\citep{leitner_2011_aa,agertz_2015_aa},
exoplanet hosts
\citep{kilic_2009_aa},
ages of clusters
\citep{kalirai_2008_aa,kalirai_2009_aa},
age of the Galactic halo 
\citep{kalirai_2012_aa},
and supernova Type Ia rates
\citep{pritchet_2008_aa,greggio_2010_aa,kistler_2013_aa}.

In \S\ref{sec:method} we describe the input physics and model choices.
In \S\ref{sec:baseline} we describe the characteristics of the 3 \msun
model using the nominal reaction rates. In \S\ref{sec:sampling} we
discuss the Monte Carlo sampling procedure of the stellar reaction
rates.  In \S\ref{sec:mcstars} and \S\ref{sec:mcstars_fixed_c12ag} 
we build the Monte Carlo stellar evolution models and analyze the 
statistical properties of the results.  In \S\ref{sec:imfm} we discuss
the impact of our results on the IFMR from a 
grid of \MESA models. 

\section{Input Physics}
\label{sec:method}
Models of 1-6 \msun masses in 0.1 \msun increments are evolved using the
Modules for Experiments in Stellar Astrophysics software instrument
\citep[henceforth \MESA, version 7624,][]{paxton_2011_aa,paxton_2013_aa,paxton_2015_aa}.  
All models begin with a metallicity of $Z$=0.02 and a solar
distribution from \citet{grevesse_1998_aa}.  The models are evolved
with the Reimer mass loss prescription \citep{reimer_1975_aa} with
$\eta$=0.5 on the Red Giant Branch (RGB) and a Bl\"ocker mass loss prescription
\citep{blocker_1995_aa} with $\eta$=0.1 on the Asymptotic Giant Branch (AGB). 
The effects of rotation and magnetic fields are neglected for this
study. Each stellar model is evolved from the pre-MS until the
envelope mass on the CO WD remnant is less than \mbox{0.01 \msun}.

\MESA has a variety of controls to specify the mass resolution of a
stellar model.  Sufficient mass resolution is required to prevent 
gradients of the stellar structure from becoming under-resolved, but an
excessive number of grid cells impacts performance.  One method of
changing the mass resolution is the \texttt{mesh\_delta\_coeff}
(\mesh) parameter, which acts a global scale factor limiting the change in
stellar properties between two grid points. A smaller value of \mesh
will increase the number of grid points. Another method of controlling
the mass resolution is to set the minimum number of cells in a model 
independent of any remeshing through the \texttt{max\_dq} parameter. 
We use \hbox{\mesh= 0.5} and \texttt{max\_dq} = 0.01. 
There are $\approx$ 3000 cells in the stellar models 
at ZAMS and \mbox{$\approx$ 5700} cells during the thermally-pulsing AGB phase.

\MESA also offers a rich set of timestep controls.  The parameter 
\texttt{varcontrol\_target}, \var, broadly controls the temporal resolution 
by modulating the relative variation in the structure from one model to the next.
We use the default \var=10$^{-4}$. At a finer level of
granularity, timestep controls such as \texttt{delta\_lg\_XH\_cntr\_min}=10$^{-6}$
regulate depletion of H fuel in the core which aids resolving 
evolution from the ZAMS to the terminal-age main sequence (TAMS),
while the timestep control \texttt{delta\_HR\_limit}=0.0025 limits the 
path length in the Hertzsprung-Russell (HR) diagram between timesteps. 
For our chosen temporal resolution parameters, a timestep of $\Delta t$
$\approx$ 1-40 yr is achieved during the TP-AGB phase. 
The sensitivity of our results to the choice of spatial and temporal parameters 
is discussed further in \S\ref{sec:baseline}.
We also adopt the baseline mixing parameters described in
\citet{farmer_2015_aa}: convective overshoot of \mbox{ \overshoot = 0.016},
thermohaline mixing of {\mbox \thermo = 1.0}, and semiconvection mixing of
{\mbox \semi = 0.01}. All the \MESA inlists and 
many of the stellar models are available\footnote{\url{http://mesastar.org/results}}.

\begin{figure*}[!htb]
\centering{\includegraphics[width=1.5\columnwidth]{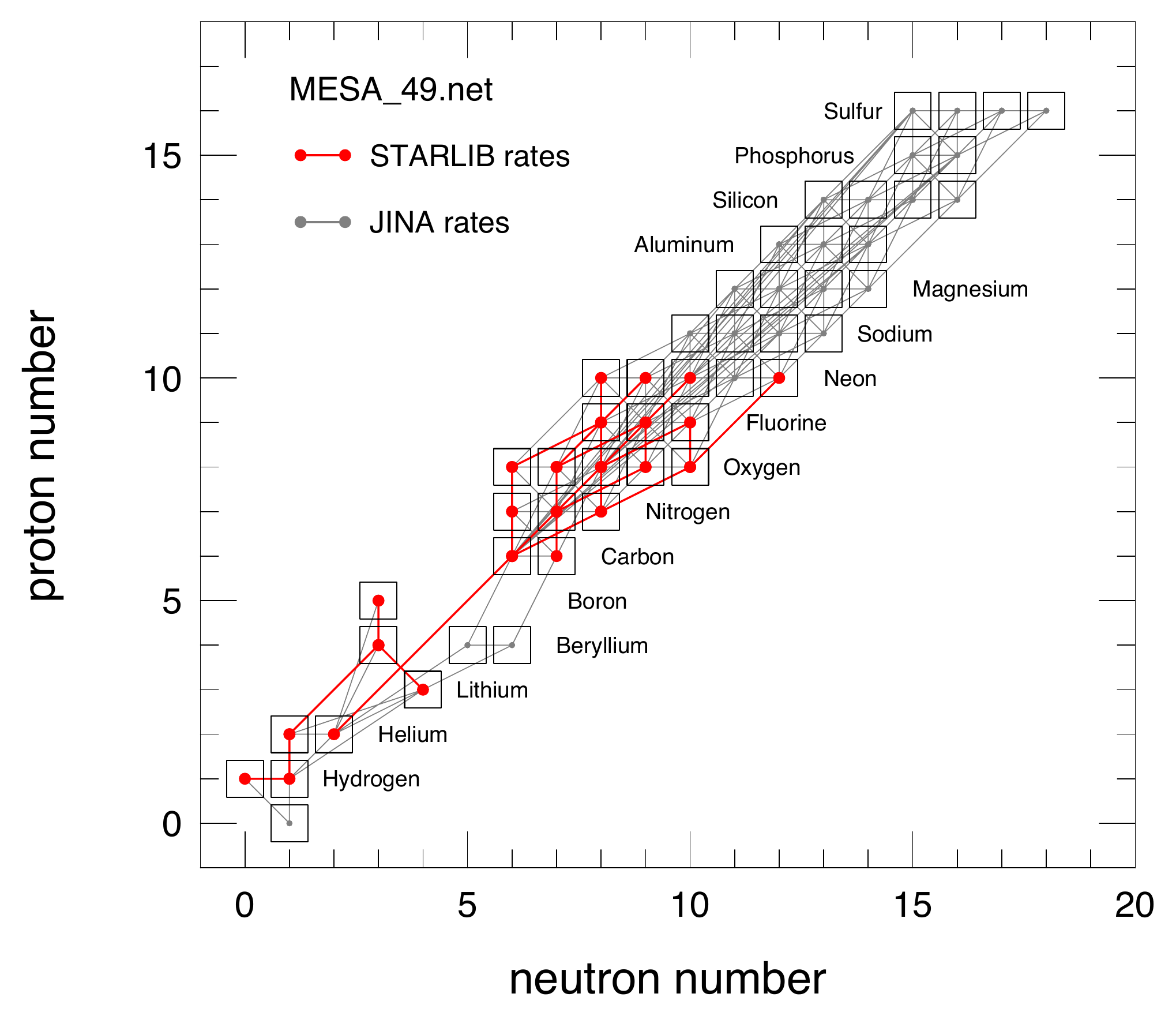}}
\caption{
The \net{49} reaction network used to quantify the variance of
CO WD properties.
This network uses 405 reaction rates to couple 49 isotopes to track hydrogen (\emph{pp}, CNO-,
NeNa-, MgAl-cycles) and helium burning. Reactions between isotopes
that use the STARLIB probability functions are shown in
red while reactions that use the JINA REACLIB are shown in gray.
}\label{fig:nzplane}
\end{figure*}

\begin{deluxetable}{ccc}{!htb}
\tablecolumns{3}
\tablewidth{1.0\linewidth}
\tablecaption{Sampled Nuclear Reaction Rates}
\tablehead{
\colhead{Nuclear Reaction} & 
\colhead{Reference}  & 
\colhead{Rate Identifier}}
\startdata
$^{1}$H(p,$\gamma$)$^{2}$H            & \texttt{nacr} (Exp.) & 1 \\
$^{2}$H(p,$\gamma$)$^{3}$He           & \texttt{de04} (Exp.) & 2  \\
$^{3}$He($\alpha$,$\gamma$)$^{7}$Be   & \texttt{cy08} (Exp.) & 3 \\
$^{7}$Li(p,n)$^{7}$Be                 & \texttt{de04} (Exp.) & 4  \\
$^{7}$Be(p,$\gamma$)$^{8}$B           & \texttt{nacr} (Exp.) & 5  \\
$^{12}$C(p,$\gamma$)$^{13}$N          & \texttt{nacr} (Exp.) & 6  \\
$^{13}$C(p,$\gamma$)$^{14}$N          & \texttt{nacr} (Exp.) & 7  \\
$^{13}$N(p,$\gamma$)$^{14}$O          & \texttt{nacr} (Exp.) & 8  \\
$^{14}$N(p,$\gamma$)$^{15}$O          & \texttt{im05} (Exp.) & 9  \\
$^{15}$N(p,$\alpha$)$^{12}$C          & \texttt{nacr} (Exp.) & 10  \\
$^{15}$N(p,$\gamma$)$^{16}$O          & \texttt{nacr} (Exp.) & 11 \\
$^{14}$O($\alpha$,p)$^{17}$F          & \texttt{taly} (Theory) & 12 \\
$^{15}$O($\alpha$,$\gamma$)$^{19}$Ne  & \texttt{mc10} (MC) & 13  \\
$^{16}$O(p,$\gamma$)$^{17}$F          & \texttt{mc10} (MC) & 14  \\
$^{17}$O(p,$\alpha$)$^{14}$N          & \texttt{bu15} (Exp.) & 15    \\
$^{17}$O(p,$\gamma$)$^{18}$F          & \texttt{bu15} (Exp.) & 16 \\
$^{18}$O(p,$\alpha$)$^{15}$N          & \texttt{mc13} (MC) & 17  \\
$^{18}$O(p,$\gamma$)$^{19}$F          & \texttt{mc13} (MC) & 18  \\
$^{17}$F(p,$\gamma$)$^{18}$Ne         & \texttt{mc10} (MC) & 19  \\
$^{18}$F(p,$\alpha$)$^{15}$O          & \texttt{mc10}  (MC) & 20  \\
$^{19}$F(p,$\alpha$)$^{16}$O          & \texttt{nacr} (Exp.) & 21 \\
$^{16}$O($\alpha$,$\gamma$)$^{20}$Ne  & \texttt{mc10} (MC) & 22  \\
$^{14}$N($\alpha$,$\gamma$)$^{18}$F   & \texttt{mc10} (MC) & 23  \\
$^{18}$O($\alpha$,$\gamma$)$^{22}$Ne  & \texttt{mc10} (MC) & 24  \\
$^{12}$C($\alpha$,$\gamma$)$^{16}$O   & \texttt{ku02} (Exp.) & 25  \\
Triple-$\alpha$                       & \texttt{nacr} (Exp.) & 26
\enddata
\tablecomments{Experimental reaction rates with approximate uncertainties are labeled "Exp", 
while experimental reaction rates with statistically rigorous uncertainties are labeled "MC". 
References are denoted as follows: 
\texttt{nacr} - \citet{angulo_1999_aa},
\texttt{de04} - \citet{descouvemont_2004_aa},
\texttt{cy08} - \citet{cyburt_2008_aa},
\texttt{im05} - \citet{imbriani_2005_aa},
\texttt{taly} - \citet{sallaska_2013_aa},
\texttt{mc10} - \citet{iliadis_2010_aa},
\texttt{bu15} - \citet{buckner_2015_aa},
\texttt{mc13} - \citet{sallaska_2013_aa},
and
\texttt{ku02} - \citet{kunz_2002_aa}.
Rate identifiers (RI) in column 3 are assigned for later reference.
}
\label{tbl:sampled_rates}
\end{deluxetable}

We evolve each stellar model with \MESA's \net{49}, a nuclear reaction 
network that follows 49 isotope from $^{1}$H to $^{34}$S. 
In Table \ref{tbl:sampled_rates}, we list 26 forward nuclear reaction 
rates known to have an impact on the final state of the CO WD
\citep{salaris_1997_aa, metcalfe_2002_aa, straniero_2003_aa}. 
For detailed reviews on the structural and evolutionary properties 
of white dwarfs, see \citet{koester_2002_aa,hansen_2003_aa,althaus_2010_aa}.
For each stellar model, these 26 forward rates are sampled simultaneously 
and independently according to their experimental uncertainties. 
Otherwise, all forward thermonuclear reaction rates are from the JINA reaclib
version V2.0 2013-04-02 \citep{cyburt_2010_aa}. 
We adopted this hybrid approach of using 26 Monte Carlo from STARLIB with
the remaining rates from JINA reaclib because STARLIB is not yet fully
integrated with \MESA, requiring each STARLIB rate and uncertainty specification
to be manually cast into \MESA's tabular format.
The 405 reaction rates linking the 49 isotopes are shown in Figure \ref{fig:nzplane}. 

Inverse rates are calculated directly from the forward rates (those
with positive $Q$-value) using detailed balance, rather than using
fitted rates. This is important for H and He burning
processes approaching equilibrium
\citep[e.g.,][]{hansen_2004_aa,iliadis_2007_aa}.  The nuclear
partition functions used to calculate the inverse rates are taken from
\citet{rauscher_2000_aa}.  Thermal neutrino energy-losses are taken
from the fitting formula of \citet{itoh_1996_aa}. Electron screening
factors for both weak and strong thermonuclear reactions are taken
from \citet{alastuey_1978_aa} with plasma parameters from
\citet{itoh_1979_aa}.  All the weak rates are based (in order of
precedence) on the tabulations of \citet{langanke_2000_aa},
\citet{oda_1994_aa}, and \citet{fuller_1985_aa}.

\section{Characteristics of the baseline 3 \msun stellar model}
\label{sec:baseline}
In this section we present characteristics of the solar metallicity,
non-rotating 3 \msun model evolved with the STARLIB rates.

\begin{figure}[!htb]        
        \begin{subfigure}{
                \includegraphics[trim = .1in .1in 0in .1in, clip,width=3.4in,height=2.75in]{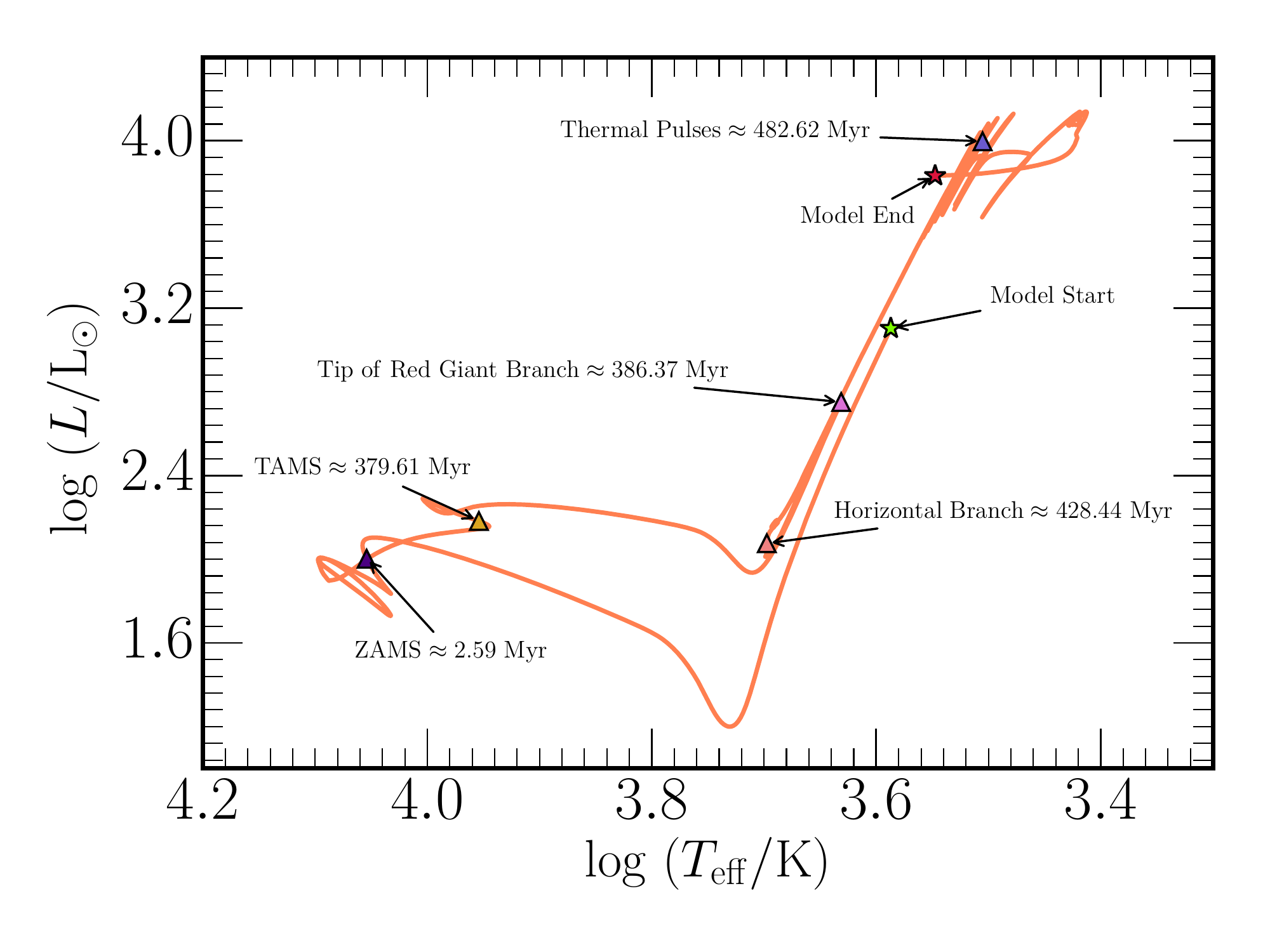}}
        \end{subfigure}                    
        \caption{
Baseline 3 \msun model with the nominal set of reaction rates in the HR diagram. Key events and ages are labeled.
        }
\label{fig:hr}
\end{figure}

\begin{figure}[!htb]        
        \begin{subfigure}{
                \includegraphics[trim = 0in 0in 0in 0in, clip,width=3.5in,height=2.75in]{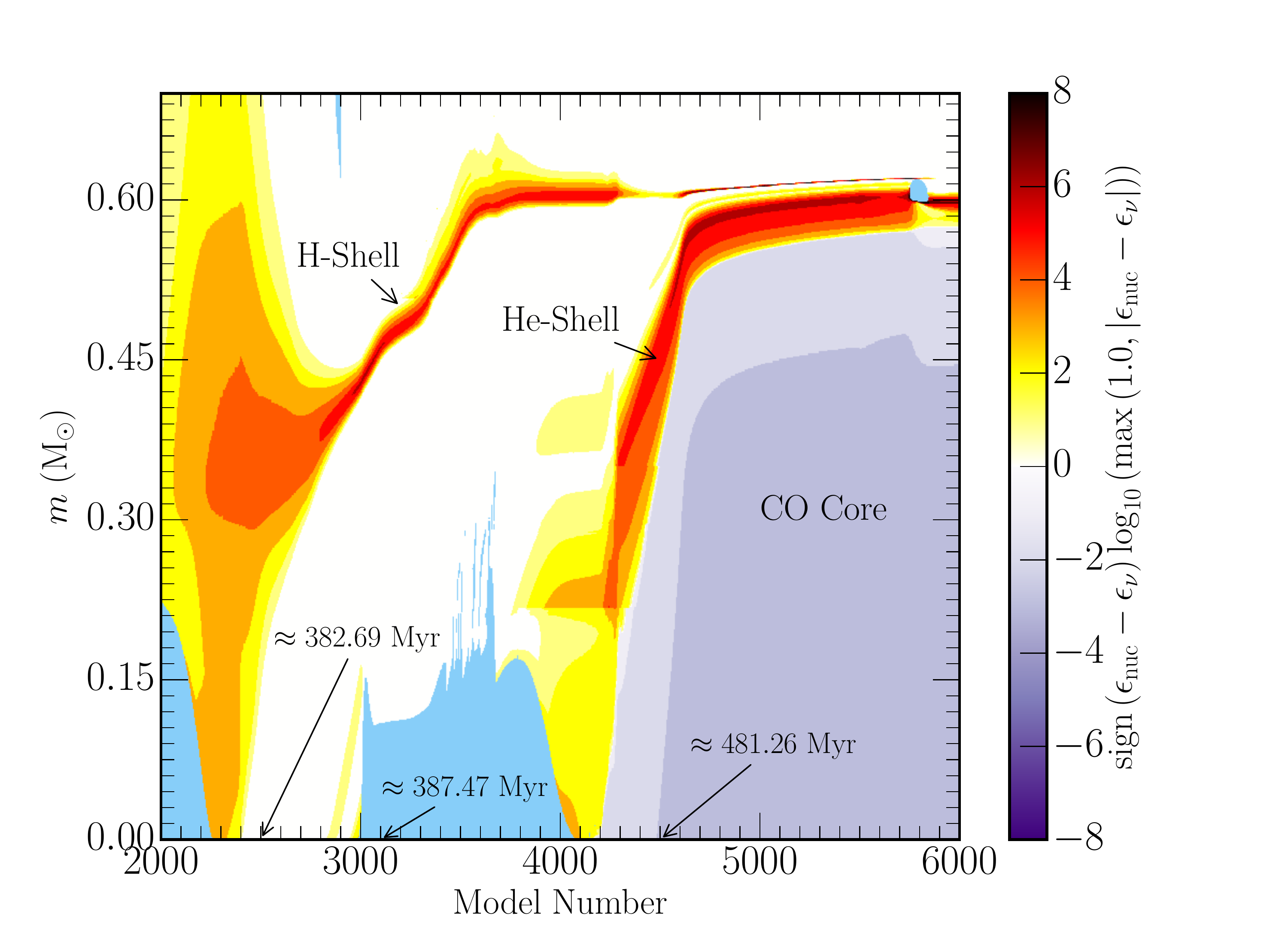}}
        \end{subfigure}                    
        \caption{
Kippenhahn diagram of the baseline 3 \msun model from the MS to formation of a $\approx$ 0.645 \msun WD.
        }
\label{fig:kipp}
\end{figure}

Figure \ref{fig:hr} shows the model star in the HR diagram.  Core
H burning is initiated at $t \approx$ 2.59 Myr with a central
pulse that lasts $t \approx$ 0.12 Myr while the star continues to
contract from log($R/R_{\odot}$) $\approx$ 0.43 to 0.37. At \mbox{$t \approx$ 3.26 Myr} 
a subsequent H pulse halts further contraction and causes a
decrease in the surface temperature and luminosity. This secondary
pulse is quenched at $t \approx$ 3.48 Myr causing the star to expand
to log($R/R_{\odot}$) $\approx$ 0.35. These unstable H
pulses at the core of the stellar model cause the fluctuations near
the labeled ZAMS location in Figure \ref{fig:hr}. Following the first
two failed H pulses, burning in the core continues at \mbox{$t \approx$ 3.80 Myr} 
initiating a further contraction of the star from log($R/R_{\odot}$)
$\approx$ 0.35 to 0.30. At this configuration, the stellar model
proceeds through stable H burning until reaching the TAMS at $t \approx$ 379.6 Myr.

Once H is depleted in the core, the core contracts, the
envelope expands, and the star ascends the RGB. The model star reaches
\mbox{log($R/R_{\odot}$) $\approx$ 1.65} at the tip of the RGB at an
age of $t \approx$ 386.37 Myr. Helium burning begins in the core.  The
stellar model maintains core He burning while contracting to a
radius of log($R/R_{\odot}$) $\approx$ 1.20.  The star populates
horizontal branch at an age of $t \approx$ 428.44 Myr.  Once the
supply of He fuel in the core is depleted, the star leaves the horizontal branch
due to further core contraction and expansion of the envelope. The
star ascends the AGB \citep[e.g.,][]{lattanzio_1991_aa,iben_1991_aa,karakas_2007_aa,shingles_2015_aa}.

On the AGB, He burning continues in a thin shell above the newly
formed CO core and the H burning shell. Figure \ref{fig:kipp}
shows a Kippenhahn diagram of our \mbox{3 \msun} non-rotating stellar
model. The color bar represents the net nuclear energy
generation rate. Dark red-orange regions denote regions of strong
nuclear burning such as the H and He burning shells
annotated in the diagram. Regions of purple indicate a logarithmic
increase in the cooling rate, such as in the CO core. Convective regions
are shown in light blue. For simplicity, regions undergoing
semi-convective, thermohaline, or convective boundary mixing are not
shown.

\begin{figure}[!htb]
         \centering
        \begin{subfigure}{
                \includegraphics[trim = .1in .1in 0in .1in, clip,width=3.4in,height=2.75in]
                {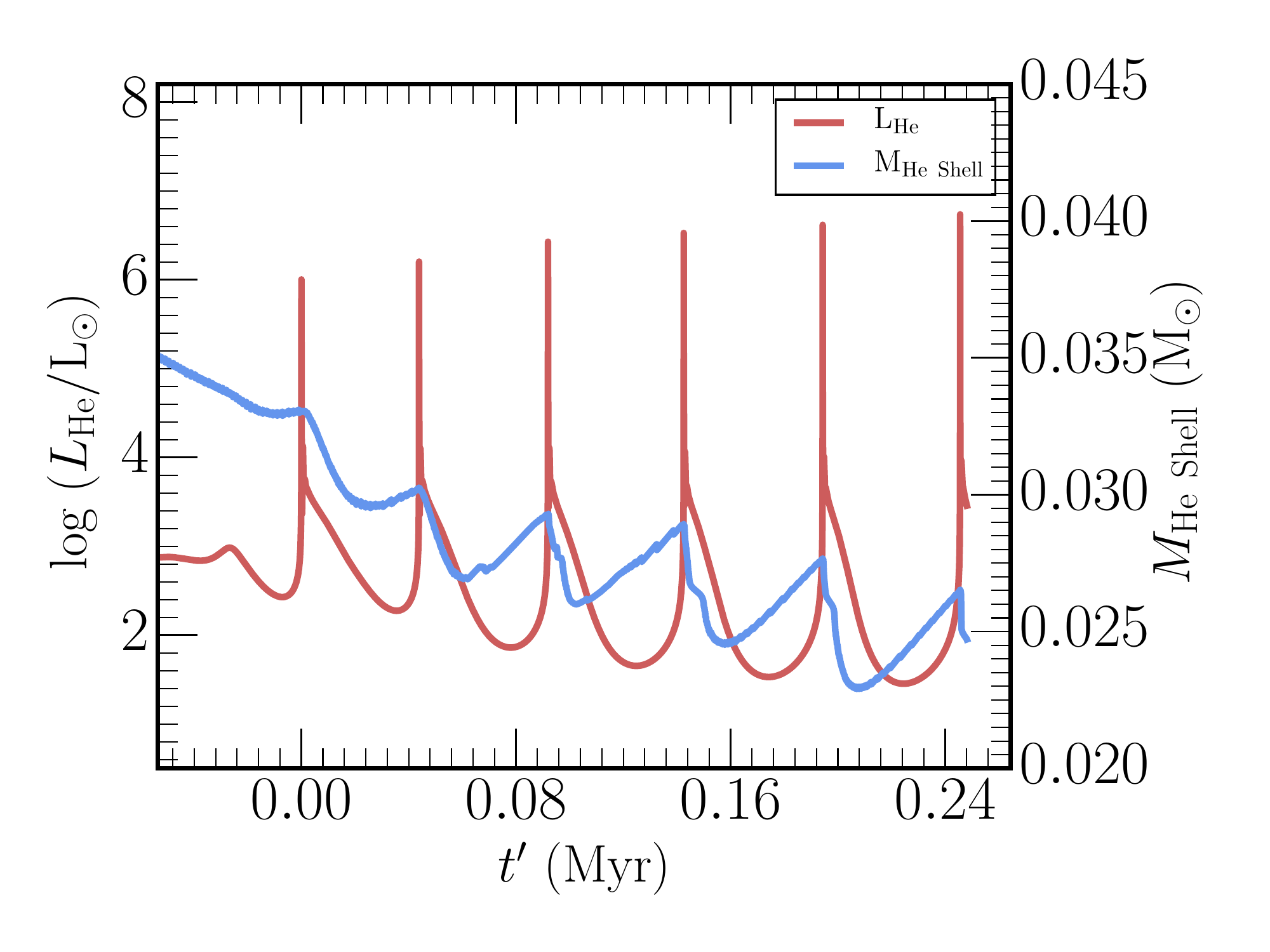}}
        \end{subfigure}
        \caption{
Helium burning luminosity during the thermal pulses relative to 
time at first thermal pulse at approximately 482.47 Myr.
}
\label{fig:lumin}
\end{figure}

The geometrically thin He shell grows as material from H burning
shell is processed, causing the He shell to increase in temperature
and pressure.  Once the mass in the He shell reaches a critical value,
helium is reignited and causes a He shell flash. This flash causes
an extraordinary release of energy and expansion of the intershell
region. The first thermal pulse is achieved in our baseline model at an
age of $t \approx$ 482.47 Myr. Figure \ref{fig:lumin} depicts the
subsequent thermal pulses by showing He burning luminosity as a
function of stellar age. The time between pulse is often referred to
as the interpulse period and has been shown to be well described by
\begin{equation}
\textup{log}~\Delta t_{\rm{TP}} = 4.5(1.689-M_{\rm{CO}})~\rm{yr}
\label{eq:intp}
\end{equation}
for $Z$ = 0.02 \citep{boothroyd_1988_aa}.

Our solar metallicity \mbox{3 \msun} model goes through a series of six thermal pulses,
$n_{\rm{TP}} = 6$, with a recurrence time of 
\mbox{log($\Delta t_{\rm{TP}}/{\rm yr}) \approx$ 4.66}. This value agrees to 
within a factor of two of Equation~(\ref{eq:intp}). The difference 
is due to the He abundance used in our model, $Y=0.28$, 
and the strong dependence of Equation~(\ref{eq:intp}) on the He abundance
\citep{boothroyd_1988_aa}.

The luminosity of a Red Supergiant with a degenerate CO core and H and 
He shell sources may be approximated by the core mass - luminosity 
relationship
\begin{equation}
\begin{split}
L/L_{\odot} = &  \ 2.38 \times 10^{5} \mu^{3}Z^{0.04} \\
& \times (M_{\rm{CO}}^{2} - 0.0305M_{\rm{CO}} - 0.1802)
\end{split}
\label{eq:lum}
\end{equation}
for core masses 0.5 \msun $\lesssim$ $M_{\rm{CO}}$ $\lesssim$ 0.66 \msun 
\citep{marigo_1996_aa,marigo_1998_aa}. 
At the peak luminosity of the final thermal pulse our stellar model reaches 
\mbox{log($L/L_{\odot}$) $\approx$ 4.14}. This value agrees to within $\approx$ 5\% 
using Equation~(\ref{eq:lum}) with \mbox{$\mu \approx$ 0.621}, $Z$=0.02, and 
$M_{\rm{CO}} \approx$ 0.585 M$_{\odot}$. 

\begin{figure}[!htb]
\centering{\includegraphics[width=1.0\columnwidth]{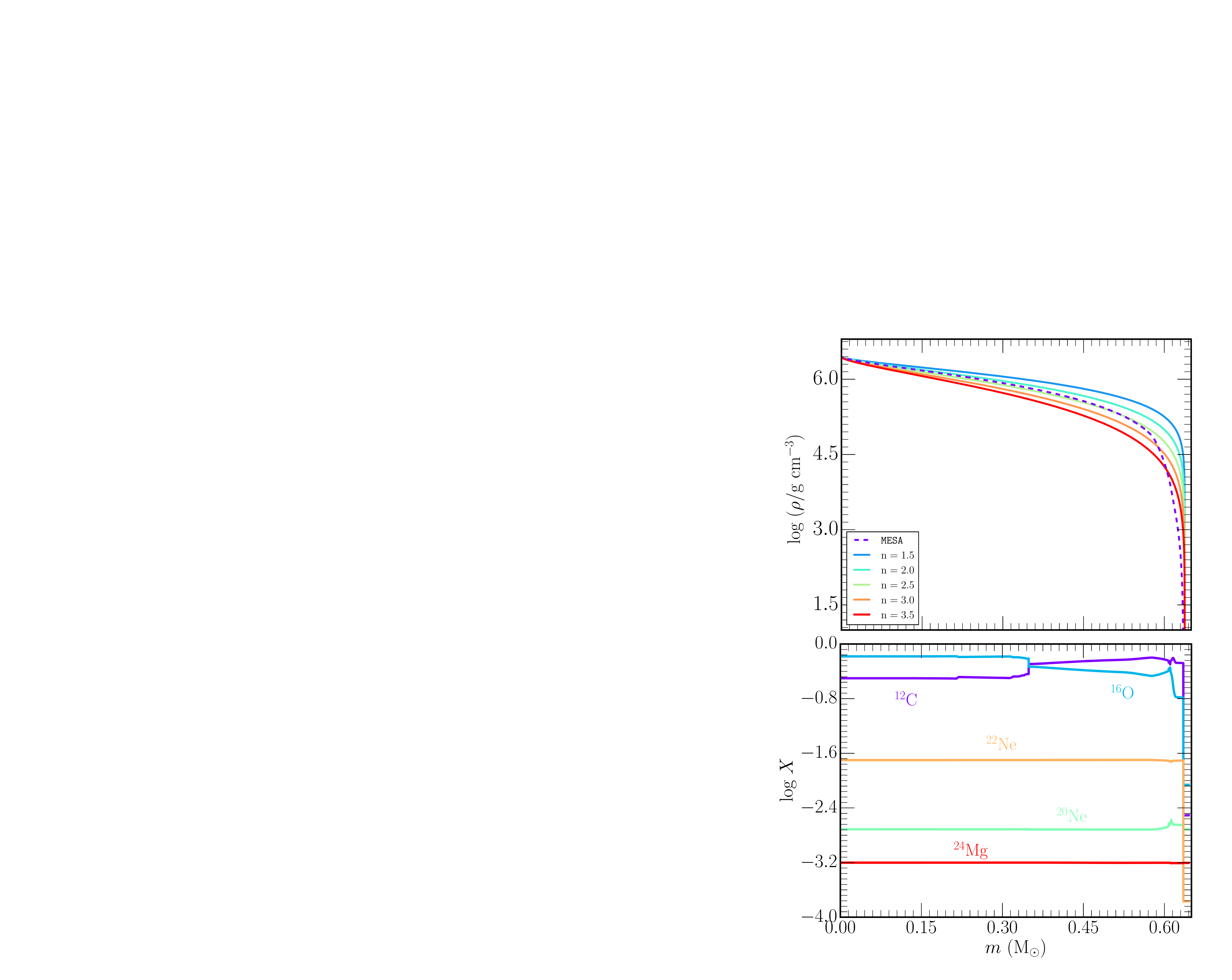}}
\caption{
Density and abundance profiles in the 0.64 \Msun CO WD remnant 
produced by the 3 \msun model. 
The \MESA density structure (dashed purple curve) is compared to 
polytropic density profiles for $n$ = 3/2, 2, 5/2, 3, and 7/2.
}\label{fig:3m_prof}
\end{figure}

Figure \ref{fig:3m_prof} shows the density and abundance
profiles of the CO WD remnant. The upper panel shows the density
profile of the \MESA model (dashed line) and polytropic density
of varying indices 
$n$\footnote{
Polytrope models were 
calculated using the open-source tool at
\url{http://cococubed.asu.edu/code\_pages/polytrope.shtml}.}.
For the nearly isothermal core temperature of $T \approx 1\times10^8$ K, 
the \MESA density profile is best fitted by a density profile of a $n$=5/2 polytrope.

In the baseline 3 \msun model, $^4$He initially burns to $^{12}$C via
the triple-$\alpha$ reaction in the core on a timescale of \mbox{$\approx$ 50 Myr}. 
Once an abundance of carbon builds up, $^{12}$C can capture an
alpha particle to become $^{16}$O. For the \mbox{3 \msun} model, this occurs
on longer timescales of \mbox{$\approx$ 100 Myr}
\citep[e.g.,][]{iben_2013_aa,iben_2013_ab}.  Late in core He
burning, the rising abundance of $^{12}$C allows alpha capture to
compete with and eventually dominate the triple-$\alpha$ flow, which
thus causes the $^{12}$C to decrease. By the end of core He
burning the mass fractions are \mbox{X($_{\rm c}$$^{12}$C) = 0.3151} and 
X$_{\rm c}$($^{16}$O) = 0.6585.  The remaining $\approx$ 2.6\% of the original He is
mostly in isotopes of Ne and Mg.

\begin{figure}[!htb]      
\centering  
\includegraphics[trim = .1in .1in 0in .1in, clip,width=3.4in,height=2.75in]{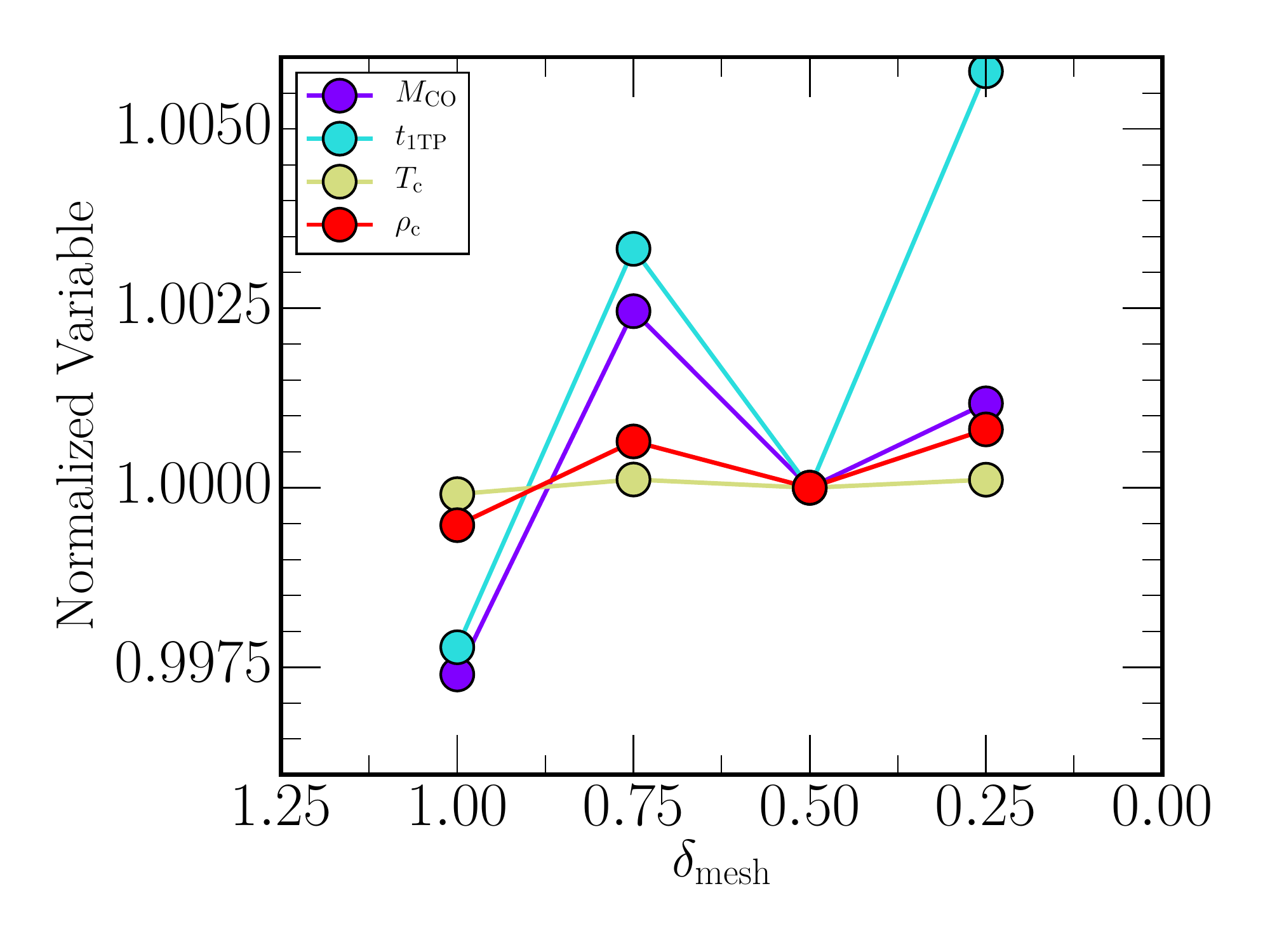}
\caption{
Mass of the CO core (purple), age (blue-green), central temperature \Tc \ (gold), 
and central density \rhoc \ (red) at the first thermal pulse, normalized 
to the values from our baseline choice of \mbox{(\mesh=0.5,\var=10$^{-4}$)}.  
The other spatial and temporal combinations shown are 
\mbox{(\mesh=1.0,\var=10$^{-2}$)},
\mbox{(\mesh=0.75,\var=10$^{-3}$)}, and
\mbox{(\mesh=0.25,\var=10$^{-5}$)}.
        }\label{fig:conv_1}
\end{figure}

The isotope $^{22}$Ne is perhaps the most interesting of these remaining 
isotopes as it carries all of the available neutrons, which can have 
significant consequences on the peak luminosity of any subsequent explosion 
of the CO WD as a supernova Type Ia
\citep[e.g.,][]{townsley_2009_aa,howell_2009_aa,neill_2009_aa,childress_2013_aa,moreno-raya_2015_aa}.
Most of a main-sequence star's initial metallicity $Z$ comes from the CNO
and $^{56}$Fe nuclei inherited from its ambient interstellar medium. The
slowest step in the hydrogen burning CNO cycle is proton capture onto
$^{14}$N. This results in all the CNO catalysts piling up into $^{14}$N when
hydrogen burning on the main sequence is completed. During He
burning the reactions
$^{14}{\rm N}(\alpha,\gamma)^{18}{\rm F}(,\beta^{+}\nu_e)^{18}{\rm O}(\alpha,\gamma)^{22}{\rm Ne}$ convert all
of the $^{14}$N into $^{22}$Ne. Thus, the final mass fraction of $^{22}$Ne in the
CO remnant
\begin{equation}
 X(^{22}{\rm Ne}) = 22 \left( 
                    \frac{X(^{12}{\rm C})}{12} 
                  + \frac{X(^{14}{\rm N})}{14} 
                  + \frac{X(^{16}{\rm O})}{16} 
                        \right)
 \label{eq:x22}
\end{equation}
is determined by the birth abundances of CNO.
For the solar mass fractions used in constructing the
initial \mbox{3 \msun model}, Equation~(\ref{eq:x22}) gives ${\rm X}(^{22}{\rm Ne})$=0.021
gives the same value shown in Figure~\ref{fig:3m_prof}. 
In Figure~\ref{fig:3m_prof} we also see a notable feature at the chemical 
transition region from $^{16}$O to $^{12}$C. These transition regions were found to 
be caused by extra mixing episodes during core He burning and have implications
for the Brunt-V$\ddot{\rm{a}}$is$\ddot{\rm{a}}$l$\ddot{\rm{a}}$ frequency profile \citep{romero_2012_aa,romero_2013_aa}.

To assess the convergence of our 3 \msun model to the choice 
of spacetime resolution parameters we evolve a series of models
from the pre-MS to the first thermal pulse and  
varying the spatial resolution parameter 
\texttt{mesh\_delta\_coeff}, \mesh, 
and temporal resolution parameter \texttt{varcontrol\_target}, \var.
The spatial and temporal resolutions explored are
\mbox{(\mesh=1.0,\var=10$^{-2}$)},
\mbox{(\mesh=0.75,\var=10$^{-3}$)},
\mbox{(\mesh=0.5,\var=10$^{-4}$)}, and 
\mbox{(\mesh=0.25,\var=10$^{-5}$)}.
Changing \mesh by a factor of two, changes the number of cells by roughly a factor of two,
and changing \var by a factor of ten changes the timestep by a factor of roughly two.

Figure \ref{fig:conv_1} shows the mass of the CO core, age, \Tc, 
and \rhoc \ at the first thermal pulse.  Each of
these quantities is normalized to the values from our baseline choice
of \mbox{(\mesh=0.5,\var=10$^{-4}$)}.  Figure \ref{fig:conv_1}
suggests these key end-of-evolution quantities are converged with regard
to the mass grid and timestep to within 3 to 4 significant figures.

\section{Reaction Rate Sampling}
\label{sec:sampling}

\begin{figure}[!htb]
\centering
{\includegraphics[trim = .1in .1in .1in .1in, clip,width=3.4in,height=2.75in]{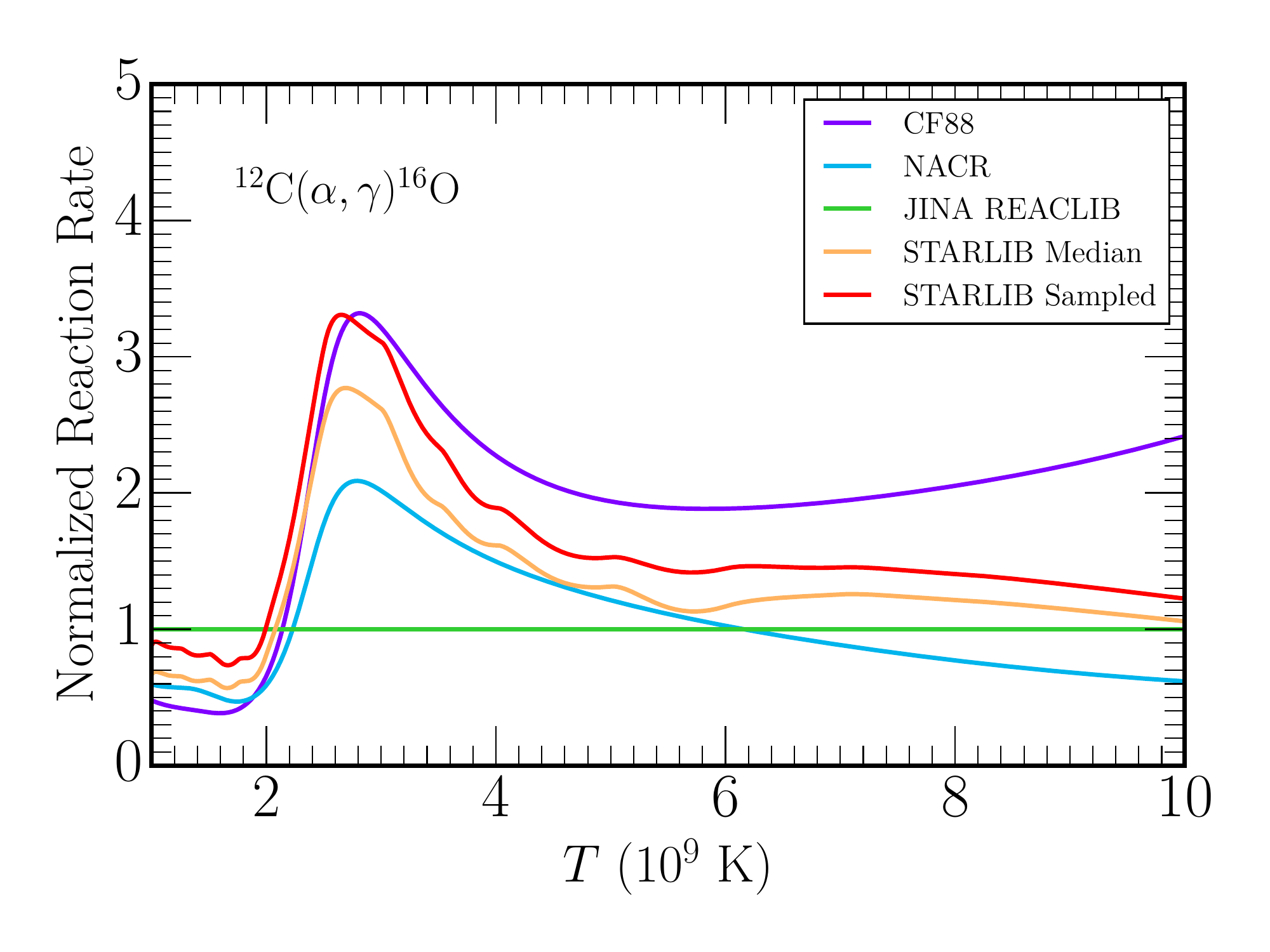}}
\caption{
The $^{12}$C$(\alpha,\gamma)^{16}$O nuclear 
reaction rate, normalized by the rate given in JINA-REACLIB, 
for \citet[][CF88]{caughlan_1988_aa}, 
\citet[][NACR]{angulo_1999_aa},
\citet[][JINA-REACLIB]{cyburt_2010_ab},
\citet[][STARLIB Median]{sallaska_2013_aa},
and a sampled rate distribution (STARLIB Sampled) computed 
for a specific Gaussian deviate of $p_i$ = 0.849741. The JINA-REACLIB
rate is based on \citet{buchmann_1996_aa} and STARLIB on
\citet{kunz_2002_aa}. Note: 
The CF88 rate used in \MESA is 1.7 times the original CF88, see \citet{weaver_1993_aa}.
}\label{fig:rr_example}
\end{figure} 

The STARLIB rate library provides the median nuclear reaction rate,
$\textup{Rate}_{\rm{med}}$, and the associated 
factor uncertainty, henceforth $f.u.$, at 
temperature ranges from $10^6$ K to $10^{10}$ K \citep{sallaska_2013_aa}.
Following \citet{longland_2010_aa}, all reaction and decay rates given 
in STARLIB are assumed to follow a log-normal probability distribution. 
Such a distribution is described by the location and spread parameters, 
$\mu$ and $\sigma$, respectively. These 
parameters can be obtained using the median rate and $f.u.$ 
tabulated in STARLIB, where
$\sigma = \textup{ln}~f.u.$, 
$\mu = \textup{ln}~\textup{Rate}_{\rm{med}}$.
These two parameters give a complete description of the reaction rate 
probability density at any temperature.

The relationship between these parameters form the basis of our 
sampling scheme. Following \citet{evans_2000_aa}, for a log-normal distribution
of an arbitrary quantity, $x$, samples are drawn as, 
\begin{equation}
x_i = e^{\mu + \sigma p_i } \equiv e^{\mu}(e^{\sigma})^{p_i}~.
\end{equation}
Using the relations for $\mu$ and $\sigma$, we can obtain a sampled rate 
distribution as a function of temperature from
\begin{equation}
\textup{Rate}_{\rm{samp}} 
= e^{\mu}(e^{\sigma})^{p_{i,j}} 
= \textup{Rate}_{\rm{med}} (f.u.)^{p_{i,j}}~,
\label{eq:sample_lambda}
\end{equation} 
where $p_{i,j}$ is standard Gaussian deviate with mean of zero and 
standard deviation of unity for the $j$th sampled reaction rate. 

We refer to $p_{i,j}$ as the rate variation factor for the $j$th
reaction, following the $j$ rate identifiers given in \mbox{Table \ref{tbl:sampled_rates}}.  
The rate variation factor $p_{i,j}$ differs
from the factor uncertainty $f.u.$.  For a rate variation factor of
$p_{i,j}$=0, the median Monte Carlo rate provided by
STARLIB is recovered. Since the factor uncertainty $f.u.$ is a
function of temperature, a sampled rate distribution follows the
temperature dependence of the uncertainty in the rate.

In this study, we independently sample 26 forward thermonuclear 
reaction rates (see Table~\ref{tbl:sampled_rates}) that describe 
the main H and He burning processes relevant to the 
evolution and subsequent formation of 
CO WDs. Because \MESA calculates inverse 
rates directly from the forward rates (those with positive $Q$-value) 
using detailed balance, we also implicitly, but not independently,
sample the corresponding 26 inverse rates.

\begin{figure}[!htb]        
\centering
\includegraphics[trim = .1in .1in 0in .1in, clip,width=3.4in,height=2.75in]{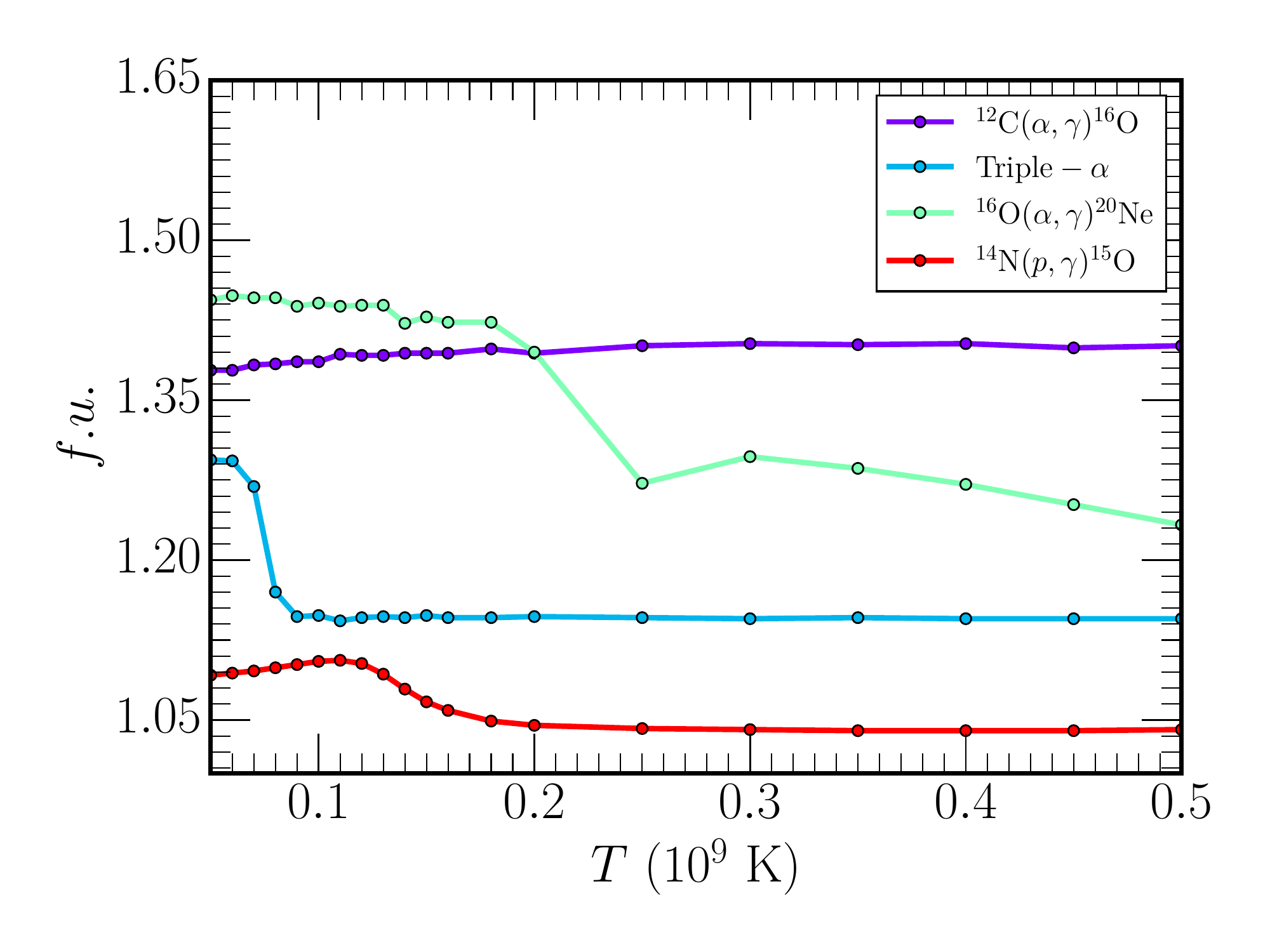}
\caption{The factor uncertainty, f.u., for 
the triple-$\alpha$, $^{12}$C($\alpha$,$\gamma$)$^{16}$O, 
$^{16}$O($\alpha$,$\gamma$)$^{20}$Ne, and $^{14}$N($p$,$\gamma$)$^{15}$O 
reaction rates
over stellar H and He burning temperatures. STARLIB data points are 
shown by the filled dots, while the lines
represent interpolations of the data.
\label{fig:fu_factors}}
\end{figure}

Experimentally-based Monte Carlo reaction rate distributions are
available for 9 out of 26 of the reactions rates included in our 
sampling scheme. For 16 reactions, such as
$^{12}\textup{C}(\alpha,\gamma)\rm{^{16}O}$ and 
$^{14}\textup{N}(p,\gamma)\rm{^{15}O}$, Monte Carlo based rate
distributions do not yet exist. This is due in part to these
reaction rates being determined by broad amplitudes rather than 
narrow resonances. For such reactions, median rate values 
and $f.u.$ are obtained from estimates of experimental 
uncertainty where available.
In the absence of experimental nuclear physics input,
such as the $^{14}\textup{O}(\alpha,p)\rm{^{17}F}$ reaction, 
theoretical median rate values are obtained using the statistical (Hauser-Feshbach)
model of nuclear reactions and computed using the nuclear reaction software instrument 
TALYS \citep{goriely_2008_aa}.

Following \citet{iliadis_2015_aa}, we assume that the deviate 
is independent of temperature, $p_i(T) =$ constant. This simplification
was shown to reproduce the same uncertainties in isotopic abundances 
as more complex sampling schemes \citep{longland_2012_aa}. Despite this 
assumption for our rate variation factors, the $f.u.$ provided by STARLIB is 
temperature dependent. This allows us to closely follow changes in the rate 
uncertainty that have been shown to occur from different resonance 
contributions.

To initialize a grid of Monte Carlo stellar models we generate 26
independent random Gaussian distributions of rate variation factors,
$p_{i,j}$, where $j$ corresponds to the rate identifiers listed in
Table \ref{tbl:sampled_rates}. A set of N stellar models is then
constructed by using the $n^{\rm{th}}$ set of random numbers, the
$p_{i,j}$, and Equation~(\ref{eq:sample_lambda}) to generate sampled
reaction rate distributions.

Once these 26 independently sampled STARLIB rate distributions are constructed, 
they are provided in tabular form to \MESA. The data are used by \MESA
to perform an interpolation over a reaction rate defined by 
10,000 points. 
Figure~\ref{fig:rr_example} shows a comparison of 
a sampled reaction rate distribution for the $^{12}$C$(\alpha,\gamma)^{16}$O 
reaction with $p_i$ = 0.849741 to fixed reaction rates from previous studies.
Figure~\ref{fig:fu_factors} plots the $f.u.$ for the 
the triple-$\alpha$, $^{12}$C($\alpha$,$\gamma$)$^{16}$O, 
$^{16}$O($\alpha$,$\gamma$)$^{20}$Ne, and $^{14}$N($p$,$\gamma$)$^{15}$O 
reaction rates
over H and He burning temperatures. In broad terms, $f.u.$ represents 
the amount a reaction rate is shifted relative
to the median reaction rate at each temperature.

\section{Monte Carlo Stellar Models}
\label{sec:mcstars}

\begin{figure*}[!htb]
         \centering
        \begin{subfigure}{
                \includegraphics[trim = .25in .25in .25in .25in, clip,width=2.75in,height=2.1in]{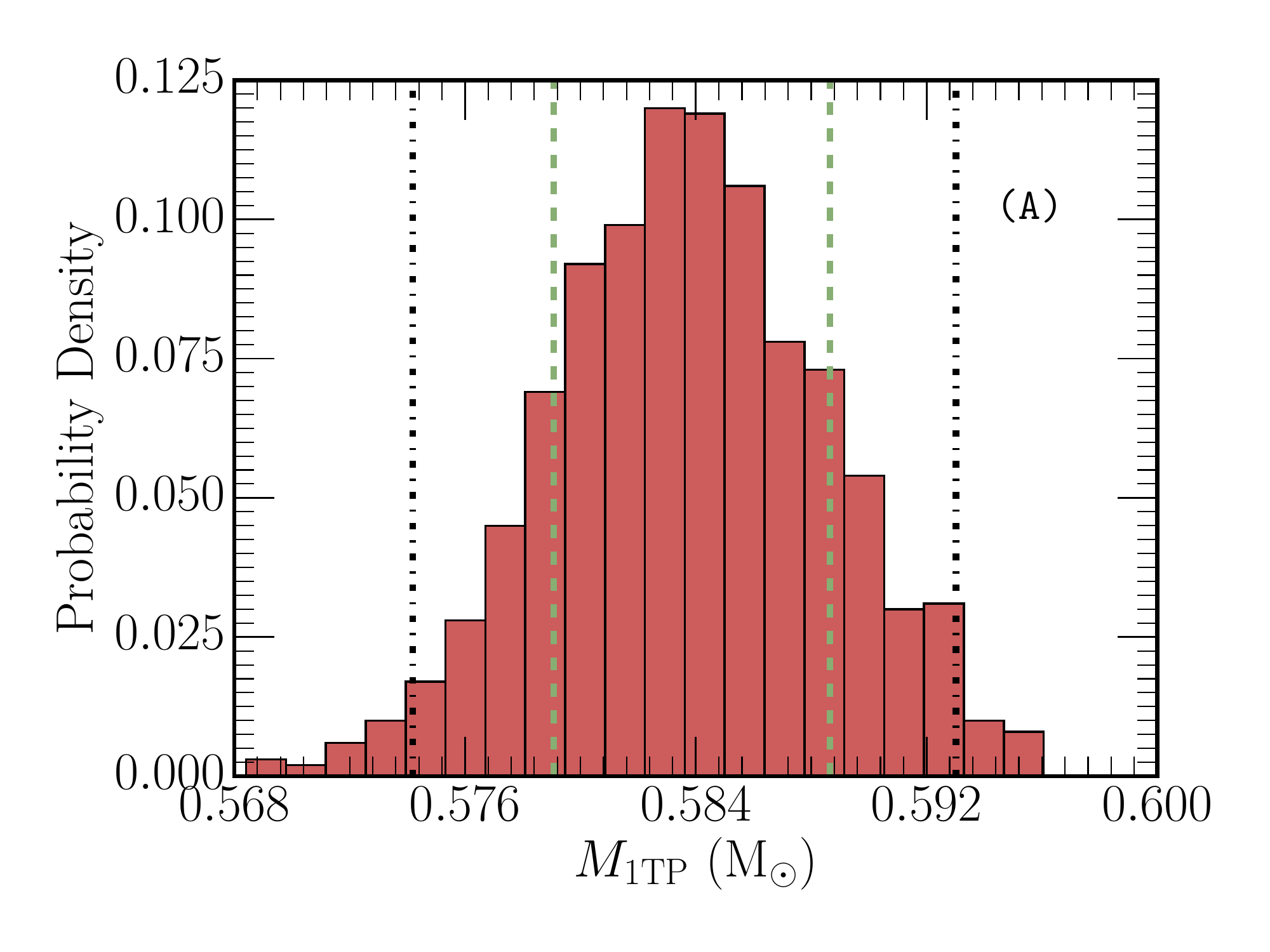}}
        \end{subfigure}
        \begin{subfigure}{
                \includegraphics[trim = .25in .25in .25in .25in, clip,width=2.75in,height=2.1in]{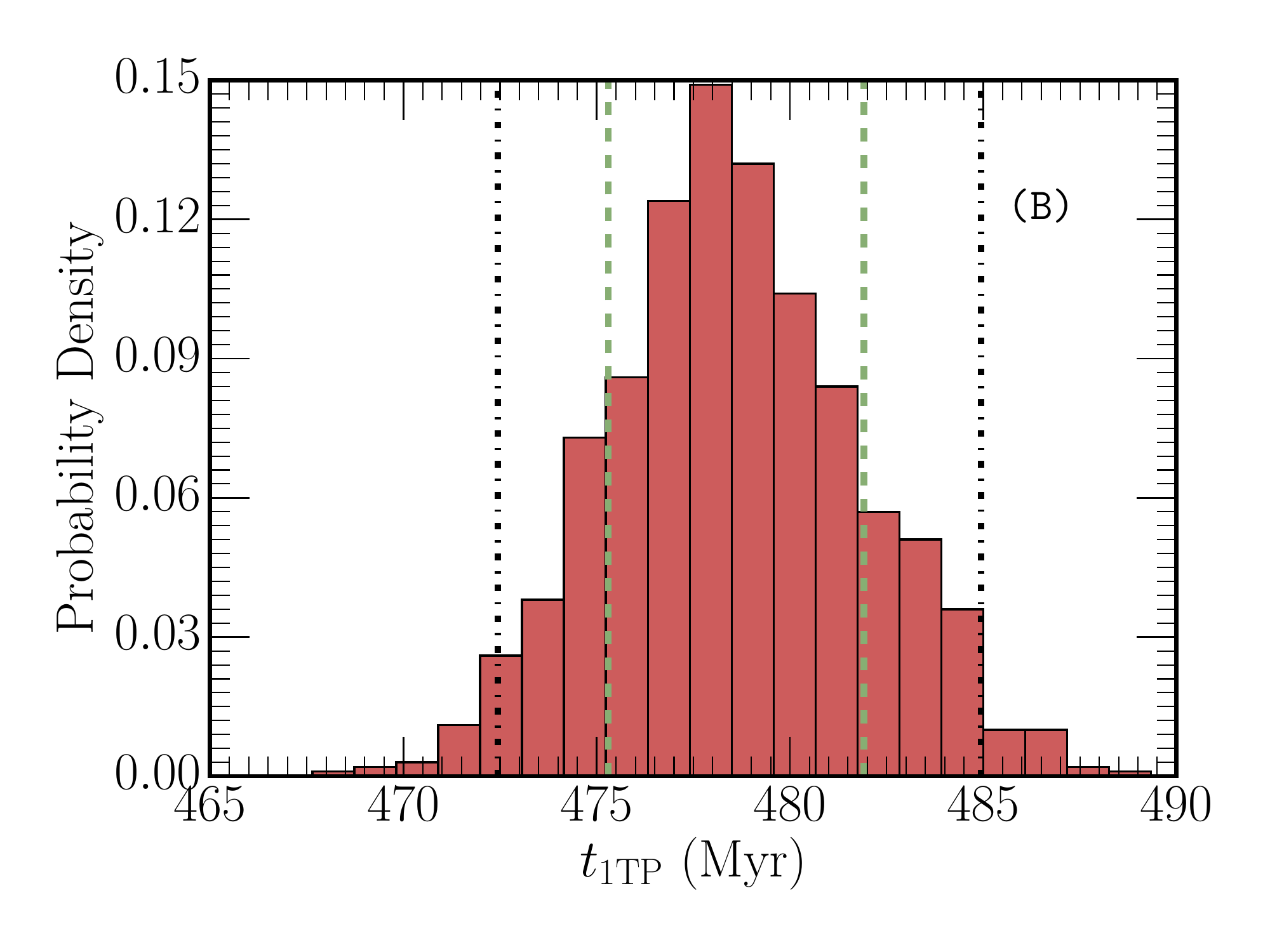}}
        \end{subfigure}          
        \begin{subfigure}{
                \includegraphics[trim = .25in .25in .25in .25in, clip,width=2.75in,height=2.1in]{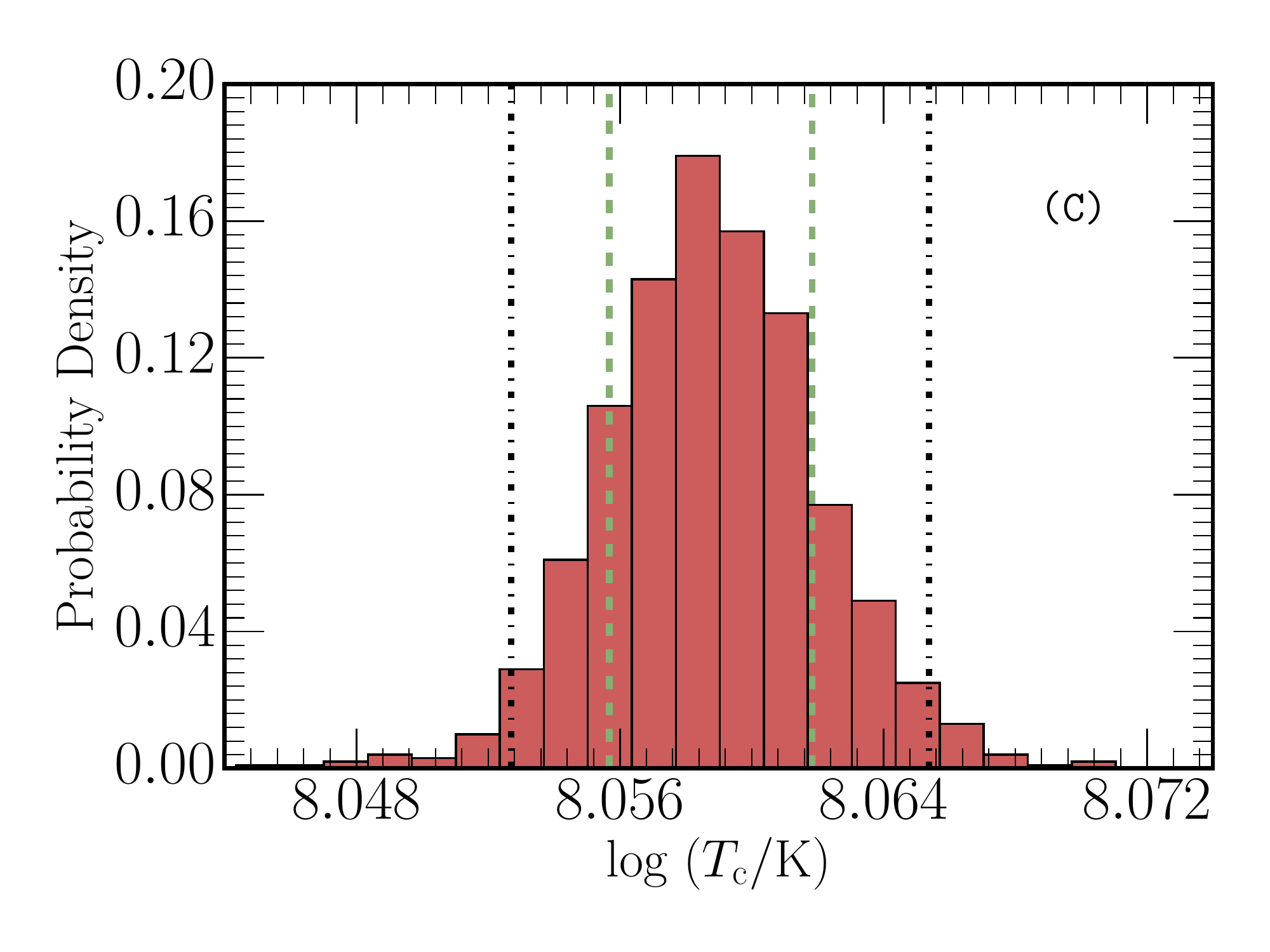}}
        \end{subfigure}
        \begin{subfigure}{
                \includegraphics[trim = .25in .25in .25in .25in, clip,width=2.75in,height=2.1in]{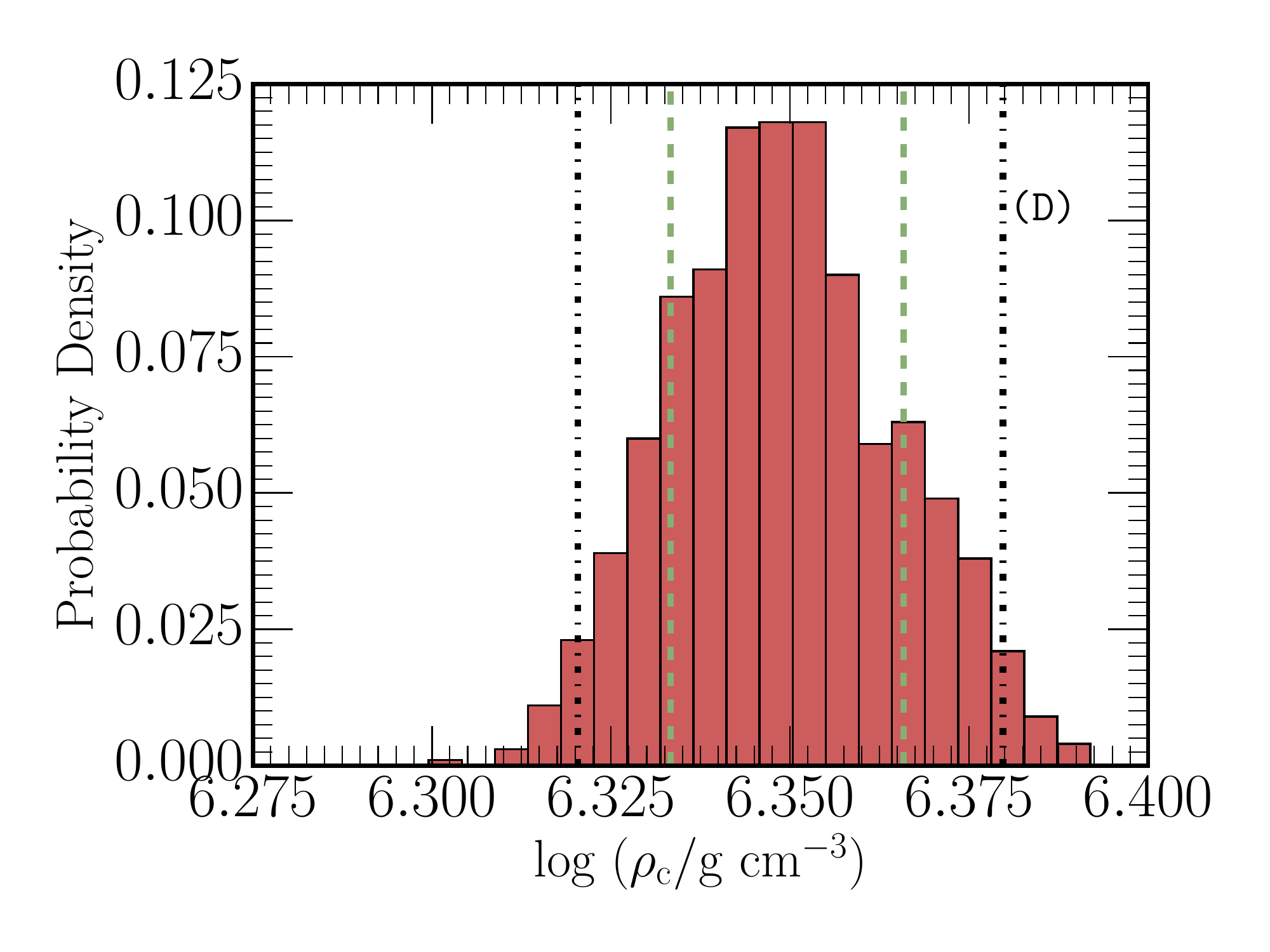}}
        \end{subfigure}
        \begin{subfigure}{
                \includegraphics[trim = .25in .25in .25in .25in, clip,width=2.75in,height=2.1in]{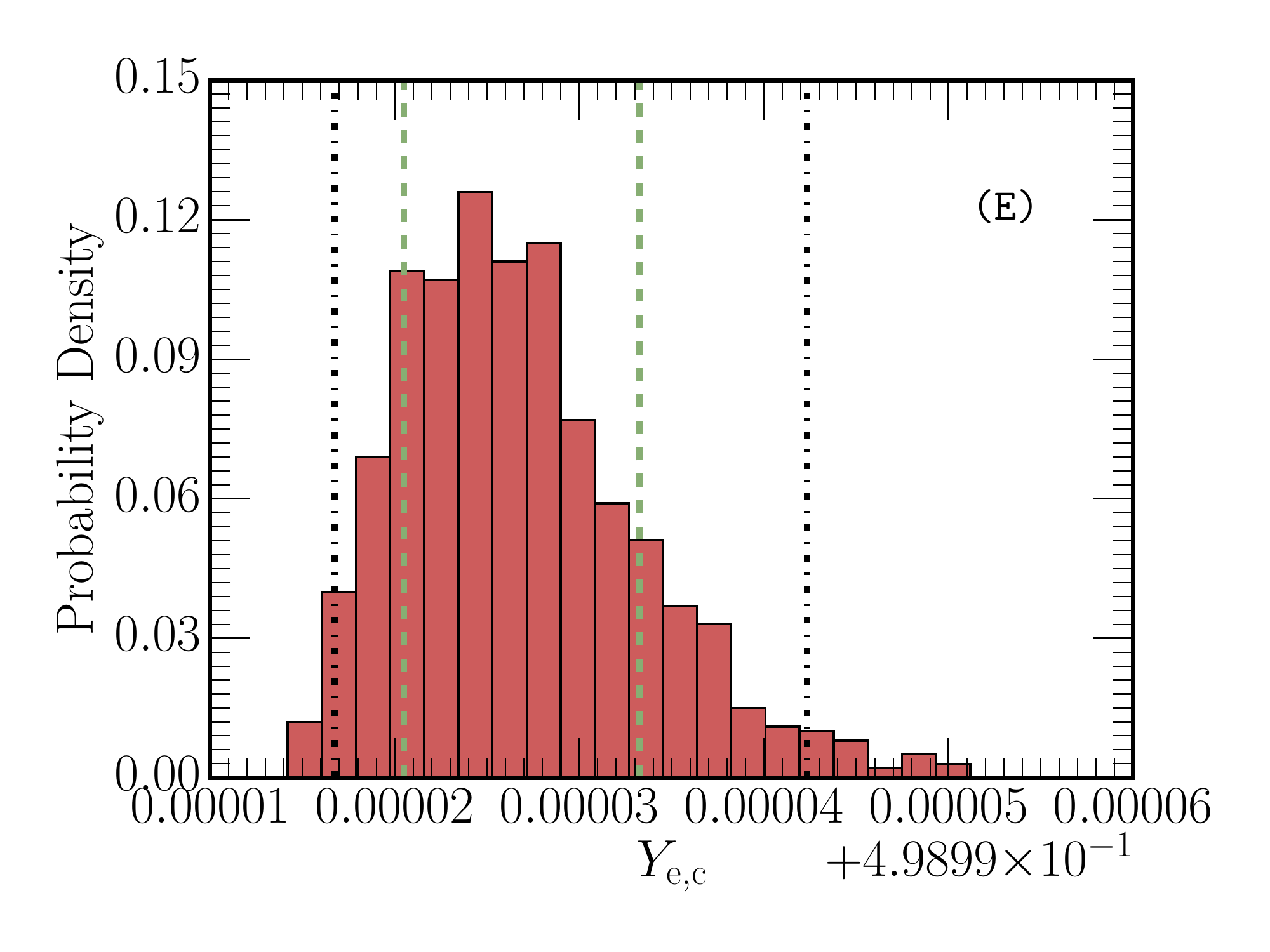}}
        \end{subfigure}
                \begin{subfigure}{
                \includegraphics[trim = .25in .25in .25in .25in, clip,width=2.75in,height=2.1in]{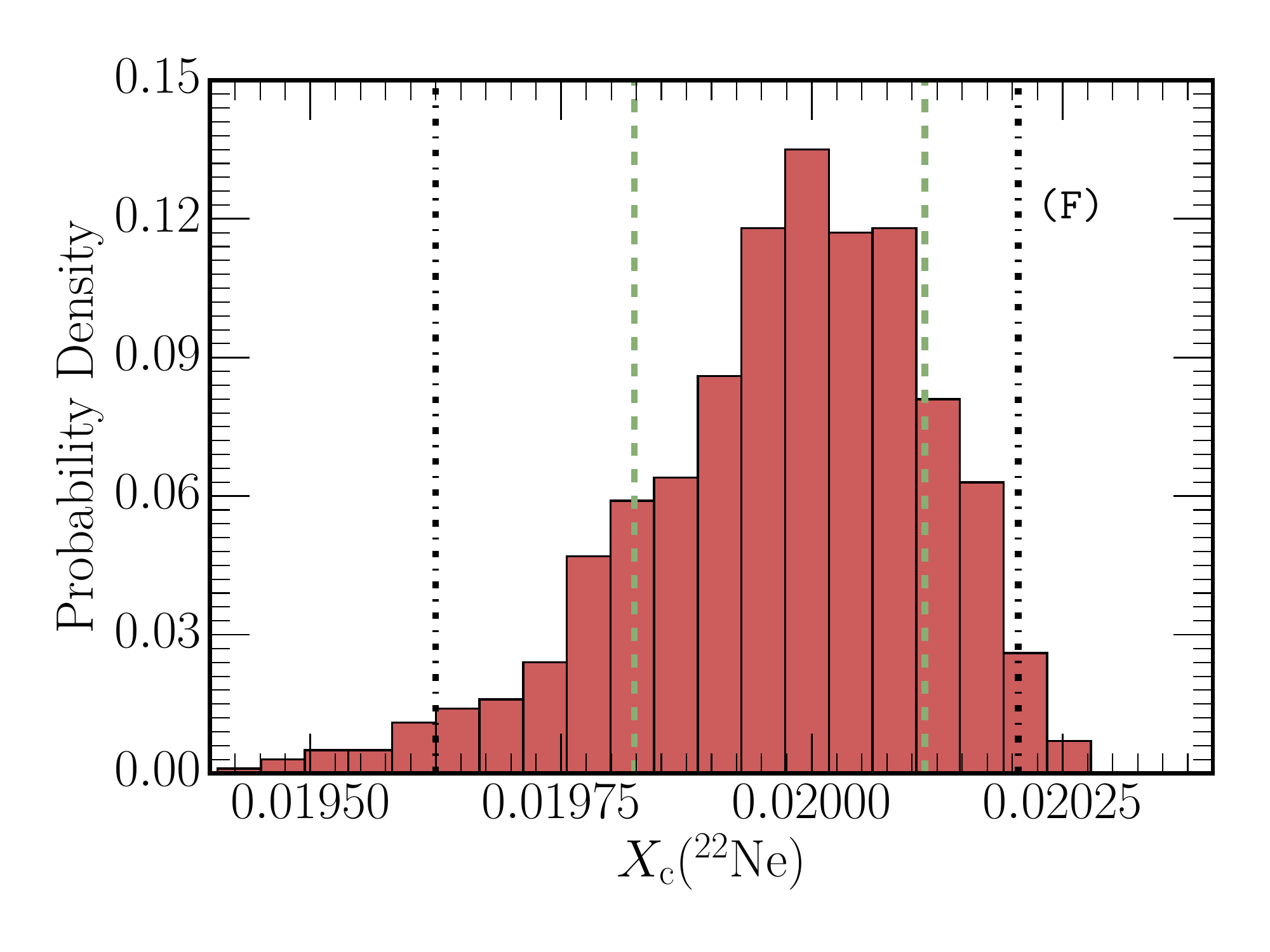}}
        \end{subfigure}
                        \begin{subfigure}{
                \includegraphics[trim = .25in .4in .25in .25in, clip,width=2.75in,height=2.1in]{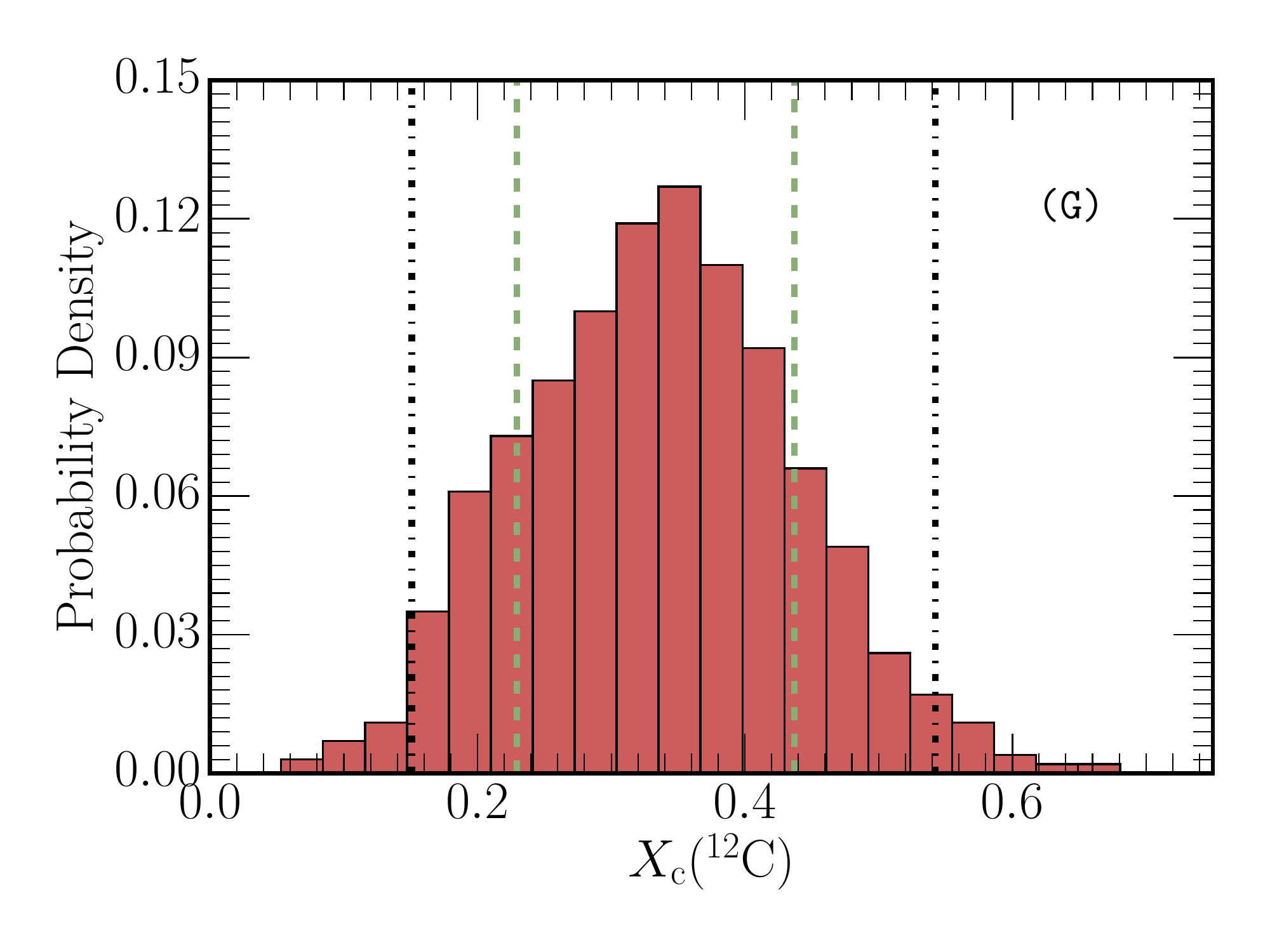}}
        \end{subfigure}
        \begin{subfigure}{
                \includegraphics[trim = .25in .4in .25in .25in, clip,width=2.75in,height=2.1in]{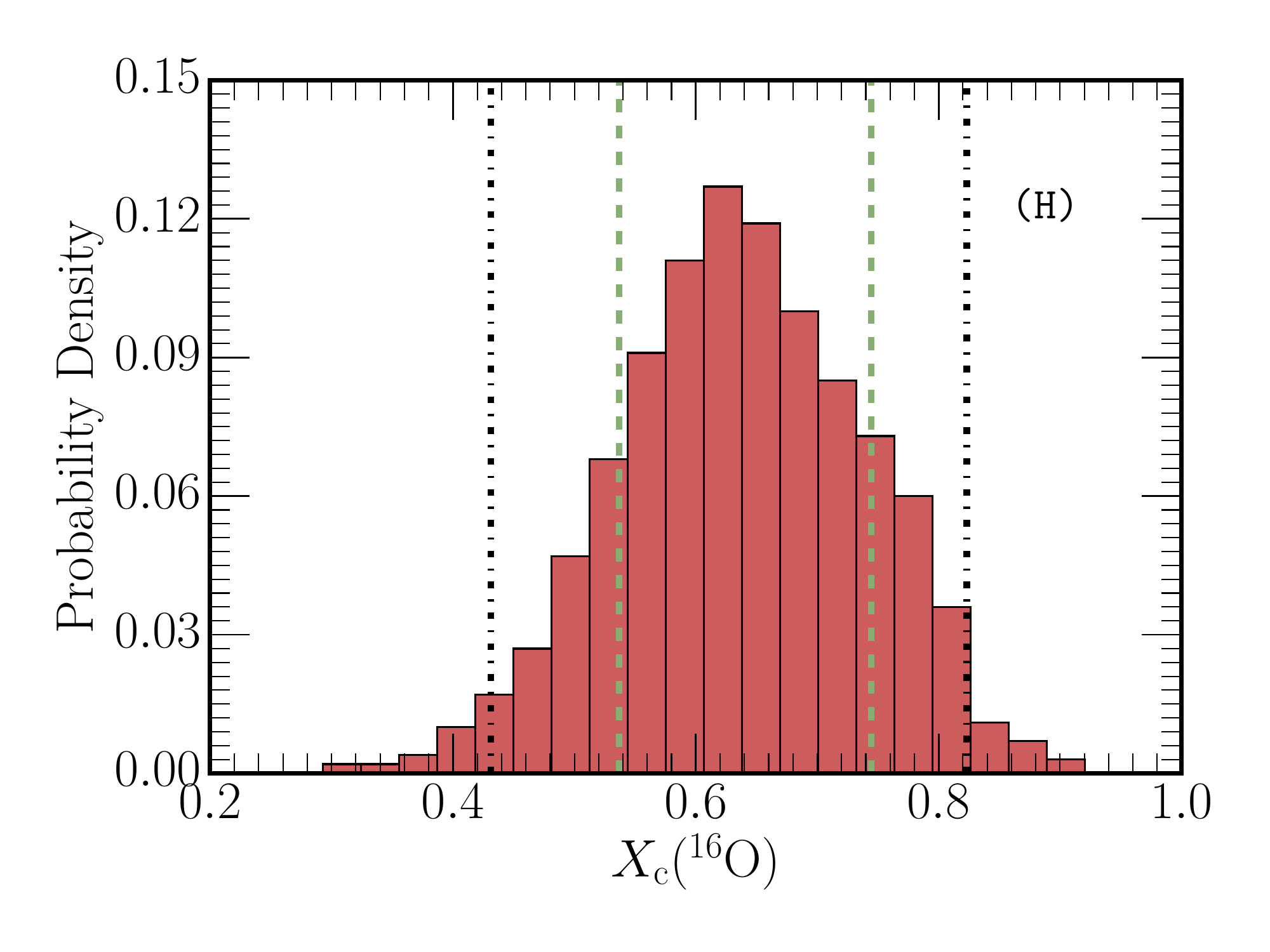}}
        \end{subfigure}
        \caption{
        Histograms of the 1,000 3 \msun Monte Carlo stellar 
        model grid sampling 26 STARLIB reaction rates. Global and
        core properties shown include 
        the mass of the CO core \texttt{(A)}, 
        age \texttt{(B)}, 
        central temperature \texttt{(C)}, 
        central density \texttt{(D)},  
        central electron fraction \texttt{(E)}, 
        central $^{22}$Ne mass fraction \texttt{(F)},
        central $^{12}$C mass fraction \texttt{(G)}, and
        central $^{16}$O mass fraction \texttt{(H)}, all at first thermal pulse. 
        The green dashed lines and the dot dashed red lines denote the 
        68\% and 95\% confidence intervals, respectively. The mean
        values for the eight quantities shown here, are enumerated in 
        Table~\ref{tbl:mcsm_props_1}.
        }\label{fig:grid1_hist}
\end{figure*}

We evolve 1,000 3 \msun models from the pre-MS to the
first thermal pulse (1TP), with each model using a different set of
sampled nuclear reaction rates from STARLIB. \MESA defines the 1TP
as the evolutionary point at which the central He
mass fraction, X$_{\rm c}(^{4}\rm{He})$, is depleted, the He
shell mass is $\le$ 0.2 \msun, and their is presence
of a convective zone with helium burning. Once a 3 \msun model has reached this
evolutionary point, the newly formed convective core is composed
primarily of CO.  Evolution to this key event allows a robust grid of
1,000 stellar models to be computationally feasible; each stellar
model in the grid took $\approx$ 24 core-hours to reach the
1TP.  Additionally, stopping the stellar model at the 1TP allows an
unbiased means to identify critical reaction rates in the H and He 
burning processes leading to the formation of a CO WD. For a 
Gaussian distribution with standard deviation of unity, we expect a 
standard error of $\sigma / \sqrt{1000} \approx 3\%$ for each independently 
sampled reaction rate.

\begin{deluxetable}{ccccc}{!htb}
\tablecolumns{5}
\tablewidth{0.95\linewidth}
\tablecaption{Variations in core quantities}
\tablehead{\colhead{Variable} & \colhead{Mean} & \colhead{68\% C.I.}  & \colhead{95\% C.I.}
& \colhead{\% Change}}
\startdata
$M_{\rm{1TP}}$ (M$_{\odot}$) 
& 5.838(-1) & 9.582(-3) & 1.884(-2) & 3.227 \\

$t_{\rm{1TP}}$ (Myr) 
& 4.786(2) & 6.611 & 1.250(1) & 2.612 \\

log($T_{{\rm c}}/{\rm K})$  
& 8.059 & 6.164(-3) & 1.269(-2) & 0.157 \\

log($\rho_{{\rm c}}/{\rm g \ cm}^{-3}$)
& 6.349 & 3.252(-2) & 5.938(-2) & 0.935 \\

$Y_{\rm{e,c}}$ 
& 4.990(-1) & 1.277(-5) & 2.557(-5) & 0.005 \\

X$_{\rm{c}}(^{22}$Ne) 
& 1.997(-2) & 2.898(-4) & 5.812(-4) & 2.910 \\

X$_{\rm{c}}(^{12}$C) 
& 3.365(-1) & 2.076(-1) & 3.916(-1) & 116.4 \\

X$_{\rm{c}}(^{16}$O) 
& 6.372(-1) & 2.076(-1) & 3.920(-1) & 61.52
\enddata
\tablecomments{The ($n$) after a given value, A, denotes A $\times$ 10$^{n}$.}
\label{tbl:mcsm_props_1}
\end{deluxetable}

Figure~\ref{fig:grid1_hist} shows histograms of 
$M_{\rm{1TP}}$ (A), 
$t_{{\rm 1TP}}$ (B), 
\Tc (C), 
\rhoc\ (D),  
$Y_{\rm e,c}$ (E), 
X$_{\rm c}$($^{22}$Ne) (F),
X$_{\rm c}$($^{12}$C) (G), and
X$_{\rm c}$($^{16}$O) (H), all at the 1TP. 
The number of bins is chosen according to the Rice Rule, $k=[2n^{1/3}]$, where 
$k$ is the number of bins and $n$ is the number of data points \citep{rice_rule}.
The green dashed lines and the dot dashed black lines denote the 
68\% and 95\% confidence intervals (C.I.), respectively. 
Relative to the arithmetic mean value, we find the width of the
95\% confidence interval to be
\mbox{$\Delta M_{{\rm 1TP}}$ $\approx$ 0.019 M$_{\odot}$} for the core mass
at the 1TP, 
\mbox{$\Delta$$t_{\rm{1TP}}$ $\approx$ 12.50} Myr for the age,
$\Delta \log(T_{{\rm c}}/{\rm K}) \approx$ 0.013 for the central temperature,
$\Delta \log(\rho_{{\rm c}}/{\rm g \ cm}^{-3}) \approx$ 0.060 for the central density,
$\Delta Y_{\rm{e,c}} \approx$ 2.6$\times$10$^{-5}$ for the central electron fraction,
\mbox{$\Delta X_{\rm{c}}(^{22}\rm{Ne}) \approx$ 5.8$\times$10$^{-4}$}, 
\mbox{$\Delta X_{\rm{c}}(^{12}\rm{C}) \approx$ 0.392}, and
\mbox{$\Delta X_{\rm{c}}(^{16}\rm{O}) \approx$ 0.392}.
In Table~\ref{tbl:mcsm_props_1} we compile the arithmetic mean values
of the eight quantities shown in Figure~\ref{fig:grid1_hist}, the
width of the 68\% and 95\% confidence intervals, and the percentage change
from the arithmetic mean using the 95\% confidence interval. To find the main sources 
of these global variations we use a Principal Component Analysis and 
Spearman Rank-Order Correlation.

\subsection{Principal Component Analysis}
\label{sec:pca}

\begin{figure*}[!ht]
\centering
\includegraphics[width=1.5\columnwidth]{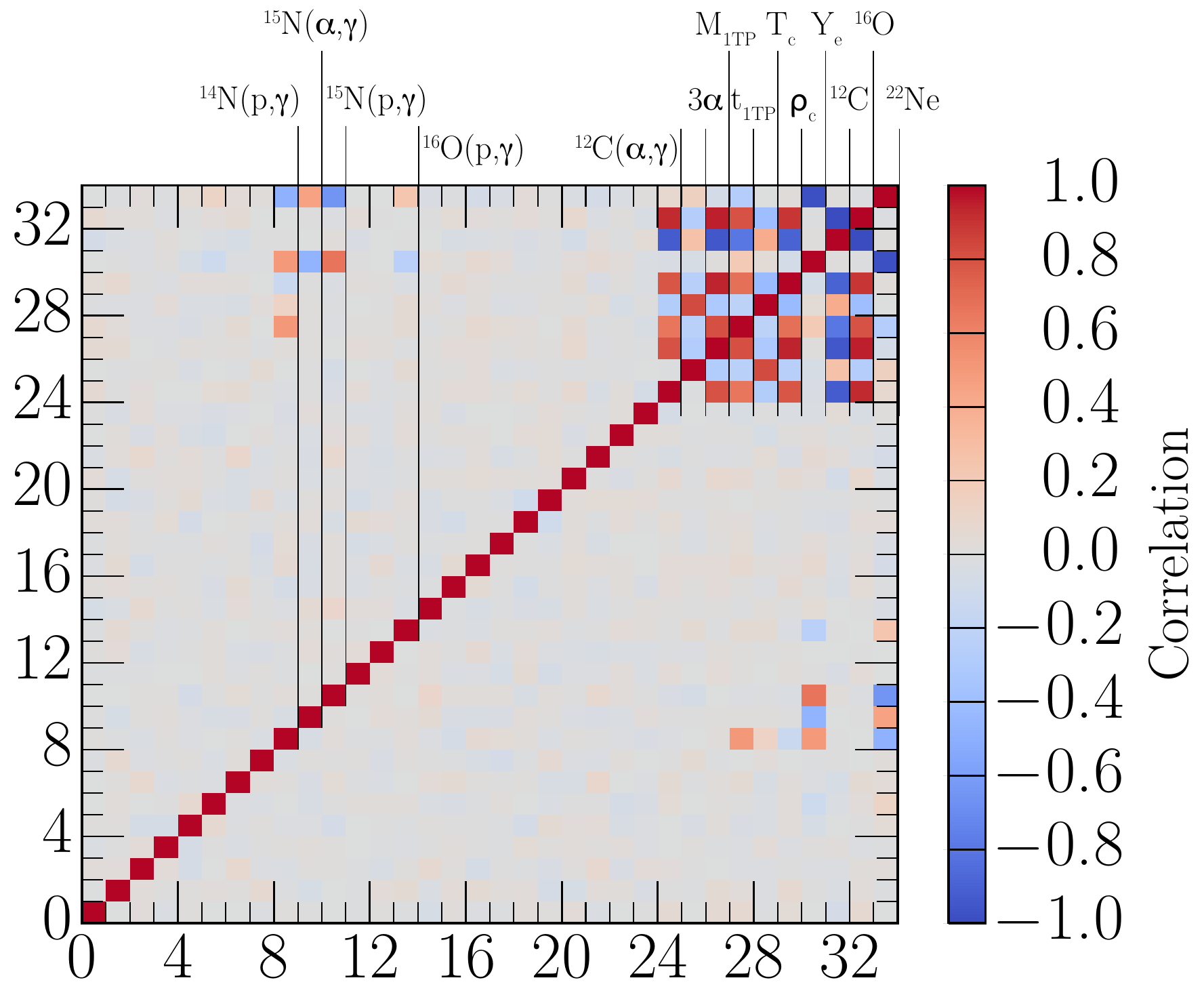}
        \caption{
        PCA correlation matrix for the 26 STARLIB reaction rates and 8 quantities
        of the CO remnant for the 1,000 MC 3 \msun stellar models. Labeled are the reaction rates whose
        experimental uncertainties show a significant correlation with the
        8 quantities measured at the 1TP. The colored box to the left to a vertical line
        is the labeled quantity.
        }\label{fig:corr_matrix}
\end{figure*}

Principal Component Analysis (PCA) is a statistical method used 
to transform a set of possibly correlated observations to a 
set linearly uncorrelated variables, referred to as principal
components 
\citep{jolliffe_2002_aa,jackson_2005}.
Among numerous applications, a PCA has been used to 
analyze large scale spectra of the Interstellar Medium 
\citep{heyer_1997_aa}, to classify optical stellar spectra of 
O to M type stars \citep{singh_1998_aa}, and 
investigate the far infrared spectral density of simulated 
galaxies \citep{safarzadeh_2016_aa}. 
Our goal with a PCA is to determine 
which reaction rate uncertainties account for most of the global variability in 
the \mbox{3 \msun}  MC stellar evolution models.
We consider 34 quantities in total $-$ the 26 STARLIB reaction rate variation factors
in the order listed in Table~\ref{tbl:sampled_rates},
$M_{{\rm 1TP}}$, 
$t_{\rm 1TP}$, 
$T_{\rm{c}}$,
\rhoc, 
$Y_{\rm e,c}$, 
X$_{\rm c}$($^{12}$C),
X$_{\rm c}$($^{16}$O), and 
X$_{\rm c}$($^{22}$Ne)
all at the 1TP.

We use the Python NumPy module {\tt corrcoef} \citep{walt_2011_aa} to calculate the 34x34
correlation matrix ${\bf C}$ of the 34x1000 data matrix ${\bf A}$,
where the correlation matrix is related to the covariance matrix ${\bf Cov}$ 
by ${\bf C}_{ij} = {\bf Cov}_{ij}/(\sigma_i \sigma_j)$ where
$\sigma_i$ and $\sigma_j$ are the standard deviations.
Figure~\ref{fig:corr_matrix} shows the symmetric
correlation matrix. Reaction rate uncertainties which show a significant correlation
(red) or anti-correlation (blue) are labeled along with the 8
quantities of the CO remnant at the 1TP.  Most of the lower left part
of the correlation matrix shows no correlation showing the reaction
rates are generally independent of each other.  We note here that
$M_{{\rm 1TP}}$ and $t_{\rm 1TP}$ are correlated with
$^{12}$C($\alpha$,$\gamma$), \rhoc, and X$_{\rm c}$($^{16}$O) while being
anti-correlated with 3$\alpha$ $T_{\rm{c}}$, and X$_{\rm c}$($^{12}$C). The stellar age
$t_{\rm 1TP}$ is correlated with the slowest reaction rate of the CNO-cycle
$^{14}$N($p,\gamma$), and $Y_e$ is correlated with
$^{14}$N($p,\gamma$), $^{15}$N($p,\alpha$), $^{15}$N($p,\gamma$), and
to a lesser extent $^{16}$O($p,\gamma$). In addition $T_{\rm{c}}$ and
\rhoc~are anti-correlated as is $Y_{\rm{e}}$ and X$_{\rm c}$($^{22}$Ne). We defer a
more detailed physical interpretation of these trends to \S\ref{sec:properties}.

\begin{deluxetable}{cccc}{b}
\tablecolumns{4}
\tablewidth{0.8\linewidth}
\tablecaption{Principal Component Analysis}
\tablehead{\colhead{Component} & \colhead{Eigenvalue} & \colhead{Proportion}  & \colhead{Cumulative}}
\startdata
1 & 5.5917 & 0.1645 & 0.1645 \\
2 & 3.0743 & 0.0904 & 0.2549 \\
3 & 1.6734 & 0.0492 & 0.3041 \\
4 & 1.2910 & 0.0380 & 0.3421 \\ 
5 & 1.2502 & 0.0368 & 0.3789
\enddata
\label{tbl:pca_grid_1}
\end{deluxetable}

We use the Python NumPy module, {\tt linalg.eig} \citep{walt_2011_aa}, to calculate the
eigenvectors and eigenvalues of the correlation matrix shown in
Figure~\ref{fig:corr_matrix}. A principal component is constructed 
using the eigenvector coefficients of a given eigenvalue. The first 
principal component takes the form, 
\begin{equation}
Y_{\rm{1}}=e_{1,1}X_{1}+e_{1,2}X_{2}+ \ldots e_{1,34}X_{34}~,
\end{equation}
where $X_1$ is the first row in the data matrix {\bf{A}},  and $e_{\rm{i,j}}$ is 
the coefficient of the eigenvector for the $i^{\rm{th}}$ principal component and 
$j^{\rm{th}}$ quantity.
The proportion of variation explained 
by the $i^{\rm{th}}$ principal component can be characterized by the ratio
$i^{\rm{th}}$ eigenvalue to that of the summation of all eigenvalues. 
In Table~\ref{tbl:pca_grid_1} we show the eigenvalues, proportion of 
total variation, and cumulative proportion of total variation for the first 
five principal components of our correlation matrix, ${\bf C}$. The first
five components account for $\approx$ 38\% of the total variation with 
only $\approx$ 16\% accounted for by the first component. 

To determine which quantities have the most  impact on the total 
variation we consider the relative contributions of each coefficient. The largest 
coefficient in the first principal component, with a value of +0.517,
corresponds to \Tc. The two subsequent coefficients that
are largest in magnitude are (-0.428, -0.404), corresponding to  
X$_{\rm c}$($^{12}$C) and X$_{\rm c}$($^{16}$O), respectively. Next, we consider
the effect of the sampled reaction rates in individual physical quantities using
the method of Spearman Rank-Order Correlation.

\subsection{Spearman Rank-Order Correlation}
\label{sec:src}

\begin{figure*}[!ht]
\centering
\includegraphics[trim = .1in .1in .75in .5in, clip,width=6.5in,height=5in]{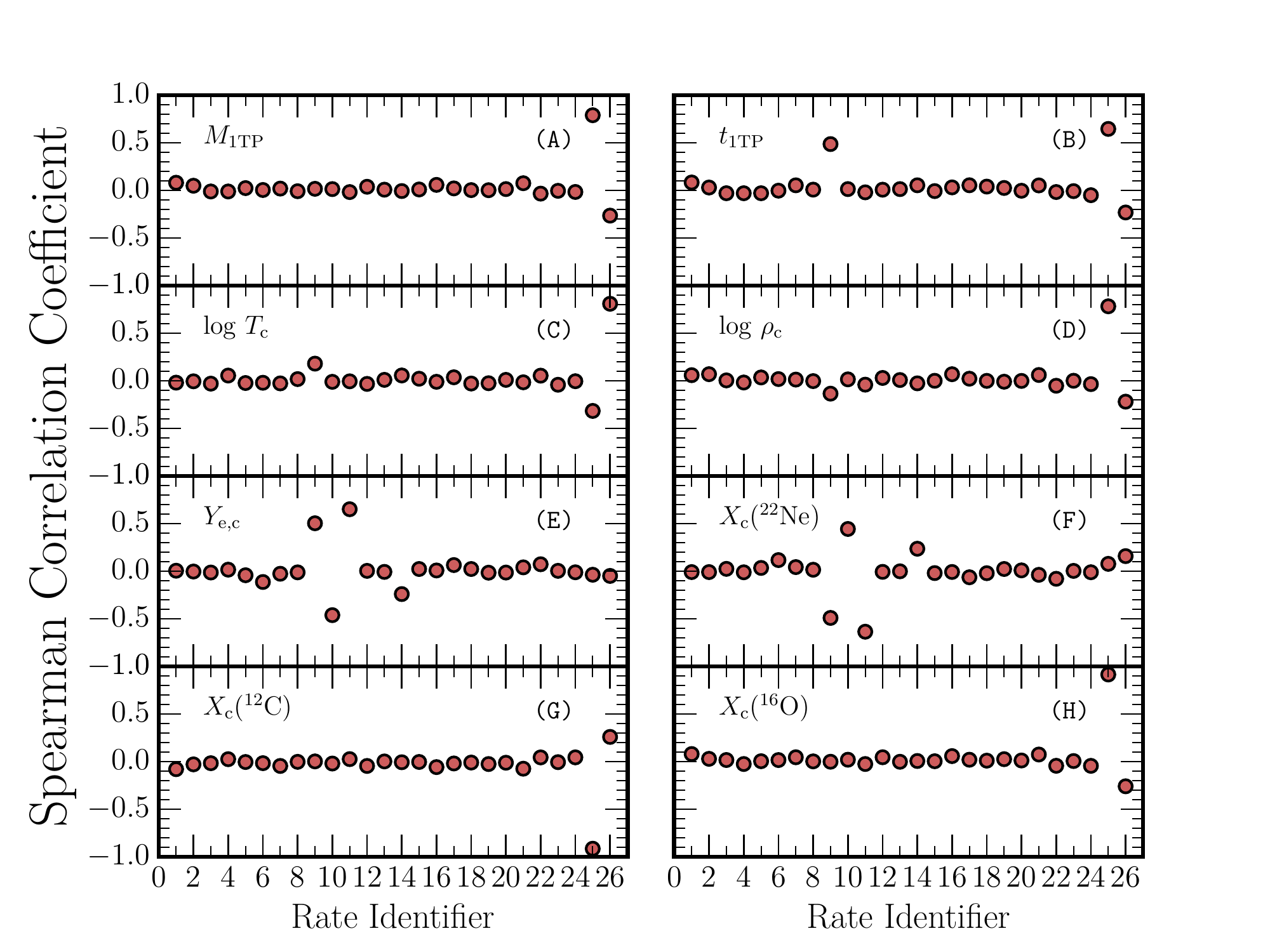}
        \caption{
        Spearman rank correlation coefficients for the 26 
        independently sampled nuclear reaction rates for 1,000 Monte
        Carlo stellar models with a ZAMS mass of 3 \msun. 
        The rate identifiers correspond to those listed in Table~\ref{tbl:sampled_rates}.
        Each sampled rate is compared individually 
        against the final mass of the CO core ($M_{{\rm 1TP}}$, \texttt{A}),
        age of stellar model ($t_{{\rm 1TP}}$, \texttt{B}),
        central temperature ($T_{{\rm c}}$, \texttt{C}),
        central density ($\rho_{{\rm c}}$, \texttt{D}),     
        central electron fraction ($Y_{{\rm e,c}}$, \texttt{E}),   
        central neon 22 mass fraction (X$_{{\rm c}}$($^{22}$Ne), \texttt{F}),   
        central $^{12}$C mass fraction (X$_{{\rm c}}$($^{12}$C), \texttt{G}), and  
        central $^{16}$O mass fraction (X$_{{\rm c}}$($^{16}$O), \texttt{H})
        at the 1TP. 
        }\label{fig:all_rhos_grid_1}
\end{figure*}

Spearman Rank-Order Correlation determines if there is a monotonically 
increasing or decreasing relationship between two scalar quantities.
The Spearman correlation coefficient, $r_{\rm{s}}$, is defined as the 
Pearson's correlation coefficient between ranked variables. For a 
sample size $N$, the data are converted to ranks and $r_{\rm{s}}$
is computed as,
\begin{equation}
\rhos = 1 - \frac{6}{N(N^2-1)}\sum\limits_{i=1}^{N} d_{\rm{i}}^2~,
\label{eq:spearman_rho}
\end{equation}
where $d_{\rm{i}}$ is the difference between the ranks of two scalar quantities.
A value of \rhos = +1 would represent a perfectly monotonically increasing relationship,
\rhos = 0, perfectly uncorrelated, and \rhos=-1, monotonically decreasing.

In Figure~\ref{fig:all_rhos_grid_1}, we show the Spearman coefficients
against the 26 independently sampled rate variation factors for
$M_{{\rm 1TP}}$ (A),
$t_{{\rm 1TP}}$ (B),
$T_{{\rm c}}$ (C),
$\rho_{{\rm c}}$ (D),     
$Y_{{\rm e,c}}$ (E),   
X$_{{\rm c}}$($^{22}$Ne) (F),   
X$_{{\rm c}}$($^{12}$C) (G), and  
X$_{{\rm c}}$($^{16}$O) (H)
at the 1TP.          

The uncertainties in the nuclear
reaction rates that have the largest effect on 
$M_{{\rm 1TP}}$ are
$^{12}\textup{C}(\alpha,\gamma)$ with \rhos=+0.79
and triple-$\alpha$ with \rhos=-0.26.  
The uncertainties in the remaining rates have a
negligible effect on $M_{{\rm 1TP}}$, $-0.03 \lesssim \rhos \lesssim
+0.08$. The same uncertainties in the reaction rates have the
largest impact on the stellar age. However, the reaction 
$^{14}\textup{N}(p,\gamma)$ also plays a significant role with 
\rhos=+0.49. 
For $T_{\rm{c}}$ the uncertainties in the rates that show the largest 
coefficients are $^{12}\textup{C}(\alpha,\gamma)$ with \rhos=-0.32, 
triple-$\alpha$ with \rhos=+0.81, and $^{14}$N(p,$\gamma$) with \rhos=+0.18. 
For \rhoc, the same uncertainties in the reaction rates dominate with
\mbox{\rhos=(+0.78, -0.22, -0.13)}, respectively.  
The largest
coefficients for $Y_{\rm{e,c}}$ correspond to the uncertainties in
$^{14}$N(p,$\gamma$) with \rhos=+0.50, $^{15}$N(p,$\alpha$) with
\rhos=-0.46, and $^{15}$N(p,$\gamma$) with \rhos=+0.65.
Similarly, variations in X$_{\rm c}$($^{22}$Ne) depend mostly on the uncertainties 
in the same three reactions, where we find \mbox{\rhos=(-0.49, +0.44, -0.64)} respectively.
Related, X$_{\rm c}$($^{12}\textup{C})$ and X$_{\rm c}$($^{16}\textup{O}$)
are dominated only by $^{12}\textup{C}(\alpha,\gamma)\rm{^{16}O}$ and triple-$\alpha$.  
For X$_{\rm c}$($^{12}$C), the coefficients are 
\rhos=(-0.91, +0.26) respectively, 
while for X$_{\rm c}$($^{12}$O)
we find \rhos=(+0.91, -0.26) respectively.
These trends in the Spearman coefficients verify the PCA results of \S\ref{sec:pca}.

\subsection{Properties of Carbon-Oxygen White Dwarfs}
\label{sec:properties}

Here we discuss the individual nuclear reaction rates which have
been identified to have the most significant impact on the WD composition 
and structural properties. In the previous section we identified 
$^{14}$N(p,$\gamma$),
$^{15}$N(p,$\alpha$),
$^{15}$N(p,$\gamma$),
\mbox{triple-$\alpha$}, and
$^{12}\textup{C}(\alpha,\gamma)$
as having the largest impact on the physical properties of the WD. 

\subsubsection{$^{14}$N(p,$\gamma$)$^{15}$O}
\label{sec:n14_pg}

The $^{14}$N($p,\gamma$)$^{15}$O was identified as having a
significant impact on the age, \Tc, $Y_{\rm e,c}$ , and 
X$_{\rm c}$($^{22}$Ne).  As noted previously in \S\ref{sec:baseline} and
\S\ref{sec:pca}, this reaction is the slowest reaction in the CNO
cycle. This result was confirmed in the seminal first direct
measurements at astrophysical energies \citep{luna_2006_aa}.

\begin{figure}[!htb]        
\centering
\includegraphics[trim = .1in .1in 0in .1in, clip,width=3.4in,height=2.75in]{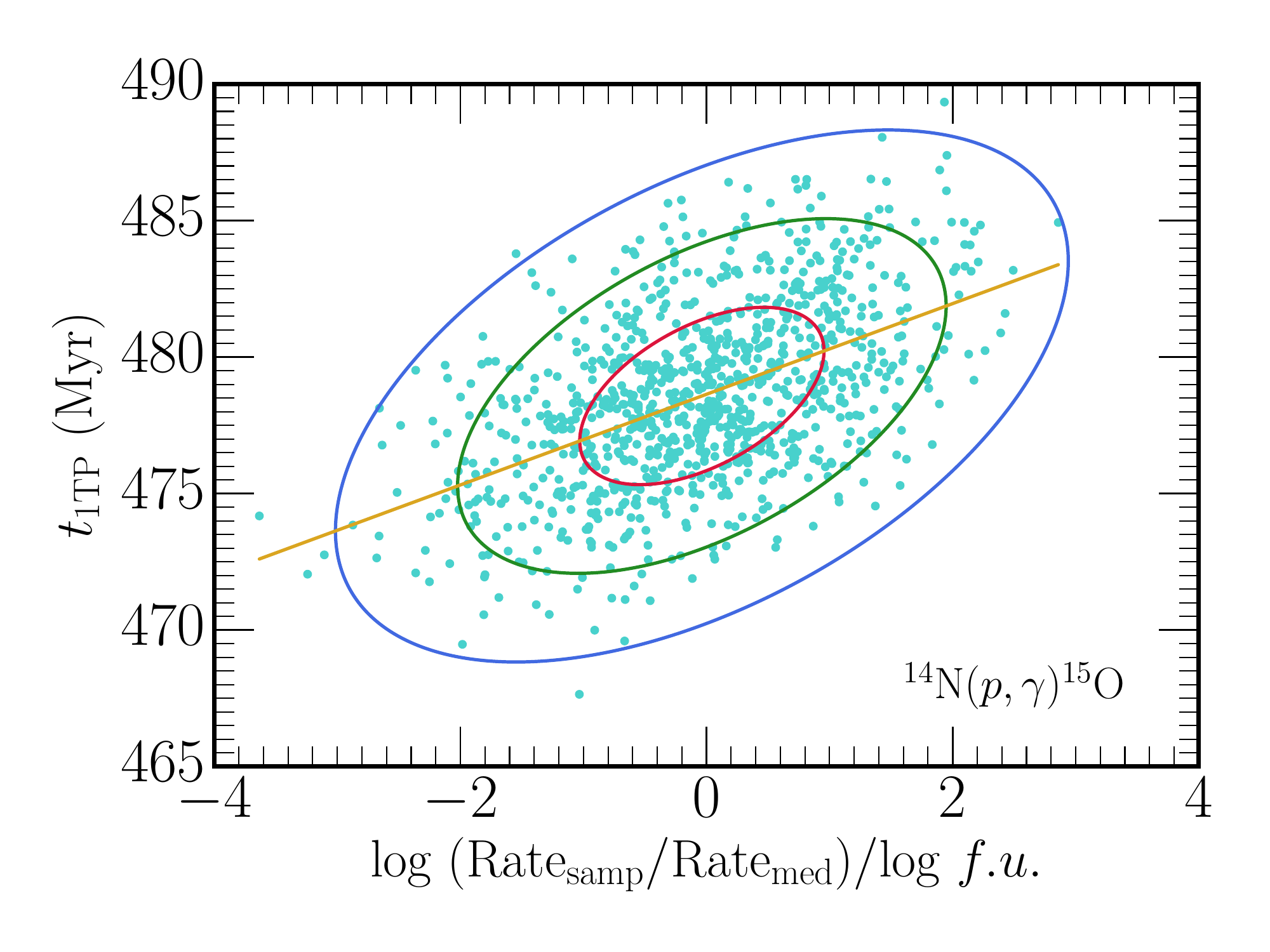}
\caption{
Rate variation factor for $^{14}$N($p,\gamma$)$^{15}$O 
versus stellar age at the 1TP, see Equation~(\ref{eq:sample_lambda}). 
Each of the 1,000 Monte Carlo 3 \msun models produces a data 
point in the figure. Overlaying the data points are the 1$\sigma$ (68\%, 
red), 2$\sigma$ (95\%, green) and 3$\sigma$ (99.7\%, blue) 
deviations about the mean of the data points. A linear regression
is performed on the raw data (gold solid line), yielding an $R^{2}$=0.257.
}\label{fig:n14_pg_age_dots}
\end{figure}

In Figure~\ref{fig:n14_pg_age_dots} we show the 
stellar age versus the effective rate variation factor. Each 
of the 1,000 \mbox{3 \msun} MC stellar models is a data
point in the figure. Also shown are the 1$\sigma$ (68\%, 
red), 2$\sigma$ (95\%, green) and 3$\sigma$ (99.7\%, blue) 
deviations about the mean. Orientation of the uncertainty 
ellipsoid is determined by the unit eigenvectors of the 2x2 covariance matrix, 
and the lengths of the semi-major and semi-minor axes of the ellipsoid 
correspond to the positive square roots of the two eigenvalues. 

To quantify how well the rate variation factor can account for the
increase in age, we perform a linear regression analysis. The
coefficient of determination ($R^2$ value) is the squared Pearson
product-moment correlation coefficient. For two quantities with a
perfectly linear relationship, $R^2$=1. The lower the $R^2$ value, the
less that the corresponding linear fit can account for the data. For
perfectly uncorrelated data, an $R^2$=0.  For the stellar age data, 
the regression analysis yields
$R^2$=0.257. This suggests that
although we find a relatively large Spearman coefficient for the rate
variation factor and the age, other rates play a significant factor in
this quantity as well.

\begin{figure}[!htb]        
\centering
\includegraphics[trim = .1in .1in 0in .1in, clip,width=3.4in,height=2.75in]{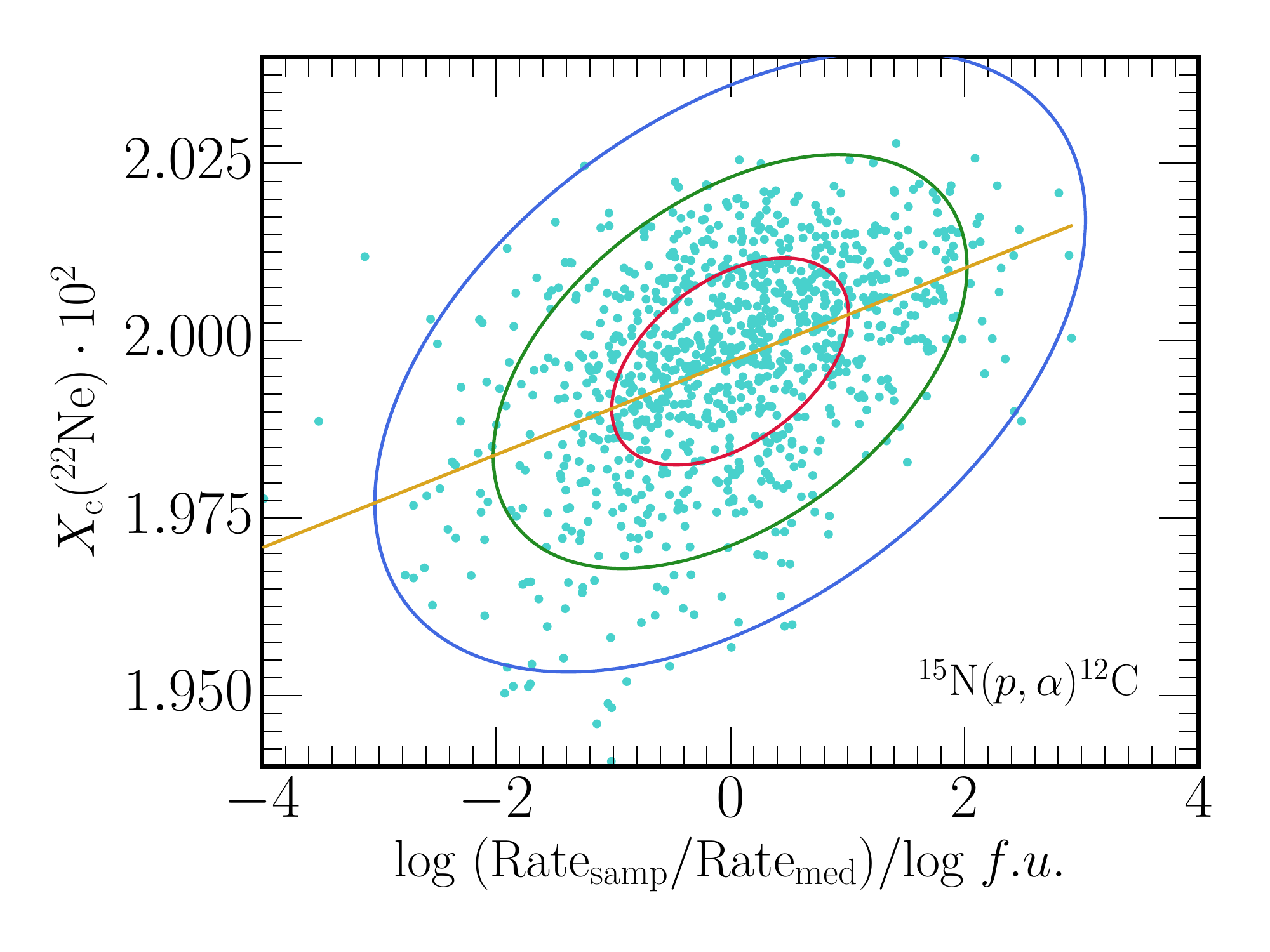}
\caption{
Same as in Figure~\ref{fig:n14_pg_age_dots} but for 
$^{15}$N($p,\alpha$)$^{12}$C 
versus X$_{\rm c}$($^{22}$Ne) at the 1TP. 
The linear regression
yields $R^{2}$ = 0.207. 
}\label{fig:n15_pa_ne22_dots}
\end{figure}

\begin{figure}[!htb]        
\centering
\includegraphics[trim = .1in .1in 0in .1in, clip,width=3.4in,height=2.75in]{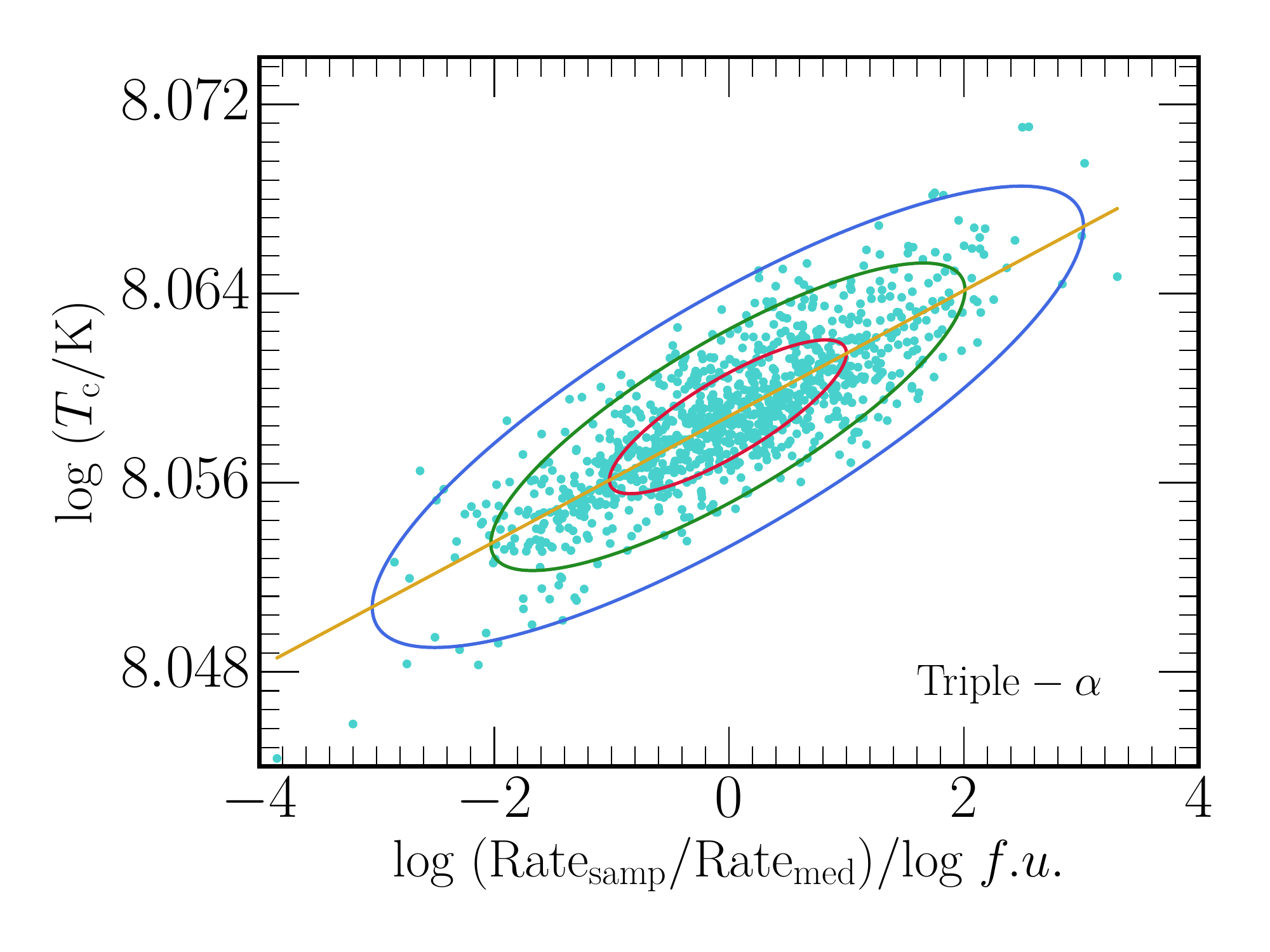}
\caption{
Same as in Figure~\ref{fig:n15_pa_ne22_dots} but for 
triple-$\alpha$
versus $T_{\rm{c}}$. The linear regression
performed yields an $R^{2}$=0.680. 
}\label{fig:3a_temp_dots}
\end{figure}

The first principal component can be represented by the line, the
major axis of the uncertainty ellipse, which minimizes the summed
squared distance of closest approach, i.e. the distance perpendicular
to the major axis line.  Linear least squares regression minimizes the
summed square distance in the one direction only. Thus, although the
two methods use a similar error metric, linear least squares treats
one dimension of the data preferentially, while PCA treats all
dimensions equally. Hence the major axis in Figure~\ref{fig:n14_pg_age_dots}
is not aligned with the least squares regression line.

For $T_{\rm{c}}$, $Y_{\rm{e,c}}$, and X$_{\rm{c}}$($^{22}$Ne) we 
find $R^{2}$=0.017, 0.246, and 0.239, respectively. The central $Y_{\rm e,c}$
and X$_{\rm c}$($^{22}$Ne) are anti-correlated
because X$_{\rm c}$($^{22}$Ne) largely determines $Y_{\rm e,c}$ 
\begin{equation}
Y_{\rm e,c} = \frac{1}{2} - \frac{X_{\rm c}(^{22}{\rm Ne})}{22} \ ~.
\label{eq:ye_ne22}
\end{equation}
An increase in X$_{\rm c}$($^{22}$Ne) decrease $Y_e$, accounting for the 
similar magnitudes of $R^{2}$.

\subsubsection{$^{15}$N(p,$\alpha$)$^{12}$C}
\label{sec:n15_pa}
Formed from the $^{14}$N(p,$\gamma$)$^{15}$O reaction, $^{15}$O
has a half-life of $\tau_{1/2} \approx$122 seconds undergoing beta decay
to form $^{15}$N. The creation of $^{15}$N signals the end of the 
catalytic CNO cycle to form $^{4}$He and $^{12}$C. We previously identified 
the $^{15}$N(p,$\alpha$)$^{12}$C reaction as having a significant impact on 
$Y_{\rm e,c}$ and X$_{\rm c}$($^{22}$Ne) -  see Equation~(\ref{eq:ye_ne22}).

Figure~\ref{fig:n15_pa_ne22_dots} shows the rate variation factor
for the $^{15}$N(p,$\alpha$)$^{12}$C reaction versus the scaled 
X$_{\rm c}$($^{22}$Ne). The ellipsoid width suggests 
the uncertainties from multiple
reactions impact X$_{\rm c}$($^{22}$Ne). The regression fit
gives $R^{2}$ = 0.207. For 3$\sigma$ changes in the
$^{15}$N(p,$\alpha$)$^{12}$C rate variation factor, only small changes in
X$_{\rm c}$($^{22}$Ne) are induced, with $\Delta$X$_{\rm c}$($^{22}$Ne)$\approx \pm 4 \times 10^{-4}$
and $\Delta Y_{\rm{e}} \approx \pm 4 \times 10^{-4}$.

\subsubsection{$^{15}$N(p,$\gamma$)$^{16}$O}
\label{sec:n15_pg}

The central $Y_{\rm e,c}$ and X$_{\rm c}$($^{22}$Ne) pair 
depends on the efficiency of the $^{15}$N(p,$\gamma$)$^{16}$O reaction.
In \S\ref{sec:baseline}, X$_{\rm c}$($^{22}$Ne) was shown to be 
well approximated by Equation~(\ref{eq:x22}). The Spearman correlation 
coefficient between the rate variation factor for $^{15}$N(p,$\gamma$)$^{16}$O 
and X$_{\rm{c}}$($^{22}$Ne) suggested that for a decrease in rate efficiency, 
there would be a decrease in X$_{\rm c}$($^{22}$Ne). This is confirmed through Equation~(\ref{eq:x22}).
As a result, we have a larger $Y_{\rm e,c}$, as suggested by Figure~\ref{fig:all_rhos_grid_1}.

\subsubsection{Triple-$\alpha$}
\label{sec:3a}

The triple-$\alpha$ reaction is one of the key nuclear reactions for the
synthesis of the elements and is the main energy source
during He burning. The reaction rate is dominated by resonances,
the best known being the Hoyle state at 7.65 MeV \citep{hoyle_1954_aa}, 
but there remains considerable interest in determining all
resonances with high precision \citep{chernykh_2010_aa}.
The triple-$\alpha$ rate directly affects $M_{{\rm 1TP}}$,
$t_{{\rm 1TP}}$, \Tc, \rhoc, X$_{\rm c}$($^{12}$C) and X$_{\rm c}$($^{16}$O).

Triple-$\alpha$ rate variation factors of $\approx$ 1, 2, and 3 produce 
variations of $\approx$ 0.008, 0.0192, and 0.03 \msun in the mass of the 
CO remnant. Increasing the triple-$\alpha$ rate generally leads to smaller 
CO core masses.
This is largely due to larger triple-$\alpha$ rates depleting He fuel
faster during shell He burning, leading to thinner He shell
masses with shorter shell He lifetimes and thus less massive CO cores
(see Figure~\ref{fig:kipp}). For the stellar age, the
mean trend is for larger triple-$\alpha$ rates to produce shorter
evolutionary times from the pre-MS to the 1TP, due to larger
triple-$\alpha$ rates depleting the available core He fuel
faster. A rate variation factor between -2.0 and 2.0 produces a
\mbox{$\approx$ 12 Myr} change in the evolutionary time to reach the 1TP
relative to the \mbox{478 Myr} using the median reaction rate in our 
\mbox{3 \msun} model. For the triple-$\alpha$ rate  versus $M_{{\rm 1TP}}$
and $t_{{\rm 1TP}}$, we find $R^2$=0.080 and 0.060, respectively. 
While the Spearman correlation coefficients obtained suggested a 
monotonically decreasing relationship for these two quantities, 
the majority of the variation cannot be accounted for by the triple-$\alpha$ 
reaction.

Figure~\ref{fig:3a_temp_dots} shows the strongly correlated 
relation between $T_{\rm{c}}$ at the 1TP and the triple-$\alpha$ rate.
Larger rates tend to produce larger $T_{\rm{c}}$, although the magnitude 
of the effect is relatively small, with a rate variation factor between -3.0
and 3.0 (corresponding to the extrema loci of the 3$\sigma$ ellipse), 
producing a $\approx$ 5.5$\times$10$^6$ K change in $T_{\rm{c}}$.

The central density \rhoc\ at the 1TP is less
correlated with triple-$\alpha$ rate than $T_{\rm{c}}$.
Larger rates tend to yield smaller \rhoc. A rate variation factor between 
-3.0 and 3.0 produces a $\approx$ 5.1$\times$10$^5$ g cm$^{-3}$ change
in \rhoc, a 23\% difference. 

\begin{figure}[!htb]        
\centering
\includegraphics[trim = .1in .1in 0in .1in, clip,width=3.4in,height=2.75in]{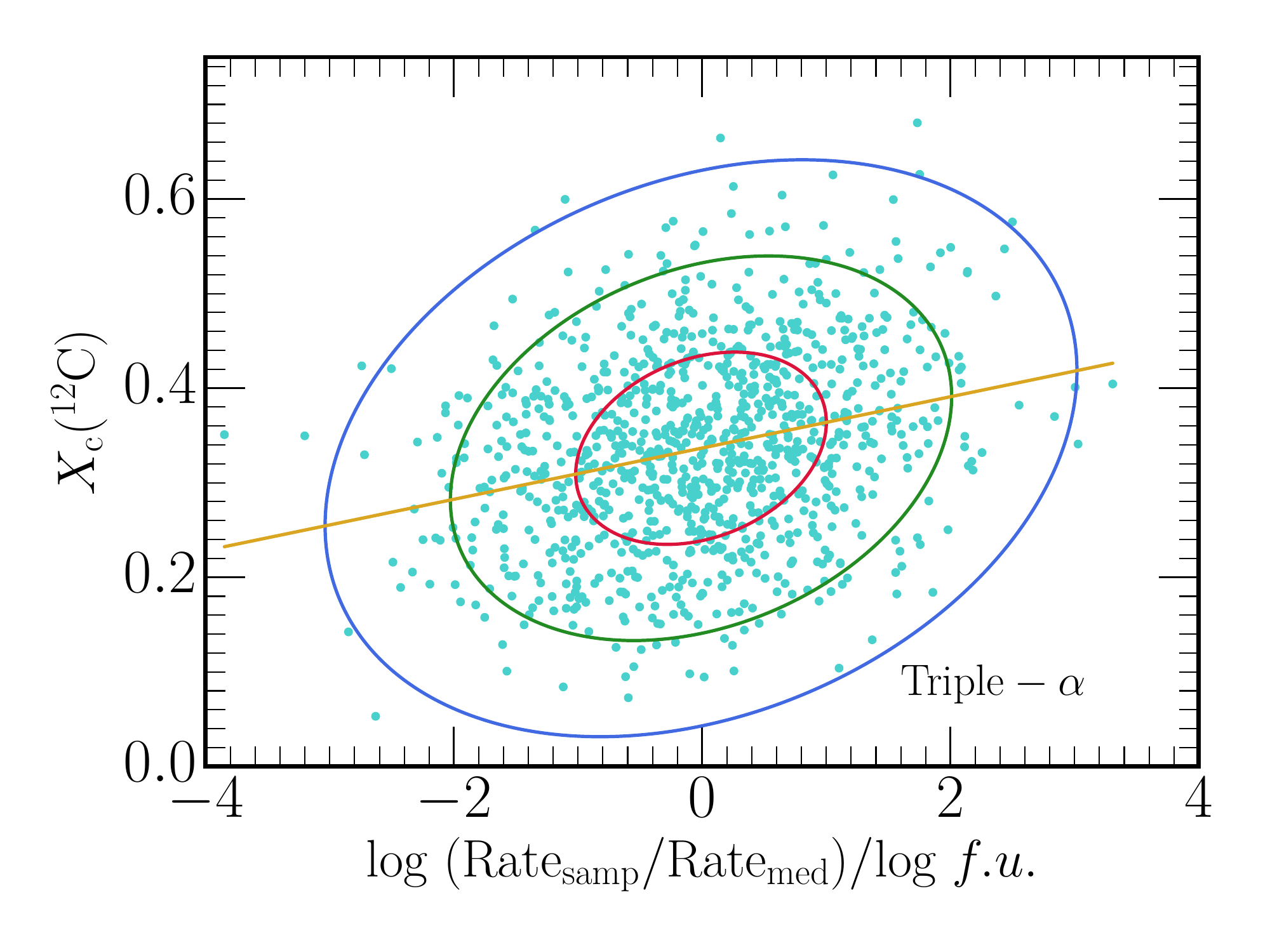}
\caption{
Same as in Figure~\ref{fig:3a_temp_dots} but for 
triple-$\alpha$
versus X$_{\rm{c}}$($^{12}$C). The linear regression
yields $R^{2}$ = 0.072. 
}\label{fig:3a_c12_dots}
\end{figure}

\begin{figure}[!htb]        
\centering
\includegraphics[trim = .1in .1in 0in .1in, clip,width=3.4in,height=2.75in]{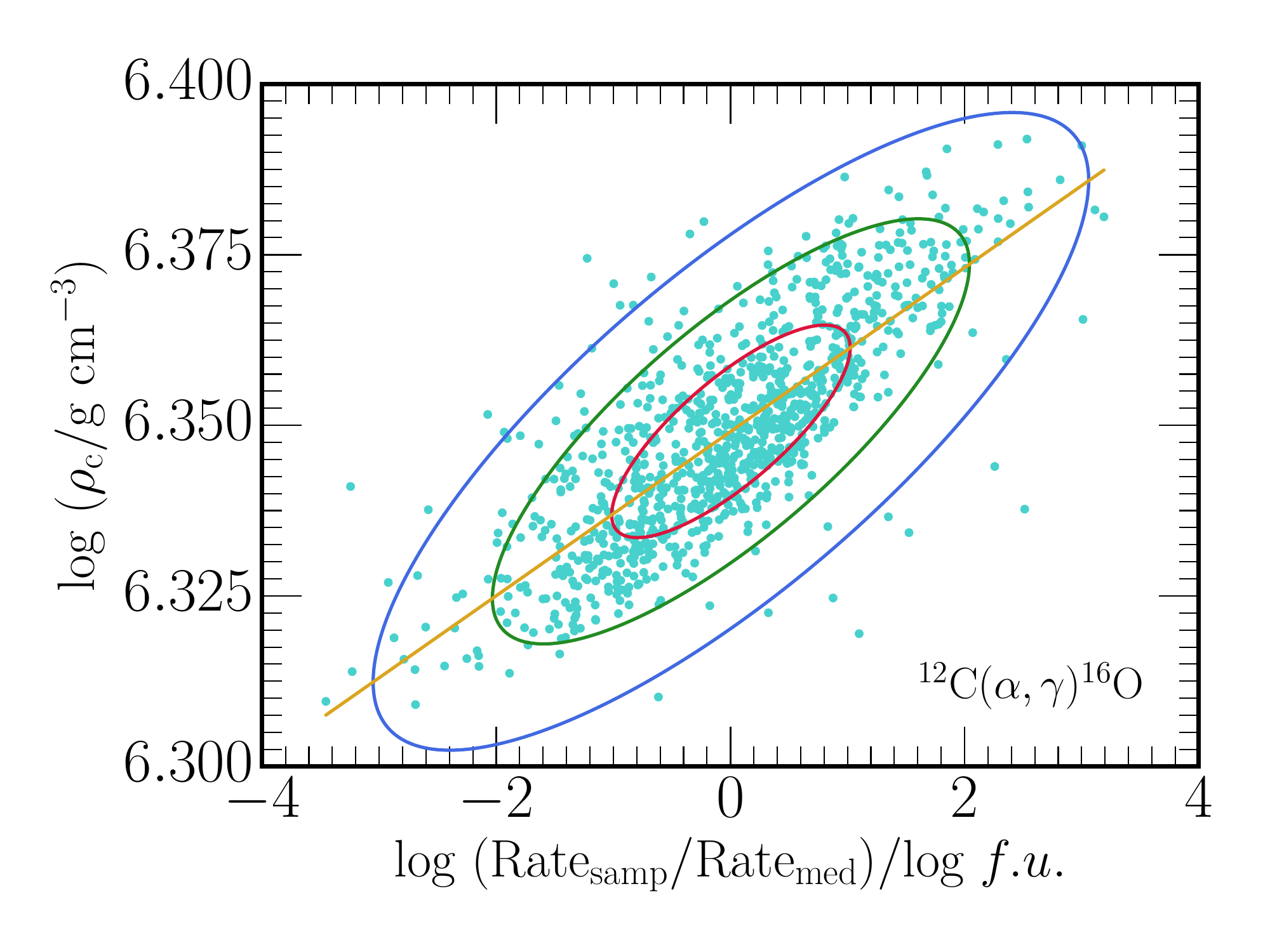}
\caption{
Same as in Figure~\ref{fig:3a_c12_dots} but for 
$^{12}$C($\alpha$,$\gamma$)$^{16}$O versus \rhoc \ at the 1TP.
The linear regression yields $R^{2}$ = 0.618. 
}\label{fig:c12_ag_rho_dots}
\end{figure}

\begin{figure}[!htb]        
\centering
\includegraphics[trim = .1in .1in 0in .1in, clip,width=3.4in,height=2.75in]{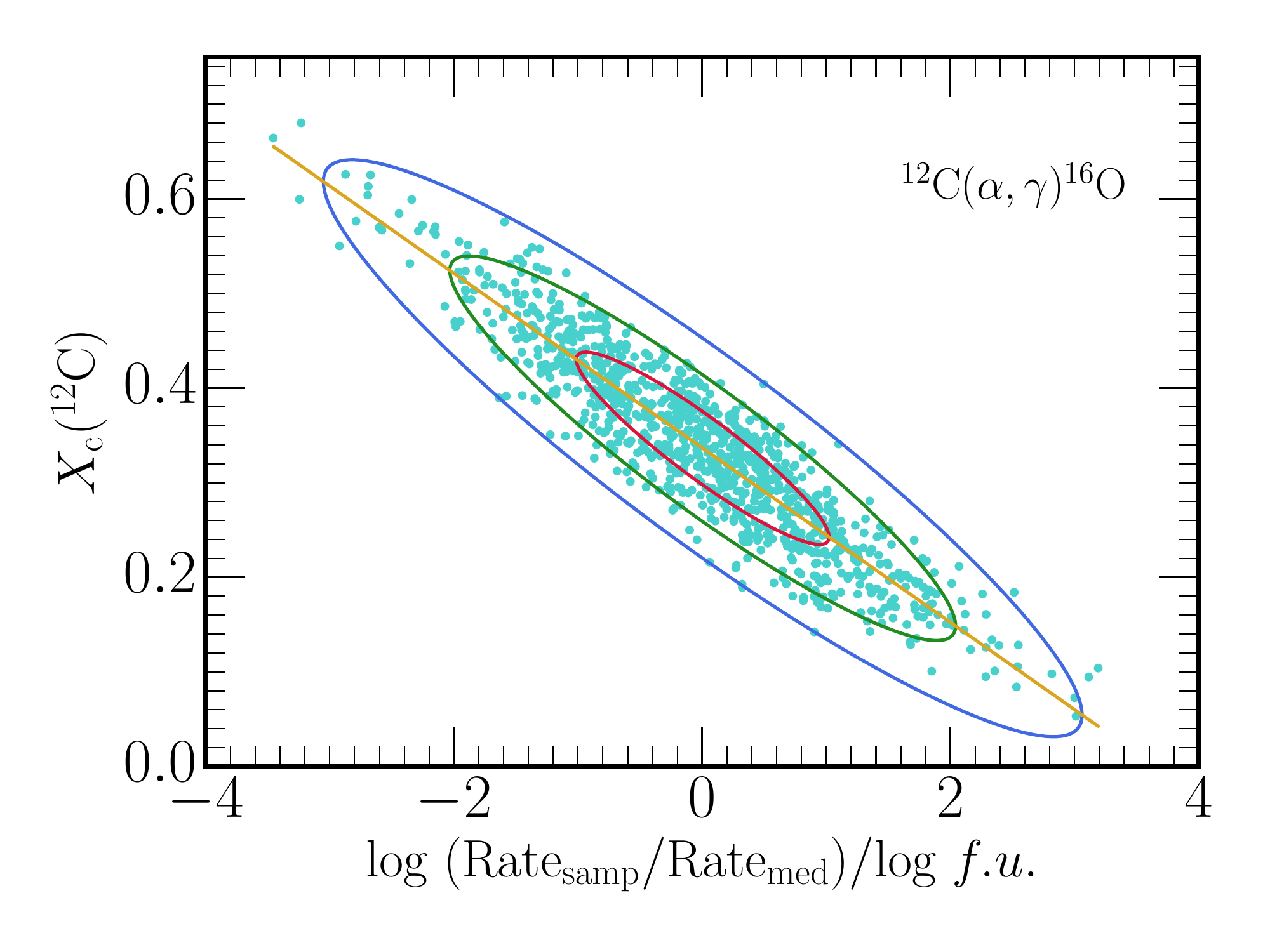}
\caption{
Same as in Figure~\ref{fig:c12_ag_rho_dots} but for 
$^{12}$C($\alpha$,$\gamma$)$^{16}$O
versus X$_{\rm c}$($^{12}$C0 at the 1TP.
The linear regression yields $R^{2}$ = 0.854. 
}\label{fig:c12_ag_c12_dots}
\end{figure}

The central X$_{\rm c}$($^{12}$C) and X$_{\rm c}$($^{16}$O) have 1$\sigma$ variations of 
$\Delta X_{\rm{c}}$($^{12}\textup{C}$)$\approx \pm$0.12 and 
$\Delta X_{\rm{c}}$($^{16}$O)$\approx \pm$ 0.15. Figure~\ref{fig:3a_c12_dots} 
shows the rate variation factor for triple-$\alpha$ versus 
X$_{\rm{c}}$($^{12}$C) at the 1TP. The linear regression 
provides an $R^{2}$=0.072. We see large scatter in the 
data about the linear fit suggesting that other reactions have
larger affects on the final $^{12}$C abundance.

\subsubsection{$^{12}$C($\alpha$,$\gamma$)$^{16}$O}
\label{sec:rcag}

The $^{12}$C($\alpha$,$\gamma$)$^{16}$O reaction is one of the most
fundamental yet complex reactions.  Experimental uncertainties arise
from difficulties in measuring the astrophysical S-factor due to the
small cross section at He burning temperatures and the complicated
level structure of the $^{16}$O nucleus
\citep{doboer_2013_aa,an_2015_aa}.  Moreover, a lack of resonances
near the Gamow window and strong interference between the ground state
captures introduce further uncertainties.

Our Monte Carlo stellar models utilize the \citet{kunz_2002_aa} rate
distribution for $^{12}$C($\alpha$,$\gamma$)$^{16}$O. Previous studies
have investigated the effect of varying this rate using a
multiplicative factor of $\pm$ 3 in the context of nucleosynthesis of
X-Ray Bursts \citep{parikh_2008_aa}. Our study differs in that
we vary our rate according to Equation~(\ref{eq:sample_lambda}). The
median rate value at a given temperature is multiplied by
$f.u.^{p_{\rm{i}}}$, allowing us to account for temperature dependent
changes in experimental uncertainty.

In general, an increase of the $^{12}$C($\alpha$,$\gamma$)$^{16}$O
reaction leads to a more massive core, prolonged stellar lifetimes,
and a decrease in $T_{\rm{c}}$ at the 1TP.  In contrast,
\rhoc\ increases with the $^{4}$He abundance. In
Figure~\ref{fig:c12_ag_rho_dots} we show the rate variation factor for
$^{12}$C($\alpha$,$\gamma$)$^{16}$O versus \rhoc\ at the 1TP, and find
$R^{2}$=0.618.  We find that $\approx$ 62\% of a 3$\sigma$ change in
the efficiency of the rate can alter the central density by 
$\Delta \log$(\rhoc/g cm$^{-3}$) $\approx$ $\pm$ 0.05.

Helium burning is initiated by the triple-$\alpha$ reaction. As isotopes of 
$^{12}$C are formed, the proportion of carbon to oxygen is determined 
by the competition between the carbon-producing triple-$\alpha$ reaction 
and the carbon-depleting, oxygen-producing radiative capture reaction.
The central X$_{\rm c}$($^{12}$C) and X$_{\rm c}$($^{16}$O) are strongly anti-correlated 
with $^{12}$C($\alpha$,$\gamma$)$^{16}$O reaction. In 
Figure~\ref{fig:c12_ag_c12_dots} we show the rate variation factor versus
the final $^{12}$C mass fraction at the 1TP. The linear regression
yields $R^2$=0.854. The production of $^{12}$C favors a larger
\rhoc\ while the opposite is true for $^{16}$O.

In most cases, the $^{12}\textup{C}(\alpha,\gamma)\rm{^{16}O}$
reaction dominates the CO core properties. It is only challenged
by the \mbox{triple-$\alpha$}
reaction in a vicious battle to consume the remaining He fuel in the
final stages of He burning.
The relative strength of these channels 
directly affects the amount $^{4}$He
processed either to $^{12}$C or $^{16}$O. Recent studies have investigated
the affect of altering $^{12}\textup{C}(\alpha,\gamma)\rm{^{16}O}$ 
in massive stars using multiplicative rate enhancement factors
\citep{west_2013_aa,tur_2007_aa}. Because of the overwhelming impact of this
reaction rate, the effect of other H and He burning rates may be overlooked
in the assessment of uncertainties in low mass stars. To address
this possibility, we consider a grid of Monte Carlo stellar models 
with a fixed rate distribution for $^{12}\textup{C}(\alpha,\gamma)\rm{^{16}O}$.

\section{With $^{12}$C($\alpha,\gamma$)$^{16}$O Fixed}
\label{sec:mcstars_fixed_c12ag}

We evolve an additional grid of 1,000 3 \msun stellar models from the 
pre-MS to the 1TP, using the same input physics as the grid in 
\S\ref{sec:mcstars}. Each model uses a fixed rate distribution for 
$^{12}\textup{C}(\alpha,\gamma)\rm{^{16}O}$ reaction, the STARLIB 
median rate distribution, while sampling the 25 remaining 
reaction rates listed in Table~\ref{tbl:sampled_rates}. 

\begin{figure*}[!htb]
         \centering
        \begin{subfigure}{
                \includegraphics[trim = .25in .25in .25in .25in, clip,width=2.75in,height=2.1in]{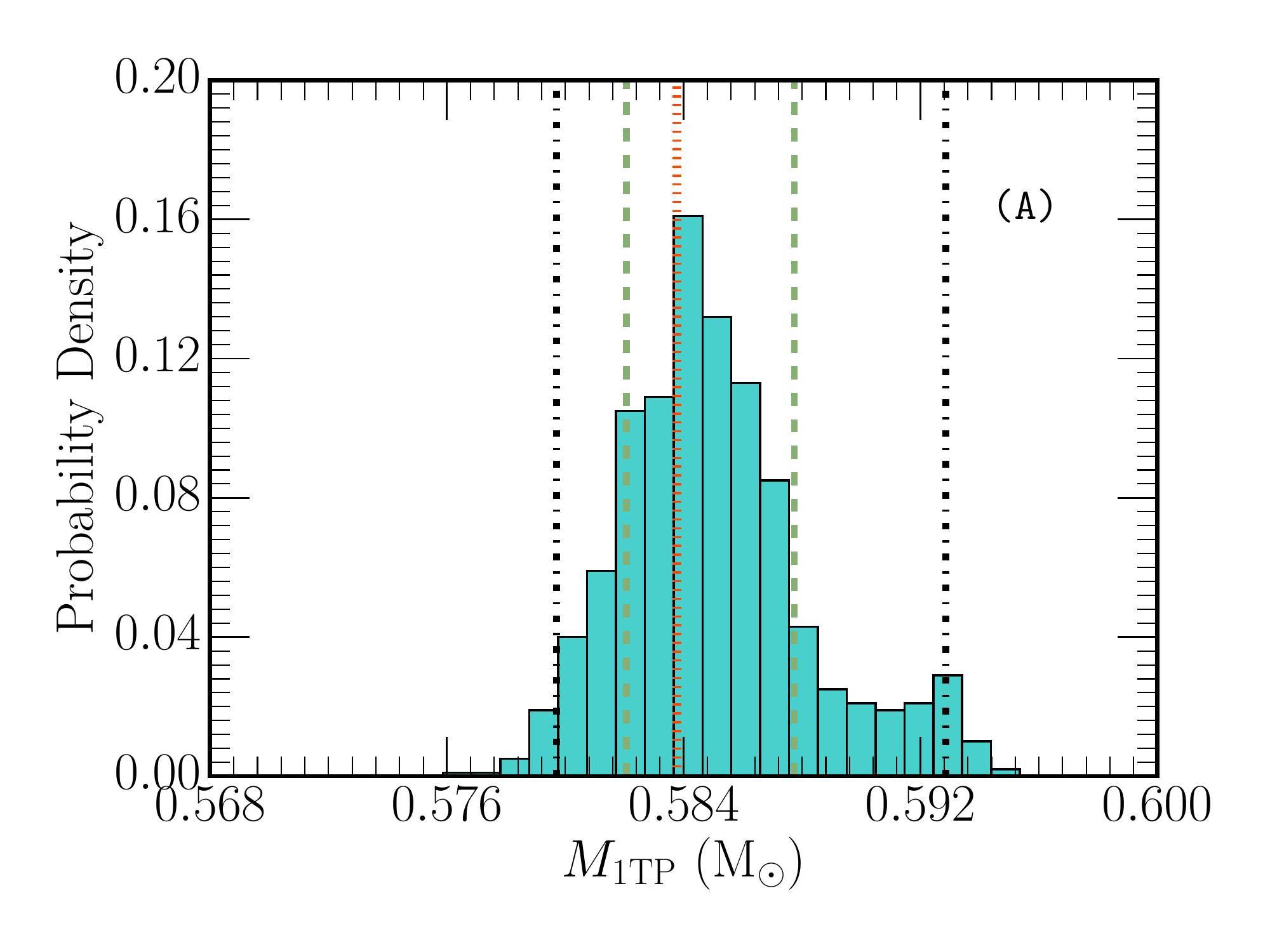}}
        \end{subfigure}
        \begin{subfigure}{
                \includegraphics[trim = .25in .25in .25in .25in, clip,width=2.75in,height=2.1in]{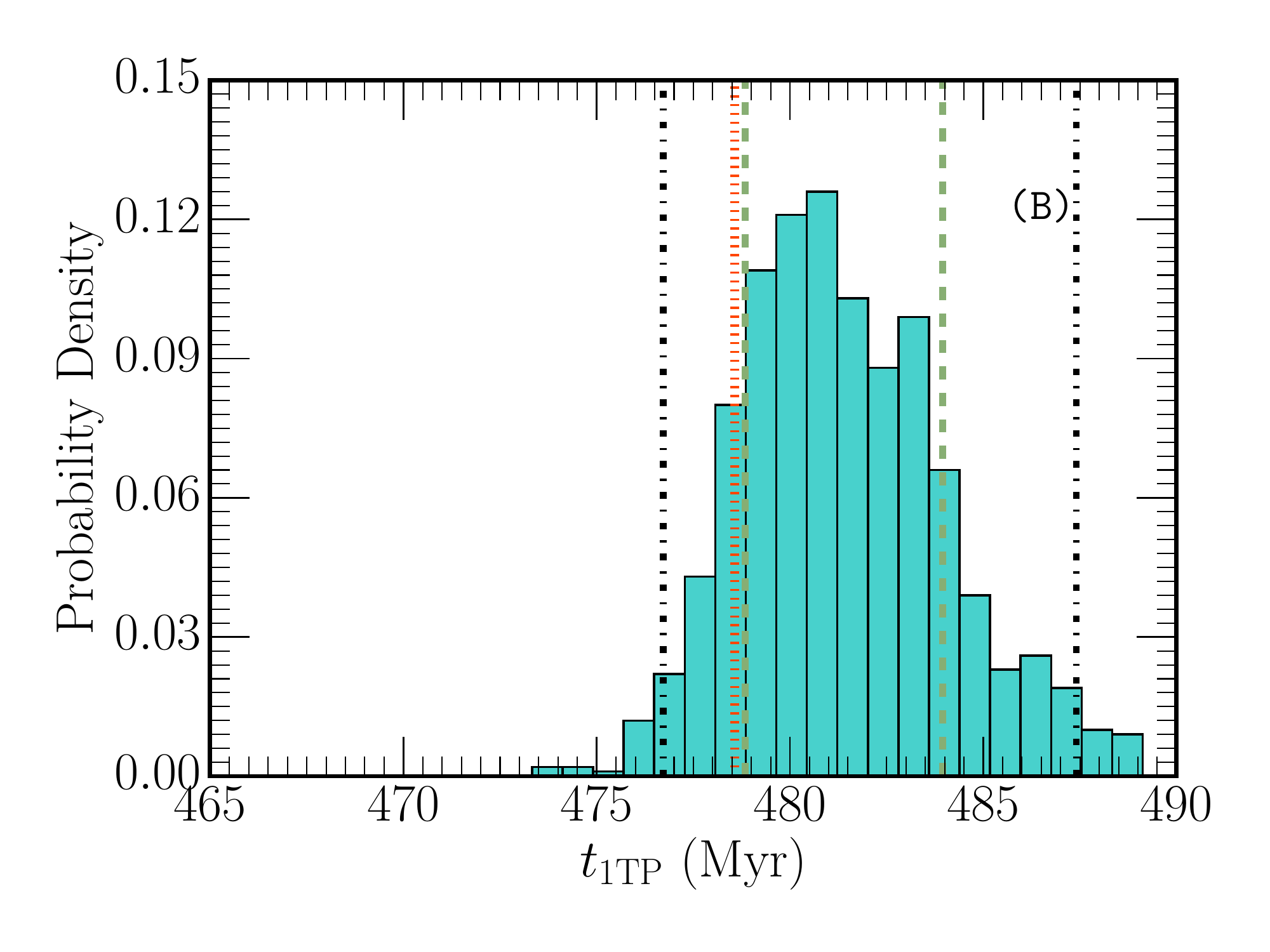}}
        \end{subfigure}          
        \begin{subfigure}{
                \includegraphics[trim = .25in .25in .25in .25in, clip,width=2.75in,height=2.1in]{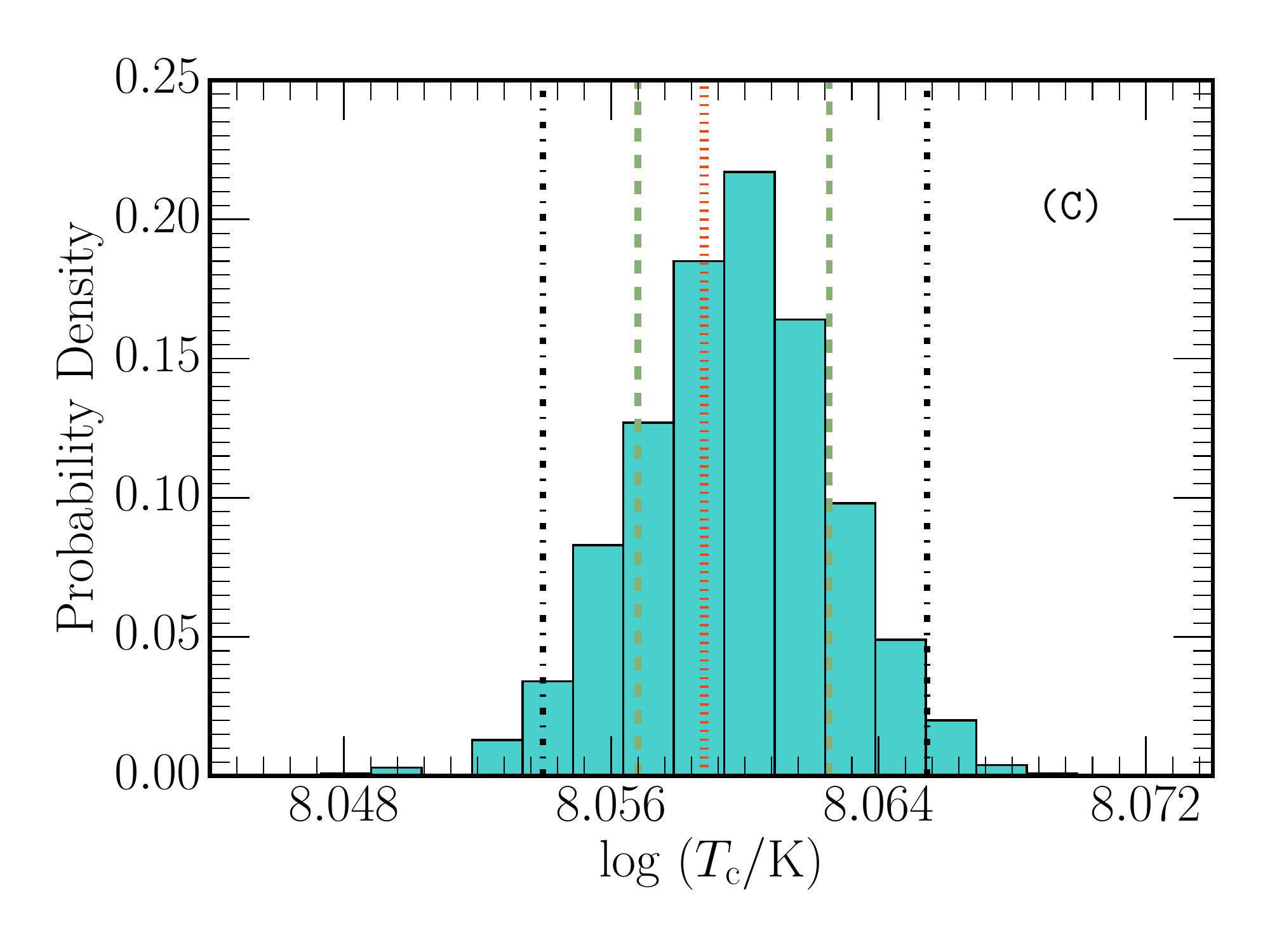}}
        \end{subfigure}
        \begin{subfigure}{
                \includegraphics[trim = .25in .25in .25in .25in, clip,width=2.75in,height=2.1in]{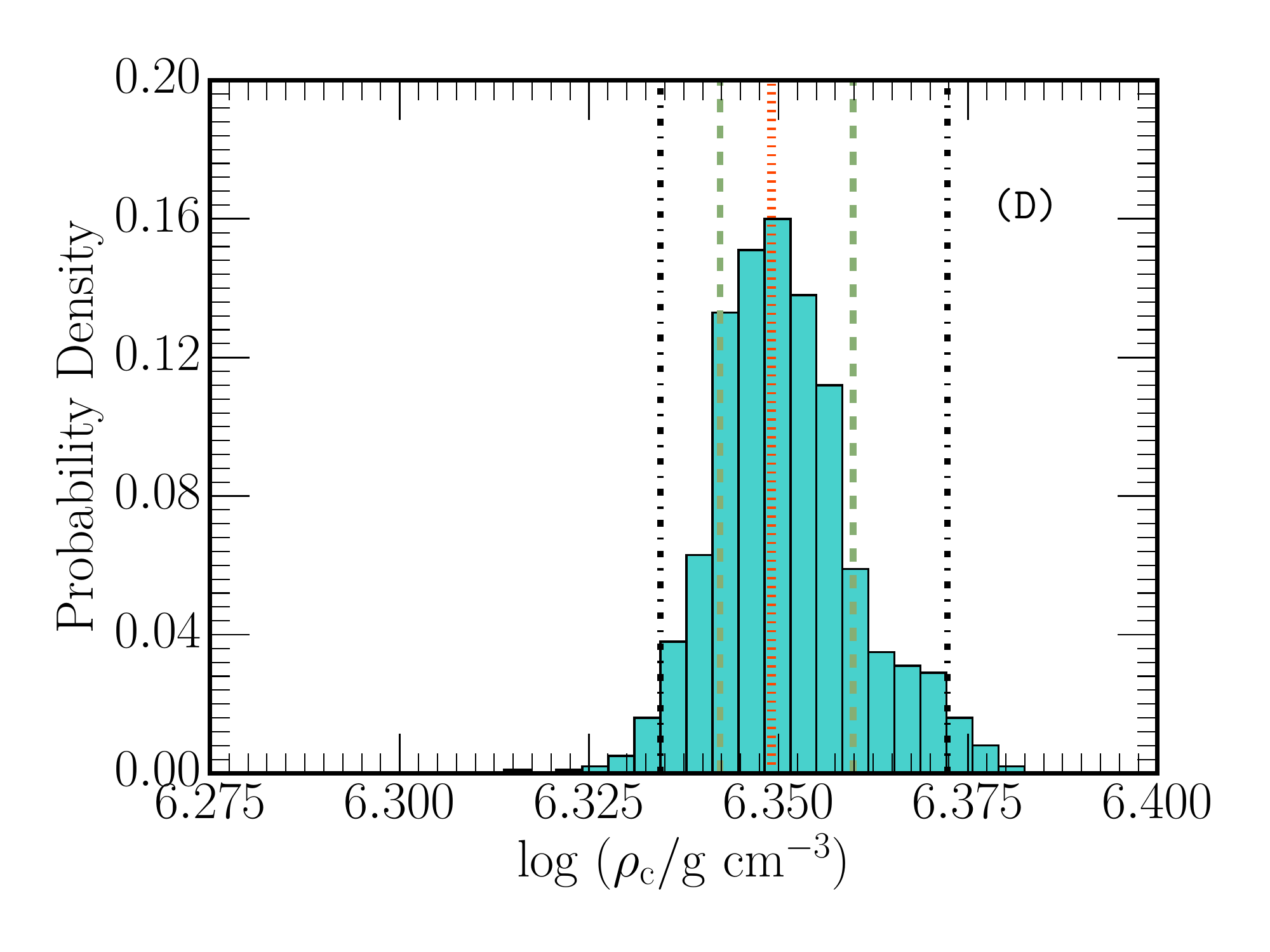}}
        \end{subfigure}
        \begin{subfigure}{
                \includegraphics[trim = .25in .25in .25in .25in, clip,width=2.75in,height=2.1in]{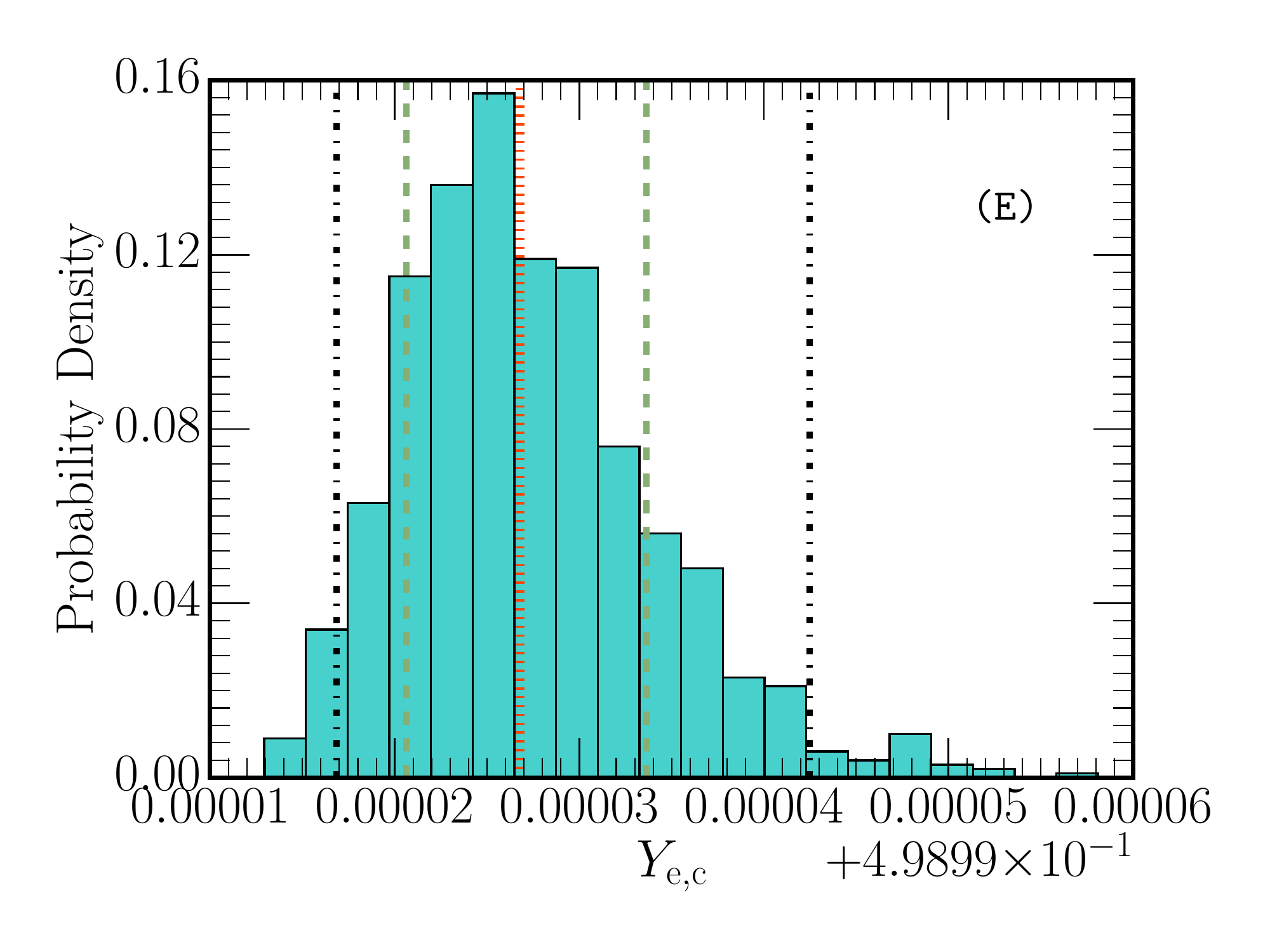}}
        \end{subfigure}
                \begin{subfigure}{
                \includegraphics[trim = .25in .25in .25in .25in, clip,width=2.75in,height=2.1in]{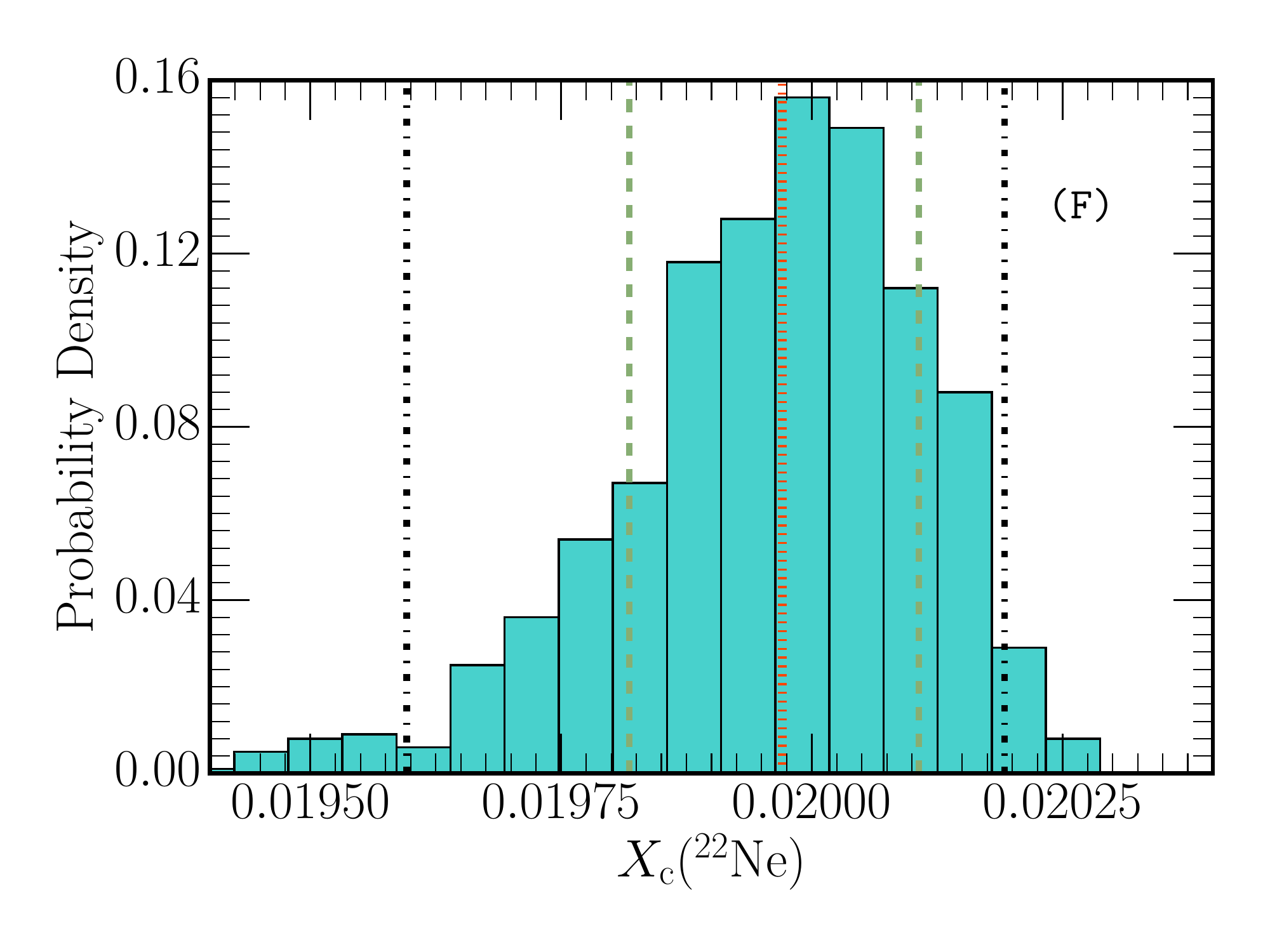}}
        \end{subfigure}
                        \begin{subfigure}{
                \includegraphics[trim = .25in .25in .25in .25in, clip,width=2.75in,height=2.1in]{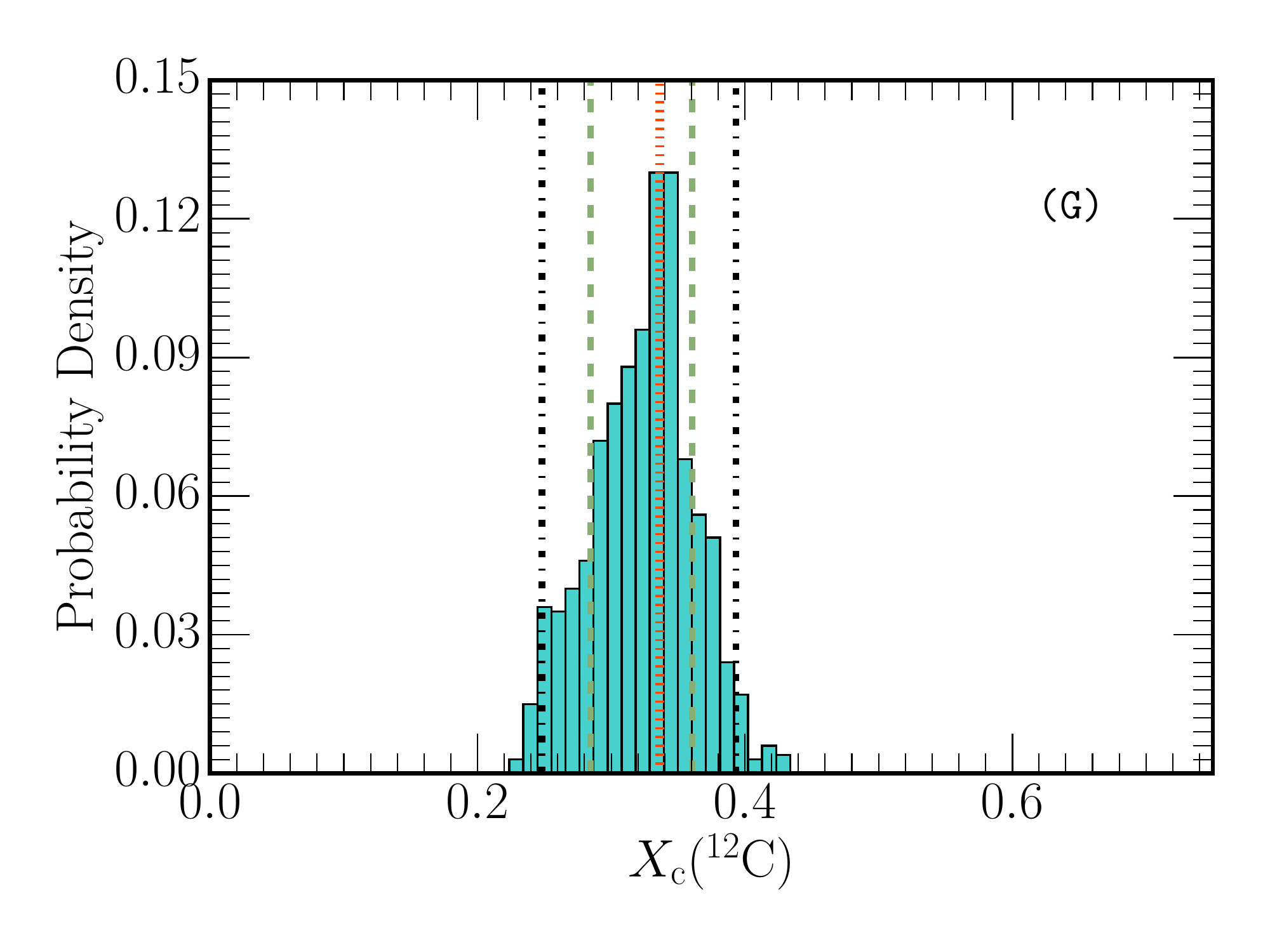}}
        \end{subfigure}
        \begin{subfigure}{
                \includegraphics[trim = .25in .25in .25in .25in, clip,width=2.75in,height=2.1in]{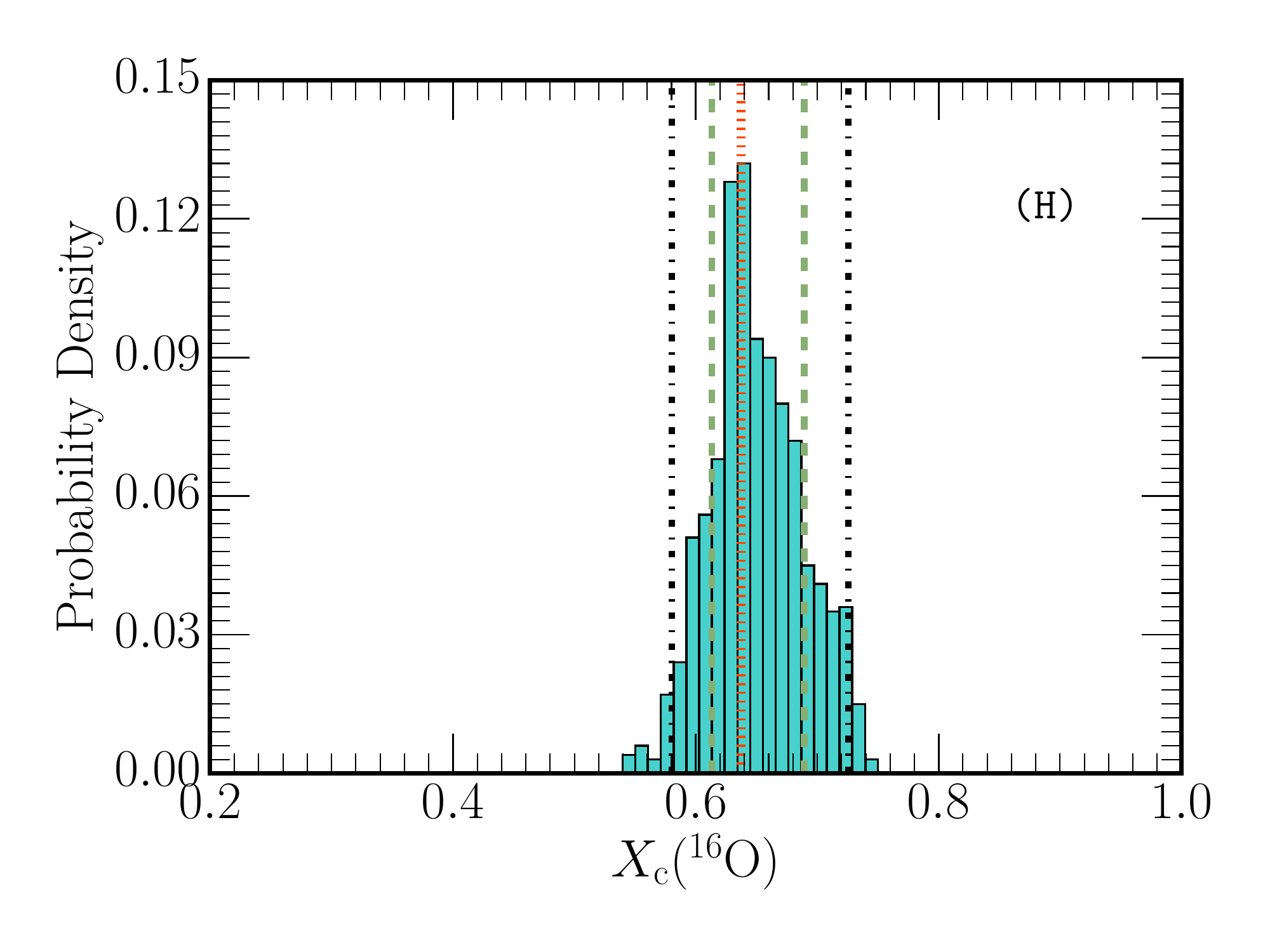}}
        \end{subfigure}
        \caption{
        Histogram of the 1,000 3 \msun Monte Carlo stellar 
        model grid sampling 25 STARLIB reaction rates. Global and
        core properties shown include 
        the mass of the CO core \texttt{(A)}, 
        age \texttt{(B)}, 
        central temperature \texttt{(C)}, 
        central density \texttt{(D)},  
        central electron fraction \texttt{(E)}, 
        central $^{22}$Ne mass fraction \texttt{(F)},
        central $^{12}$C mass fraction \texttt{(G)}, and
        central $^{16}$O mass fraction \texttt{(H)}, all at the 1TP. 
        The green dashed lines and the dot dashed red lines denote the 
        68\% and 95\% confidence intervals, respectively. The red dashed
        line denotes the mean value obtained from Table~\ref{tbl:mcsm_props_1} 
        for comparison. The x-axis ranges are the same as in Figure~\ref{fig:grid1_hist}.
        }\label{fig:grid2_hist}
\end{figure*}

Figure~\ref{fig:grid2_hist} shows histograms of the 
the mass of the CO core (A), 
age (B), 
\Tc \ (C), 
\rhoc \ (D),  
$Y_{\rm e,c}$ (E), 
X$_{\rm c}$($^{22}$Ne) (F),
X$_{\rm c}$($^{12}$C) (G), and
X$_{\rm c}$($^{16}$O) (H) at the 1TP. 
The green dashed lines and the dot-dashed black lines denote the 
same intervals as in Figure~\ref{fig:grid1_hist}. 
Relative to the arithmetic mean value, we find the width of the
95\% confidence interval to be
\mbox{$\Delta M_{{\rm 1TP}}$ $\approx$ 0.013 M$_{\odot}$} for the core mass,
$\Delta t_{{ \rm 1TP}} \approx$ 10.69 Myr for the age,
\mbox{$\Delta \log(T_{{\rm c}}/K) \approx$ 0.011} for the central temperature,
$\Delta \log(\rho_{{ \rm c}}/{\rm g cm}{-3}) \approx$ 0.038 for the central density,
$\Delta$$Y_{\rm{e,c}}$ $\approx$ 2.6$\times$10$^{-5}$ for the central electron fraction,
\mbox{$\Delta$$X_{\rm{c}}(^{22}\rm{Ne})$ $\approx$ 6.0$\times$10$^{-4}$}
\mbox{$\Delta$$X_{\rm{c}}(^{12}\rm{C})$ $\approx$ 0.145}, and
\mbox{$\Delta$$X_{\rm{c}}(^{16}\rm{O})$ $\approx$ 0.145}.
Table~\ref{tbl:mcsm_props_grid2} lists the arithmetic mean values
of the eight quantities shown in Figure~\ref{fig:grid2_hist}, the
width of the 68\% and 95\% confidence intervals, and the percentage change
from the arithmetic mean using the 95\% confidence interval. We use the same 
methods as in \S\ref{sec:mcstars} to compare the differences. 

\begin{deluxetable}{ccccc}{!htb}
\tablecolumns{5}
\tablewidth{0.95\linewidth}
\tablecaption{Variations in core quantities - Fixed $^{12}$C($\alpha,\gamma$) }
\tablehead{\colhead{Variable} & \colhead{Mean} & \colhead{68\% C.I.}  & \colhead{95\% C.I.}
& \colhead{\% Change} }
\startdata
$M_{\rm{1TP}}$ (M$_{\odot}$) 
& 5.850(-1) & 5.672(-3) & 1.315(-2)  &  2.248  \\

$t_{\rm{1TP}}$ (Myr) 
& 4.814(2) & 5.110 & 1.069(1) &  2.221   \\

log $T_{\rm{c}}$ (K) 
& 8.060 & 5.723(-3) & 1.150(-2) &  0.143 \\

log $\rho_{\rm{c}}$ (g cm$^{-3}$)
& 6.351 & 1.754(-2) & 3.790(-2) &  0.597  \\ 

$Y_{\rm{e,c}}$ 
& 4.990(-1) & 1.301(-5) & 2.562(-5) &  0.005  \\

X$_{\rm{c}}(^{22}$Ne)
& 1.996(-2) & 2.891(-4) & 5.961(-4) &  2.986 \\

X$_{\rm{c}}(^{12}$C) 
& 3.235(-1) & 7.597(-2) & 1.451(-1) &  44.85  \\

X$_{\rm{c}}(^{16}$O) 
& 6.501(-1) & 7.610(-2) & 1.452(-1) & 22.34
\enddata
\label{tbl:mcsm_props_grid2}
\end{deluxetable}

We compute the eigenvalues and eigenvectors of the
correlation matrix as in
\S\ref{sec:pca}. Here we consider 33 quantities in total $-$ 
the 25 STARLIB reaction rate variation factors
in the order listed in Table~\ref{tbl:sampled_rates},
$M_{{\rm 1TP}}$, 
$t_{\rm 1TP}$, 
$T_{\rm{c}}$,
\rhoc, 
$Y_{\rm e,c}$, 
X$_{\rm c}$($^{12}$C),
X$_{\rm c}$($^{16}$O), and 
X$_{\rm c}$($^{22}$Ne) at the 1TP.

\begin{deluxetable}{cccc}{b}
\tablecolumns{4}
\tablewidth{0.8\linewidth}
\tablecaption{Principal Component Analysis - Fixed $^{12}$C($\alpha,\gamma$)}
\tablehead{\colhead{Component} & \colhead{Eigenvalue} & \colhead{Proportion}  & \colhead{Cumulative}}
\startdata
1 & 4.782 & 0.1449 & 0.1449 \\
2 & 3.045 & 0.0923 & 0.2372 \\
3 & 1.802 & 0.0546 & 0.2918 \\
4 & 1.408 & 0.0427 & 0.3345 \\
5 & 1.246 & 0.0378 & 0.3722
\enddata
\label{tbl:pca_grid_2}
\end{deluxetable}

In Table~\ref{tbl:pca_grid_2} we show the eigenvalues, proportion of 
total variation, and cumulative proportion of total variation for the first 
five principal components. The first five components account for 
$\approx$ 37\% of the total variation with only $\approx$ 14\% accounted 
for by the first component. The amount of variation attributed to the first
component is less than that accounted for when the 
$^{12}$C($\alpha,\gamma$)$^{16}$O is included in the sampling scheme
(see Table~\ref{tbl:pca_grid_1}). With a fixed rate
distribution, we find less variation. The magnitude of the 
decrease explained by the first principal component
for this grid compared to that found in Table~\ref{tbl:pca_grid_1} further suggests
that the $^{12}$C($\alpha,\gamma$)$^{16}$O dominates  the WD composition 
and structural properties
properties. In the limit that all 33 physical quantities are perfectly uncorrelated, 
the amount of variation accounted for by the first principal component would 
approach zero. The contrary is true for a system of 33 quantities in which 
variation among all the quantities can be accounted for by one variable in the set.

The largest positive coefficients in the first principal component are the
$^{18}$O($\alpha,\gamma$)$^{22}$Ne rate variation factor with a value of +0.473,
$^{18}$O($p,\alpha$)$^{15}$N with +0.330, and X$_{\rm c}$($^{16}$O) with +0.308. The largest negative coefficients are -0.335 for
$^{18}$F($p,\alpha$)$^{15}$O and -0.326 for $^{15}$O($\alpha,\gamma$)$^{19}$Ne. 
For the second principal component, the largest positive coefficients are +0.413
for X$_{\rm c}$($^{12}$C), +0.412 for X$_{\rm c}$($^{16}$O) and +0.371 for 
$^{18}$F($p,\alpha$)$^{15}$O. The largest negative coefficients are
-0.214 for \rhoc\ and -0.156 for $^{16}$O($p,\gamma$)$^{17}$F.

We also consider a subset of the data and use a PCA 
to visualize the differences between the two grids. For example, we 
consider X$_{\rm c}$($^{12}$C) and the rate variation
factor for the triple-$\alpha$ reaction. Using the same steps
as in \S\ref{sec:pca} we construct a feature vector, $\textbf{w}$, with the obtained eigenvectors
as the columns. Next, we create a new matrix, $\textbf{x}$, containing our standardized
data, where a vector $a$ is transformed to a standardized vector $a'$, where 
$\textbf{a\textprime}=[\textbf{a}-\bar{a}]/\sigma_{a}$, where $\bar{a}$ is the arithmetic 
mean and $\sigma_{a}$ is the standard deviation of the vector $\textbf{a}$. 
We then compute the transformed data using 
\begin{equation}
\textbf{y} = \textbf{w}^{\rm{T}}\times \textbf{x}~.
\label{eq:lin_tran}
\end{equation}

\begin{figure}[!htb]        
\centering
\includegraphics[trim = .1in .1in 0in .1in, clip,width=3.4in,height=2.75in]{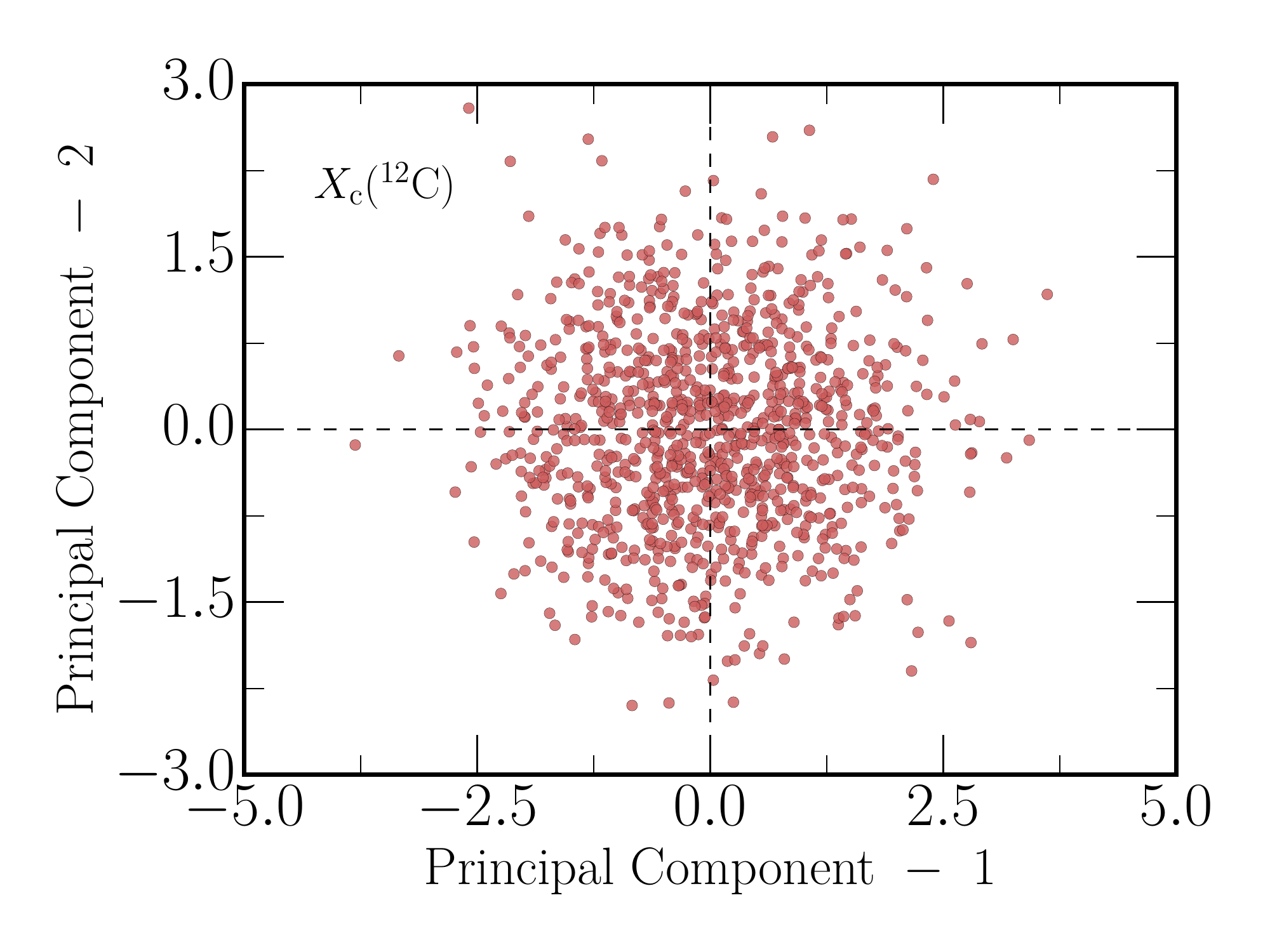}
\caption{
Transformed data for X$_{\rm c}$($^{12}$C) 
versus the rate variation factor for triple-$\alpha$ for the 
grid of 1,000 3 \msun Monte Carlo stellar models sampling 
26 STARLIB reaction rates shown along the first and
second principal components. The dashed lines
denote the eigenvectors of the correlation matrix.
}\label{fig:pca_proj}
\end{figure}

\begin{figure}[!htb]        
\centering
\includegraphics[trim = .1in .1in 0in .1in, clip,width=3.4in,height=2.75in]{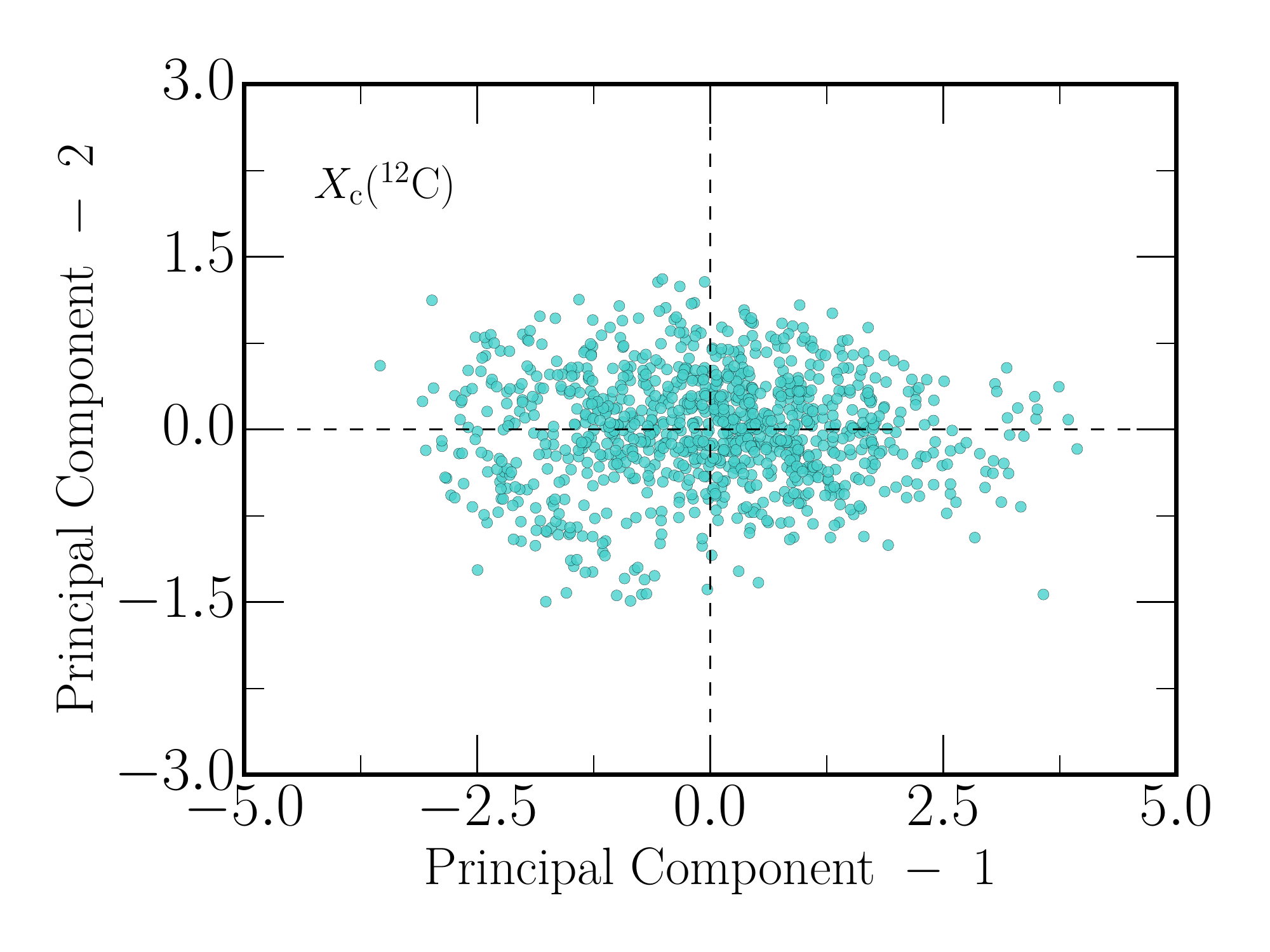}
\caption{
Same as in Figure~\ref{fig:pca_proj} but for the grid
of Monte Carlo stellar models sampling 25 STARLIB
reaction rates and fixed rate distribution for 
$^{12}$C($\alpha,\gamma$)$^{16}$O.
}\label{fig:pca_proj_fixed_c12_ag}
\end{figure}

\begin{figure*}[!ht]
\centering
\includegraphics[trim = .1in .1in .75in .5in, clip,width=6.5in,height=5in]{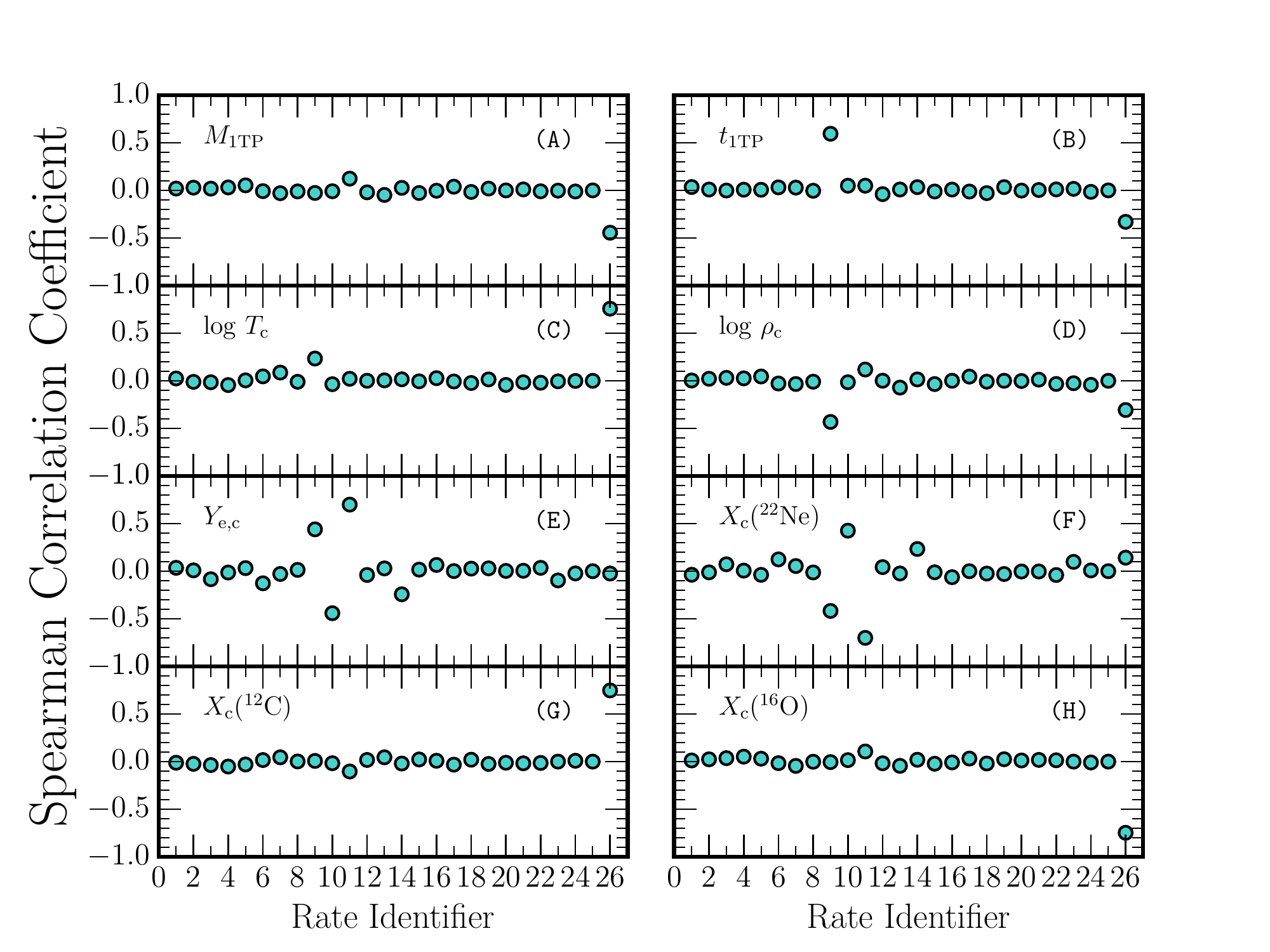}
        \caption{
        Spearman rank correlation coefficients for the 25 
        independently sampled nuclear reaction rates for 1,000 Monte
        Carlo stellar models with a ZAMS mass of 3 \msun. 
        The rate identifiers correspond to those listed in Table~\ref{tbl:sampled_rates}.
        Each sampled rate is compared individually 
        against the final mass of the CO core ($M_{{\rm 1TP}}$, \texttt{A}),
        age of stellar model ($t_{{\rm 1TP}}$, \texttt{B}),
        central temperature ($T_{{\rm c}}$, \texttt{C}),
        central density ($\rho_{{\rm c}}$, \texttt{D}),     
        central electron fraction ($T_{{\rm e,c}}$, \texttt{E}),   
        central $^{22}$Ne mass fraction (X$_{{\rm c}}$($^{22}$Ne), \texttt{F}),   
        central $^{12}$C  mass fraction (X$_{{\rm c}}$($^{12}$C), \texttt{G}), and  
        central $^{16}$O  mass fraction (X$_{{\rm c}}$($^{16}$O), \texttt{H})
        at the 1TP.
        }\label{fig:all_rhos_grid_2}
\end{figure*}

Figure~\ref{fig:pca_proj} shows the linear transformation for the rate
variation factor for the triple-$\alpha$ reaction and X$_{\rm c}$($^{12}$C) 
for the grid of stellar models that sampled 26 STARLIB rates (including the 
$^{12}$C($\alpha,\gamma$)$^{16}$O reaction). We find 
considerable variation along the first and second principal component. The 
proportion of variation explained by the first component shown is $\approx$
63\%.

Figure~\ref{fig:pca_proj_fixed_c12_ag} shows the linear transformation 
for the triple-$\alpha$ rate variation factor and X$_{\rm c}$($^{12}$C) 
for the grid of stellar models that sampled 25 STARLIB rates while using the
\emph{median} STARLIB rate distribution for $^{12}$C($\alpha,\gamma$)$^{16}$O.
In contrast to Figure~\ref{fig:pca_proj}, the first principal component holds the 
majority of the variation, $\approx$ 88\%. This result suggests that when the 
$^{12}$C($\alpha,\gamma$)$^{16}$O
reaction is considered in the sampling scheme, it can overpower triple-$\alpha$. 
Specifically, the first principal component accounts for 
only a $\approx$ 63\% proportion of variation between the triple-$\alpha$ reaction
and X$_{\rm c}$($^{12}$C) - see Figure~\ref{fig:pca_proj_fixed_c12_ag}.

\begin{figure*}[!htb]
\centering{\includegraphics[trim = .1in .3in 0in .1in, clip,width=1.4\columnwidth]{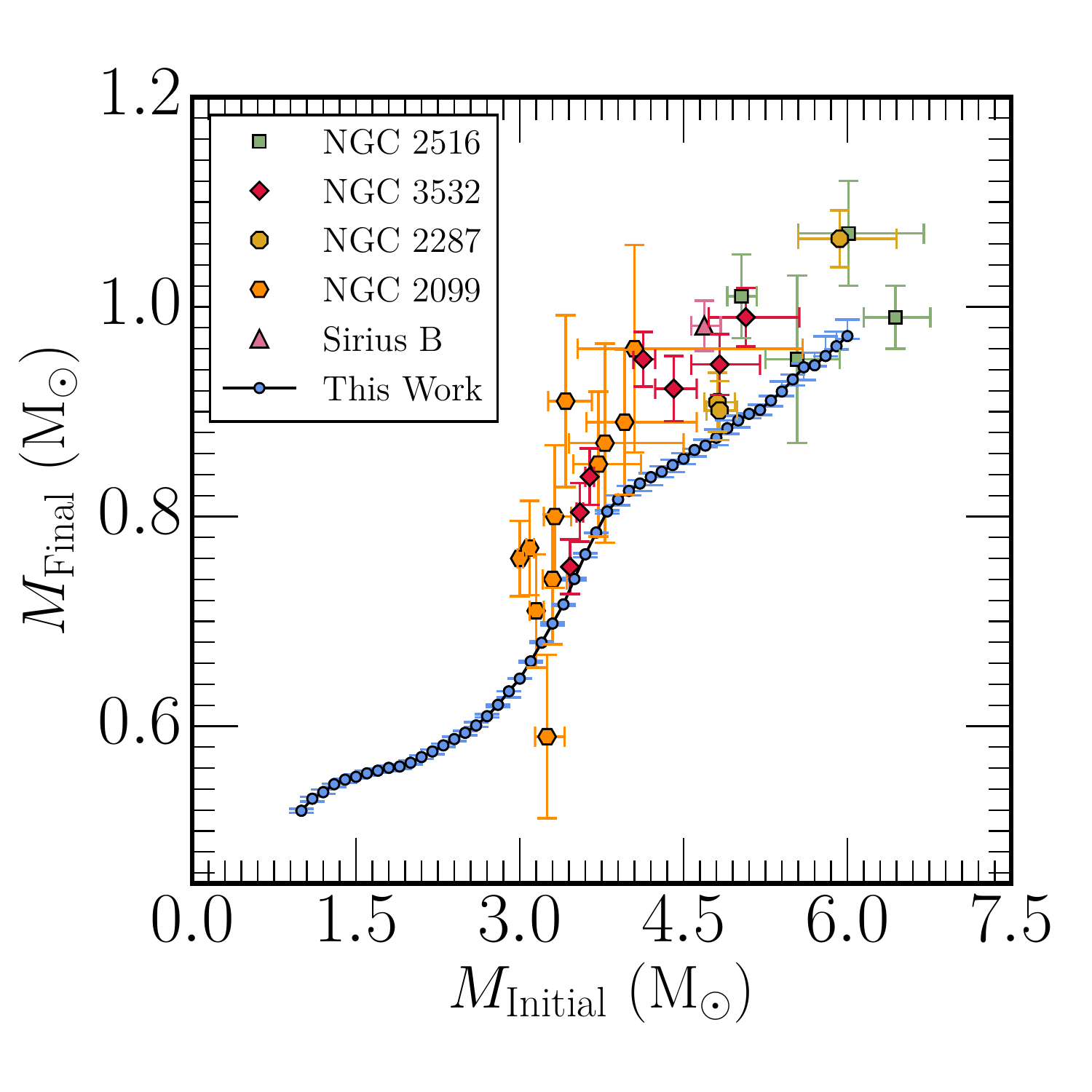}}
\caption{
IFMR of 153 stellar models evolved 
from the pre-MS to the final WD stage.
Initial masses ranged between 1-6 \msun in steps of 0.1 \msun utilizing the 
\emph{low}, \emph{median}, and \emph{high} reaction rate distributions for 
the 26 STARLIB sampled rates given in Table~\ref{tbl:sampled_rates}. Filled circles 
show the median final mass value, while the error bars denote the low and upper limits 
found from the remaining two models. Progenitor masses from the observations
were inferred from PARSEC isochrones. See \citet{bressan_2012_aa} and
\citet{dotter_2016_aa} for details.
}\label{fig:imfm}
\end{figure*}

\begin{figure}[!htb]
\centering{\includegraphics[trim = .1in .1in 0in .1in, clip,width=3.4in,height=2.75in]{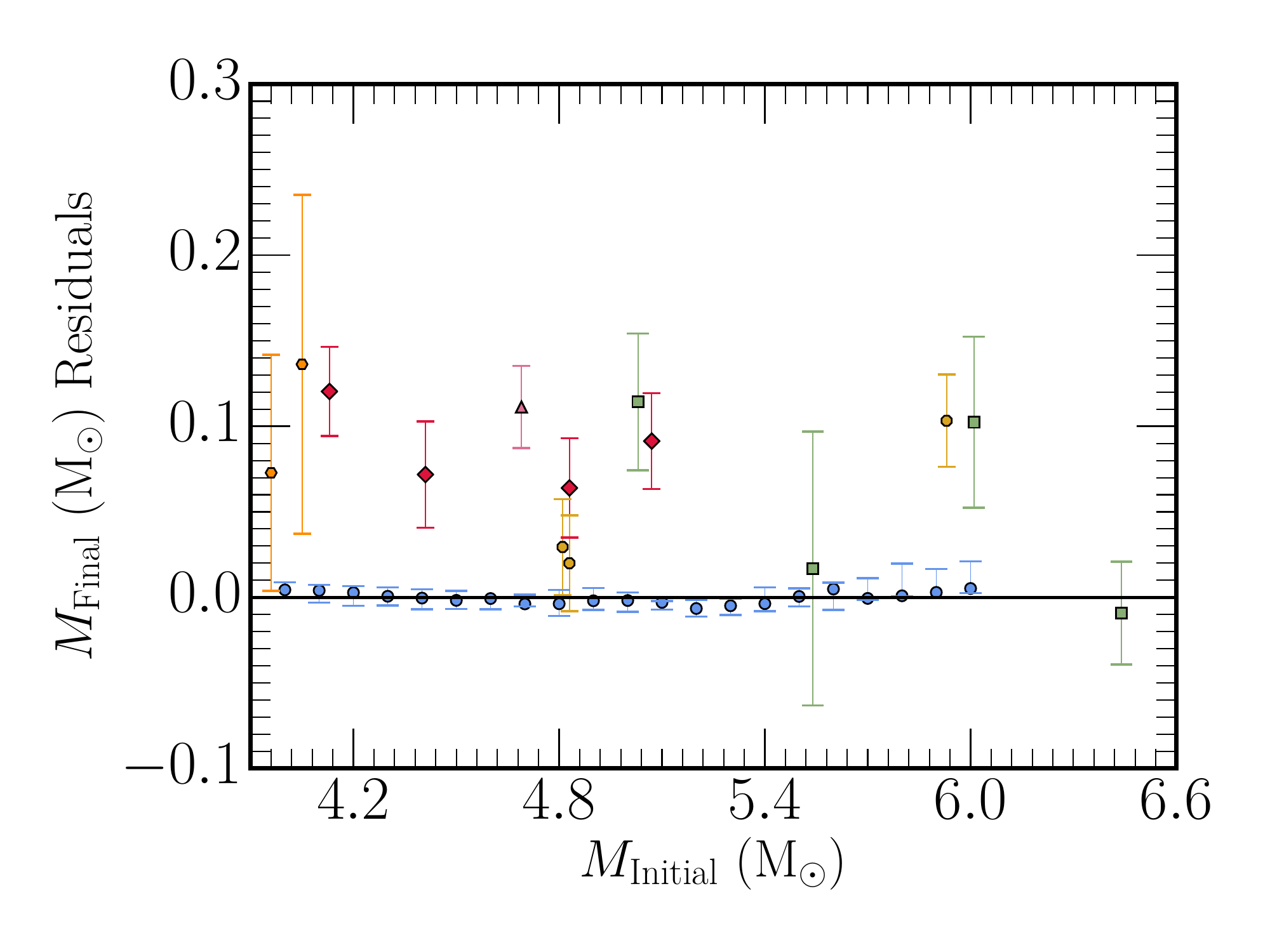}}
\caption{
IFMR residuals 
for 4.0 \msun $\lesssim$ $M_{i}\lesssim $ 6.6 \msun.
The legend is the same as in Figure~\ref{fig:imfm}.
}\label{fig:imfm_resid}
\end{figure}

In Figure~\ref{fig:all_rhos_grid_2}, we show the Spearman coefficients
against the 25 independently sampled rate variation factors for
$M_{{\rm 1TP}}$,
$t_{{\rm 1TP}}$,
$T_{{\rm c}}$,
$\rho_{{\rm c}}$,
$Y_{\rm e,c}$,
X$_{\rm c}$($^{22}$Ne),
X$_{\rm c}$($^{12}$C), and  
X$_{\rm c}$($^{16}$O), 
at the 1TP.
In the previous section, $M_{{\rm 1TP}}$ was dominated by 
the uncertainties in the $^{12}\textup{C}(\alpha,\gamma)\rm{^{16}O}$ and 
triple-$\alpha$ reaction. Here, the mass of the CO core at the 
1TP is still affected by the triple-$\alpha$ reaction with a Spearman coefficient of 
\rhos=-0.45. However, we also find that the $^{15}\textup{N}(p,\gamma)\rm{^{16}O}$,
has a larger effect in the absence of the
$^{12}\textup{C}(\alpha,\gamma)\rm{^{16}O}$ reaction, with a coefficient of \rhos=+0.12.

For $t_{{\rm 1TP}}$, it is still dominated
by the uncertainties in $^{14}\textup{N}(p,\gamma)\rm{^{15}O}$ and triple-$\alpha$, 
but for a fixed $^{12}\textup{C}(\alpha,\gamma)\rm{^{16}O}$ distribution 
we find increases in Spearman coefficients \rhos=(+0.60,-0.33), 
an increase of \mbox{$\approx$ 22\%} and $\approx$ 43\%, respectively.

$T_{{\rm c}}$ is again most affected by 
$^{14}\textup{N}(p,\gamma)\rm{^{15}O}$ and triple-$\alpha$ with
\rhos=(0.18,+0.76) respectively. 
Previously unidentified, $\rho_{{\rm c}}$ is correlated 
with $^{14}\textup{N}(p,\gamma)\rm{^{15}O}$
with \rhos=-0.43. Additionally, we see a dependence 
triple-$\alpha$ with \mbox{\rhos=-0.31}.
$Y_{{\rm e,c}}$, was dominated by individual reactions
of the CNO cycle in the previous section. Here we find similar results with 
\mbox{\rhos = (+0.44,-0.44,+0.70)} corresponding to $^{14}\textup{N}(p,\gamma)\rm{^{15}O}$,
$^{15}\textup{N}(p,\alpha)\rm{^{12}C}$, and $^{15}\textup{N}(p,\gamma)\rm{^{16}O}$, respectively.
X$_{\rm c}$($^{22}$Ne) is again anti-correlated with $Y_{{\rm e,c}}$
with similar coefficients  
of \mbox{\rhos=(-0.42,+0.43,-0.70)}, corresponding to the same CNO reactions.

The central X$_{\rm c}$($^{12}$C) and X$_{\rm c}$($^{16}$O) show the largest change for 
a fixed $^{12}\textup{C}(\alpha,\gamma)\rm{^{16}O}$. Previously,
we found large Spearman coefficients between the $^{12}\textup{C}(\alpha,\gamma)\rm{^{16}O}$
reaction and X$_{\rm c}$($^{12}$C) and X$_{\rm c}$($^{16}$O) with \rhos =(-0.91, +0.91), 
respectively. This suggested
X$_{\rm c}$($^{12}$C) and X$_{\rm c}$($^{16}$O) could be mostly determined by one reaction. However,
for a fixed $^{12}\textup{C}(\alpha,\gamma)\rm{^{16}O}$ rate distribution, 
the triple-$\alpha$ reaction dominates the variation. For 
X$_{\rm c}$($^{12}$C) and X$_{\rm c}$($^{16}$O) and triple-$\alpha$ we find \rhos = (+0.75,-0.75), respectively.
This represents an approximately 300\% increase in magnitude of the Spearman 
coefficients found in the first grid sampling the 26 STARLIB rates. Moreover, a value
of \rhos=0.1 is found for the rate variation factor for $^{15}\textup{N}(p,\gamma)\rm{^{16}O}$
and X$_{\rm c}$($^{12}$C) and \rhos=-0.1 for $^{15}\textup{N}(p,\gamma)\rm{^{16}O}$
and X$_{\rm c}$($^{16}$O). This suggests that when neglecting the dominant effect 
of the main radiative alpha capture reaction, $^{12}\textup{C}(\alpha,\gamma)\rm{^{16}O}$, 
the proton capture rate, $^{15}\textup{N}(p,\gamma)\rm{^{16}O}$ can have an effect on 
the final $^{12}$C and $^{16}$O abundances.

\section{Initial-Final Mass Relationship}
\label{sec:imfm}
The question then naturally arises: what is the impact of the uncertainties
in the nuclear reaction rates on the IFMR?
To address this question, 
we evolved a third grid of \MESA models from the pre-MS through the 
thermally pulsing AGB phase to the final formation of the CO WD. Our 
grid consists of 153 models from 1-6 \msun in steps of 0.1 \msun 
using the STARLIB \emph{low}, \emph{med}, and \emph{high} rate 
distributions for the 26 sampled rates, for a total of 3 models at each 
mass point. Our goal is to quantify the effect of composite uncertainties
in thermonuclear reaction rates on the properties of the final CO WD.
The models are evolved from the pre-MS until the mass 
of the H-rich envelope has reached a minimum value of 
$M_{\rm{env}} = 0.01$ \msun. The grid 
uses the same input physics as in \S\ref{sec:method}, 
with the exception that we increase the efficiency of mass loss on 
the RBG and AGB to $\eta$=0.7. This
increase in mass loss efficiency allows our grid to 
evolve to the our specified \MESA stopping condition in a computationally 
efficient manner. Moreover, increased efficiency in the mass loss has been shown 
to alter characteristics of the TP phase, but yield only modest affects 
on the final WD \citep{andrews_2015_aa}. Each model 
required $\approx$ 96 hours on 12 cores for each of 
the 153 models to go from the pre-MS to the final WD, 
a total of $\approx$ 176,000 core-hours.

In Figure~\ref{fig:imfm} we show the IFMR resulting from our 
grid of stellar models. The filled circles shown in Figure~\ref{fig:imfm}, are the 
median STARLIB values, while the lower and upper STARLIB limits are denoted by the blue
error bars. We compare our models to observational data for the solar metallicity 
cluster NGC 2516 \citep{catalan_2008_aa},
NGC 2099 \citep{cummings_2015_aa},
NGC 3532,
NGC 2287, 
and Sirius B (assuming solar metallicity) from
\citet{cummings_2016_aa}. The observational data 
focuses on WD candidates with inferred solar metallicity. However, the 
IFMR has been shown to depend significantly on the initial progenitor 
metallicity, shifting the trend towards higher final WD masses 
\citep{meng_2008_aa,romero_2015_aa}. The observational data for inferred initial progenitor
WD masses below 3 \msun have been excluded from this comparison as
the majority these WDs are from supersolar 
metallicity clusters such as NGC 6791 or the host environments remain largely uncertain 
\citep[and references therein]{kalirai_2008_aa,zhao_2012_aa}. 

In most cases, the most massive WD model for a chosen initial mass has the \emph{low} reaction rate 
distributions. For decreased nuclear burning efficiency, the 
He burning shell has a prolonged lifetime allowing for a more massive CO core to 
grow. For all three models at each mass point, the final CO masses
agree within $\approx$ $\pm$ 0.003 \msun. 

A noticeable feature of Figure~\ref{fig:imfm} is that the 
mean trend of our models tend to lie
below the observational trend. This discrepancy is likely due to our choices
for the mixing parameters. The largest assumption is probably the
efficiency of convective overshoot being uniform at all boundary
layers. To address this discrepancy, we evolved an additional 3 \msun model
without convective overshoot, $f_{\rm{ov}}=0.00$ at all boundaries, using 
the \emph{med} STARLIB rates. Our 3 \msun model with $f_{\rm{ov}}=0.016$ 
resulted in a final CO WD mass of $\approx$ 0.645 \msun. However, we 
found that the model without convective overshoot yielded a final CO WD 
mass of $\approx$ 0.684 \msun. This suggests that the discrepancy between
the mean trend of our IFMR and the observational data is a result of our of choices
for efficiency of convective overshoot.

Following \citet{andrews_2015_aa}, we fit our IFMR in three distinct regimes: 
$M_{i}\lesssim $ 2.5 \msun experience a degenerate He shell flash 
\citep{sweigart_1978_aa,suda_2011_aa},
2.5 $M_{i}\lesssim $ 4 \msun, undergo stable He core burning, while for 
$M_{i} \ge $ 4 \msun the second dredge-up becomes important 
\citep{dominguez_1999_aa,herwig_2000_aa}.
We construct a piece-wise linear function to the IFMR 
with turnover points at 2.5 \msun and 4 \msun : 
\begin{equation}
\begin{split}
M_{\rm{f}}/M_{\odot} = & (0.0422\pm0.002)M_{\rm{i}} + (0.4851\pm0.004) \\
                       & \qquad {\rm for} \ M_{\rm{i}} \lesssim 2.5 \msun \\
                     = & (0.1686\pm0.006)M_{\rm{i}} + (0.1516\pm0.019) \\
                       & \qquad  {\rm for} \ 2.5 \lesssim M_{\rm{i}} \lesssim 4.0 \msun \\
                     = &(0.0734\pm0.001)M_{\rm{i}} + (0.5265\pm0.006) \\
                       & \qquad  {\rm for} \ 4.0 \lesssim M_{\rm{i}} \lesssim 7.0 \msun 
\label{eq:imfm_high}
\end{split}
\end{equation}

Our 7.0 \msun upper limit is based on estimates of the lowest mass for
carbon ignition \citep[][and references therein]{farmer_2015_aa}. 
Figure~\ref{fig:imfm_resid} shows the final mass residuals using
Equation~(\ref{eq:imfm_high}).  The
upper and lower WD masses derived from the \mesa models agree
to \mbox{$\approx$ $\pm 0.008$ \msun}, while the observations agree to
within $\approx$ $\pm 0.12$ \msun.
\vfill

\section{Summary and Conclusions}
We have investigated the properties of CO cores evolved from the
pre-MS using the first complete Monte Carlo stellar models. We evolved
two grids of stellar models: (1) 1,000 3 \msun Monte Carlo stellar
models sampling 26 STARLIB nuclear reaction rates from pre-MS to the
1TP, (2) 1,000 3 \msun Monte Carlo stellar models using the
\emph{median} STARLIB rate distribution for $^{12}$C($\alpha,\gamma$)$^{16}$O 
while sampling the remaining 25 STARLIB rates from pre-MS to the 1TP.
We used a 
Principal Component Analysis and Spearman Rank-Order Correlation
to quantify the variation of the 
mass of the CO core, 
age, 
central temperature, 
central density,  
central electron fraction, 
central $^{22}$Ne mass fraction,
central $^{12}$C mass fraction, and
central $^{16}$O mass fraction, all at the 1TP.

When sampling 26 STARLIB reaction rates (including the 
$^{12}$C($\alpha,\gamma$)$^{16}$O reaction), we found that 
the experimental uncertainties in the $^{12}$C($\alpha,\gamma)^{16}\rm{O}$, 
triple-$\alpha$, and $^{14}$N($p,\gamma)^{15}\rm{O}$ reactions dominated the
properties of the stellar model. The largest changes were found for the 
$^{12}$C and $^{16}$O mass fractions. In particular, we found a percent 
change of $\approx$ 116\% and 
$\approx$ 62\% from the arithmetic mean value using the 95\% confidence
interval for the $^{12}$C and $^{16}$O mass fractions, respectively. 
The remaining six quantities had percent changes $\lesssim$ 4\%.
This result suggests that the relative abundances $^{12}$C and $^{16}$O can 
can vary significantly within experimental uncertainties, while other quantities
are well constrained.

Using a fixed rate distribution for the $^{12}$C($\alpha,\gamma$)$^{16}$O reaction
and sampling the remaining 25 STARLIB rates, we found that the triple-$\alpha$ and
$^{14}$N($p,\gamma)^{15}\rm{O}$ reactions still play a significant role. However, 
additional rates such as the $^{15}\textup{N}(p,\gamma)\rm{^{16}O}$ reaction were 
found to have non-negligible effects on the final core mass, $X_{\rm{c}}(^{12}$C) and 
$X_{\rm{c}}$$(^{16}$O). Moreover, we 
$\approx$ 45\% and $\approx$ 22\% changes from the arithmetic mean value using 
the 95\% confidence interval for the $^{12}$C and $^{16}$O mass fractions, 
respectively. The remaining six quantities had percent changes $\lesssim$ 3\%.
This suggests that the results of our Monte Carlo
stellar evolution studies is dependent on the reaction rates considered in
the sampling scheme. Additionally, our results suggest that the experimental 
uncertainties in the $^{12}$C($\alpha,\gamma$)$^{16}$O reaction has a
larger impact than triple-$\alpha$ on the final CO WD chemical composition.

To quantify the role of uncertainties in the nuclear reaction rates on the 
the IFMR, we evolved a third grid of \MESA models from the pre-MS through the 
thermally pulsing AGB phase to the final formation of the CO WD. Our 
grid consisted of 153 models from 1-6 \msun in steps of 0.1 \msun 
using the STARLIB \emph{low}, \emph{med}, and \emph{high} rate 
distributions for the 26 sampled rates, for a total of 3 models at each 
mass point. We showed that the final WD masses using each set of
reactions agreed to within $\approx$ $\pm$ 0.003 \msun. This result
suggests that uncertainties in the final mass relative to the nuclear reaction
rates are well constrained. 

The mean trend of our IFMR stellar models was found to lie below that of the observational trend. 
Differences were found to be attributed to the efficiency of different mixing processes 
in our models. \citet{salaris_2009_aa} found that the inclusion of
convective overshoot on the MS is needed to for agreement with
observational constraints. However, too efficient convective overshoot
during the AGB phase can also inhibit the growth of the CO core during
the TP-AGB phase.  3D hydrodynamic simulations of He-shell flash
convection suggests a more modest core overshoot value during the AGB
phase \citep{herwig_2006_aa}. We found that evolving a 3 \msun model 
with the \emph{median} STARLIB rates, but without convective 
overshoot yielded a final CO WD mass of $\approx$ 0.684 \msun while
our model with $f_{\rm{ov}}=0.016$ had a final mass of 
\mbox{$\approx$ 0.645 \msun}. Future efforts could include other parameters in the Monte 
Carlo sampling scheme, such as the efficiency of convective overshoot at different 
boundaries, to identify the largest areas of uncertainty in the evolution and formation 
of CO WDs.

\acknowledgements 
We thank Bill Paxton, Josiah Schwab, Pablo Marchant, Rich Townsend, Matteo Cantiello,
and Lars Bildsten for engaging conversations about the \MESA project. 
We also thank Michael Wiescher for initial discussions about this paper.
This project was supported by NASA under the
{\it Theoretical and Computational Astrophysics Networks} grant NNX14AB53G
and the {\it Astrophysics Theory Program} grant 14-ATP14-0007, and 
by NSF under the 
{\it Software Infrastructure for Sustained Innovation} grant 1339600, and 
grant PHY 08-022648 for the
{\it Physics Frontier Center} ``Joint Institute for Nuclear Astrophysics - Center for 
the Evolution of the Elements'' (JINA-CEE).
C.E.F. acknowledges support from Arizona State University
under a 2015 CLAS Undergraduate Summer Enrichment Award, 
a 2015-2016 Norm Perrill Origins Project Undergraduate Scholarship, and 
a ASU/NASA Space Grant award. 
F.X.T acknowledges sabbatical support from the Simons Foundation.

\bibliographystyle{aasjournal}

\bibliography{co_core}

\end{document}